\documentclass[twocolumn]{aastex701}
\usepackage{graphicx} 
\usepackage{subfig}
\usepackage{amsmath}
\usepackage{fixme}

\makeatletter

\newcommand{\overleftrightsmallarrow}{\mathpalette{\overarrowsmall@\leftrightarrowfill@}}
\newcommand{\overrightsmallarrow}{\mathpalette{\overarrowsmall@\rightarrowfill@}}
\newcommand{\overleftsmallarrow}{\mathpalette{\overarrowsmall@\leftarrowfill@}}
\newcommand{\overarrowsmall@}[3]{%
  \vbox{%
    \ialign{%
      ##\crcr
      #1{\smaller@style{#2}}\crcr
      \noalign{\nointerlineskip}%
      $\m@th\hfil#2#3\hfil$\crcr
    }%
  }%
}
\def\smaller@style#1{%
  \ifx#1\displaystyle\scriptstyle\else
    \ifx#1\textstyle\scriptstyle\else
      \scriptscriptstyle
    \fi
  \fi
}
\makeatother
\newcommand{\te}[1]{\overleftrightsmallarrow{#1}}

\fxsetup{
  status=draft,
  layout=inline,nomargin,nofootnote,
  theme=colorsig,
}
\FXRegisterAuthor{rh}{arh}{RH}
\FXRegisterAuthor{eo}{aeo}{EO}
\FXRegisterAuthor{la}{ala}{LA}
\FXRegisterAuthor{ck}{ack}{CK}

\graphicspath{{./}{figures/}{tables/}}

\begin{document}

\title{Dynamically Controlled Transport of GeV Cosmic Rays in Diverse Galactic Environments}
\author[0000-0002-7047-3730]{Ronan N. Hix}
\email{ronanhix@princeton.edu}
\affiliation{Department of Astrophysical Sciences, Princeton University, Princeton, NJ 08544, USA}
\author[0000-0002-5708-1927]{Lucia Armillotta}
\email{lucia.armillotta@inaf.it}
\affiliation{INAF Arcetri Astrophysical Observatory, Largo Enrico Fermi 5, Firenze, 50125, Italy}
\affiliation{Department of Astrophysical Sciences, Princeton University, Princeton, NJ 08544, USA}
\author[0000-0002-0509-9113]{Eve Ostriker}
\email{eco@astro.princeton.edu}
\affiliation{Department of Astrophysical Sciences, Princeton University, Princeton, NJ 08544, USA}
\author[0000-0003-2896-3725]{Chang-Goo Kim}
\email{cgkim@astro.princeton.edu}
\affiliation{Department of Astrophysical Sciences, Princeton University, Princeton, NJ 08544, USA}

\begin{abstract}
    We study transport of GeV cosmic rays (CRs) in a set of high-resolution TIGRESS magnetohydrodynamic simulations of the star-forming interstellar medium (ISM). Our local disk patch models sample a wide range of gas surface densities, gravitational potentials, and star formation rates (SFRs), and include a spiral arm simulation. Our approach incorporates CR advection by the background gas, streaming along the magnetic field limited by the local ion Alfv\'en speed, and diffusion relative to the Alfv\'en wave frame, with the diffusion coefficient set by the balance between streaming-driven Alfv\'en wave excitation and damping mediated by local gas properties. We find that dynamical transport mechanisms (streaming and advection) are almost solely responsible for GeV CR transport in the extra-planar regions of galaxies, while diffusion along the magnetic field dominates within the primarily-neutral ISM of galactic disks. We develop a simple 1D predictive model for the CR pressure $P_\mathrm{c}$, dependent only on injected CR flux and gas parameters. We demonstrate that the CR transport efficiency increases with increasing SFR, and provide a fit for the CR feedback yield $\Upsilon_\mathrm{c}~\equiv~P_\mathrm{c}/\Sigma_\mathrm{SFR}$ as a function of $\Sigma_\mathrm{SFR}$, the SFR surface density. We analyze lateral CR transport within the galactic disk, showing that CRs propagate away from feedback regions in spiral arms into interarm regions by a combination of gas advection and field-aligned transport. Lastly, we develop an empirical subgrid model for the CR scattering rate that captures the impacts of the multiphase ISM on CR transport without the numerical burden of full simulations.
\end{abstract}

\listoffixmes{}

\section{Introduction}\label{sec:intro}
Cosmic Rays (CRs) are relativistic charged particles, believed to be produced by shock acceleration in supernova remnants (SNR)\citep[e.g.][]{Blanford1978,Bell2004,Morlino2012}. While CRs are tied to the magnetic field, their quasi-collisionless nature allows them to spread throughout the interstellar medium (ISM), following field lines. Observations indicate that the kinetic energy spectrum of CR protons (the dominant CR species) is a broken power law that spans more than 10 orders of magnitude \citep{Grenier2015,Cummings2016}. The energy density of CRs with energy near 1 GeV in the local ISM is found to be $\sim 1~\mathrm{eV~cm^{-3}}$, comparable to the thermal, turbulent, and magnetic energy densities \cite{Boulares1990,Beck2001}. For this reason, it is believed that CRs can play an important dynamical role in the ISM, particularly as drivers of large-scale galactic winds \citep[e.g.][]{Recchia2020,Hanasz2021}. Recent work even suggests that strong CR feedback could have significant cosmological impacts in the outskirts of massive halos \citep[e.g.][]{Ji2020,Quataert2025,Lu2025}. \

CRs are also thought to contribute to the heating and ionization of both the ISM and the circumgalactic medium \citep[e.g.][]{Wiener2019, Kempski2020}. In particular, sub-GeV CRs are a crucial source of heating and ionization in dense, star-forming ISM gas, otherwise shielded from far ultra-violet (FUV) photons \citep[e.g.][]{Padovani2020}. \

The impact of CRs on the dynamical and thermal state of the gas depends on their spatial distribution, which is governed by the nature of CR propagation. However, the physics of CR transport is a deeply complex topic which is still not fully understood \citep[see review by][]{Amato2018}. Coupling between CRs and gas is primarily mediated by magnetic field interactions. As CRs stream down magnetic field lines, small fluctuations in the field can scatter them (this process is most effective for magnetic perturbations of the scale of CR gyroradius). This scattering allows the exchange of momentum and energy between the CR population and thermal gas, thereby coupling the two, while also limiting the effective speed of CR propagation along the magnetic field. Possible sources of these magnetic fluctuations are the topic of significant discussion, but it is likely that multiple processes play a role, with their relative contributions sensitive to CR energy. It is believed that CRs with energies below a few tens of GeV are primarily ``self-confined,'' whereby the CRs themselves induce (via the streaming instability) Alfv\'en waves off of which they scatter \citep[][]{Kulsrud1969,Wentzel1974}. For CRs with ultra-GeV energies, the magnetic perturbations are instead believed to be primarily produced by an extrinsic turbulent cascade \citep[e.g.][]{Kulsrud1971,Yan2002}. Nevertheless, fully realistic simulations of the multiphase, magnetized ISM with sufficient resolution to thoroughly test these mechanisms remain, for the moment, elusive.\

The mean free path for CRs is typically much smaller than the spatial scales of interest for the ISM, allowing a fluid treatment for CRs. Under this paradigm, CRs are treated as an additional fluid within a magnetohydrodynamic (MHD) simulation, with unresolved CR-wave interactions parameterized by use of a scattering (or diffusion) coefficient. Often, a constant value for this coefficient is adopted within a given simulation, with a value selected based on  observational constraints \citep[e.g.][]{Ruszkowski+23}. However, this approach ignores the inherently multiphase nature of the ISM, and it thereby fails to capture important aspects of CR transport, particularly for scales and regimes with significant variation in gas properties. In order to better model the sensitivity of CRs to the ISM environment, we have developed a more detailed physical prescription for CR transport, first presented in \cite{Lucia2021}. 
 
According to the self-confinement mechanism that is believed to be the dominant effect for mediating transport in the GeV regime \citep[e.g.][]{Zweibel17,Evoli+18}, a CR distribution with a super-Alfv\'enic bulk drift speed will excite Alfv\'en waves via gyroresonance. Individual CRs within the distribution will scatter off these induced Alfv\'en waves, tending to isotropize the CR distribution relative to the frame of propagating Alfv\'en waves. This physical picture has been validated in magnetohydrodynmaic-particle-in-cell (MHD-PIC) simulations, with the measured scattering rate consistent with the prediction of quasi-linear theory \citep[e.g.][]{Bai2019,Bambic2021,Bai2022}. Sufficiently large wave amplitudes can produce enough scattering to effectively prevent CRs from streaming at speeds higher than the local Alfv\'en speed. However, the wave amplitude is not merely dependent on production rate via resonant instability, but also upon the rate as which waves are damped by gas interactions. The rate of wave damping (and the dominant damping mechanisms) are found to be sensitive to the local conditions of the ISM, such that a balance between excitation and damping is expected to lead to very different scattering rates in different ISM phases  \citep[e.g.][]{Kulsrud2005,Plotnikov2021}.\

In the present study, we shall work within the CR-MHD + self-confinement paradigm, adopting two-moment equations to evolve the CR energy density and flux. CRs are advected by the background thermal gas and diffuse relative to the frame of Alfv\'en waves that propagate down the local CR gradient. Where the scattering rate is large, the bulk streaming velocity is limited to the local ion Alfv\'en speed.    
To determine the CR scattering coefficient, we locally balance the wave excitation (proportional to the local CR gradient) and damping rates. The damping rate considers both ion-neutral (IN) and nonlinear Landau (NLL) damping \citep[][]{Kulsrud1969,Kulsrud1971} as set by the local ISM properties. In principle, additional damping mechanisms could meaningfully contribute to these rates, as has been explored e.g. by \citet{Hopkins+21, Hopkins+22}. \
Other recent work adopting a similar CR-MHD model includes \citet{Sike+24, Thomas+23, Thomas+24}. However, with the exception of \citet{Hopkins+21}, these studies do not investigate how CR transport varies across different galactic environments. To date, such investigations have been limited to simulations assuming constant scattering rates \citep[e.g.][]{Chan+19,Dashyan+20,Rathjen+23}.

\cite{Lucia2021} describes the incorporation of this prescription in the algorithm for CR transport implemented by \cite{Jiang2018} in the \textit{Athena++} MHD code \citep{Stone2020}, as well as our model for ionization of warm and cold gas as set by CRs in the tens of MeV regime. In \cite{Lucia2021,Lucia2022,Lucia2024}, we applied our model as post-processing to compute the propagation of CRs in the TIGRESS MHD simulations, which represent kiloparsec-sized portions of star-forming galactic disks for a range of conditions \citep[][]{Kim2017,Kim2020b}. The advantage of the TIGRESS simulations is that star formation and feedback (including both SNe and photoelectric heating) are modeled in a self-consistent manner, thus providing a realistic representation of the multiphase star-forming ISM. Our previous work demonstrated that the transport of CRs is quite different in different thermal phases of the gas, with the CR scattering coefficient varying over more than 4 orders of magnitude depending on the properties of the underlying gas (e.g., density and ionization fraction). This challenges the common assumption of uniform scattering, and highlights the importance of an accurate representation of the multiphase ISM in CR transport modeling. \cite{Lucia2024} further developed our numerical technique, moving beyond a purely post-processed approach to include the ``MHD backreaction.'' This additional step captures the local aspect of the impact of the CR population upon the underlying thermal gas, more realistically coupling the fluids and marking an important step towards fully self-consistent simulations.\

In this work, we apply the CR transport prescription of \cite{Lucia2024} to a diverse range of TIGRESS galactic environments, including both those first investigated in \cite{Lucia2021} and \cite{Lucia2022} and a spiral arm model that makes use of the MHD simulation of \citet{Kim2020b}. Utilizing this simulation suite, we are able to gain insight into the dominant processes by which CRs are transported within the galactic disk and into the extraplanar regions, across a wide range of environments. We further use these results to develop simple analytic models that effectively capture both bulk CR transport behavior (relating the CR injection rate to pressure) and subgrid CR scattering parameters, as straightforward functions of galactic MHD properties. Our models enable us to fit a relationship between the star formation rate and the midplane CR pressure, analogous to the ``feedback yield'' fits presented in \citet{Ostriker2022,KimCG2024}. These fits may be found in \autoref{ssec:yield}.

The layout of this paper is as follows: In \S \ref{sec:methods}, we briefly review our numerical methods and introduce the TIGRESS environments studied in this work. In \S \ref{sec:Transport} we discuss our general findings regarding CR transport across this diverse range of environments. In \S \ref{sec:1Dmodel} we employ these results to develop a simple 1D model to accurately describe vertical transport of CRs and relate CR pressure to the star formation and SN rates, and in \S \ref{sec:lattrans} we discuss lateral CR transport. In \S \ref{sec:subgridmodel} we develop and calibrate an environmentally sensitive analytic prescription for the CR scattering coefficient, for the benefit of numerical modelers. Lastly, in \S \ref{sec:Conclusions} we summarize our primary results.

\section{Methods}\label{sec:methods}
Following the approach of previous work in this series \citep{Lucia2021,Lucia2022,Lucia2024}, we approach the problem of CR transport by post-processing pre-existing TIGRESS shearing box simulations, including a step to allow for the MHD backreaction on the CR pressure. Practically speaking, this means an initial stage where CRs are injected into an existing snapshot of a TIGRESS model and allowed to evolve into a steady state while the MHD variables are frozen. The resulting CR distribution is then briefly evolved together with the MHD variables to allow for MHD backreaction. The main features of the TIGRESS simulations utilized in this work are discussed in \S \ref{ssec:tigress}, while the methods used to post-process and MHD backreact the models are discussed in \S\ref{ssec:PPmeth} and \S\ref{ssec:MHDmeth}, respectively. 

\subsection{TIGRESS Models}\label{ssec:tigress}
The TIGRESS simulations are MHD shearing box representations of local patches of galactic disks. The code features self-consistent modelling of ISM evolution, including star formation and feedback, both from far-UV (FUV) heating and resolved core-collapse supernova remnant expansion; for full details, see \cite{Kim2017} and \cite{Kim2020a}. TIGRESS is built on the grid-based MHD code \textit{Athena} \citep{Stone2008}, using vertically ($z$-direction) stratified shearing boxes to simulate kiloparsec-scale patches of differentially rotating galactic disks with ideal MHD. The simulations include gas self-gravity, gravitational forces from an old stellar disk and dark-matter halo, and optically thin cooling. Star formation is treated using sink particles to represent star cluster formation and gas accretion below the cell resolution. These star clusters are assigned stellar populations drawn from a Kroupa initial mass function \citep{Kroupa2001}. Instantaneous FUV luminosity and core-collapse supernova explosion rates for each cluster are adopted from \textsc{starburst99} \citep{Leitherer1999}. The TIGRESS simulations are run for several hundred Myr, a long enough period to span several star formation/feedback cycles within an overall self-regulated condition. A realistic multiphase ISM is produced, in which supernova-driven shocks and turbulence churn and assist radiation in heating the ISM, providing the necessary thermal, turbulent, and magnetic pressure support to offset the vertical weight of the gas \citep{Kim2017, Ostriker2022}. Large-scale extraplanar gas motions driven by SN heating and acceleration are also produced, creating realistic multiphase fountain flows and winds \citep{Kim2018, Kim2020a}.\

Snapshots taken from various times in these simulation runs provide a repertoire of realistic ISM conditions, sampling a variety of epochs in the star formation/feedback cycle for each galactic environment. Most importantly in the context of CR transport, the multiphase ISM (covering a full range of temperature from cold to warm to hot) and magnetic fields (consisting of both mean and turbulent components) are self-consistently produced. This feature makes the TIGRESS models an ideal basis from which to study CR transport across diverse galactic environments. The environments studied in this work are subdivided into two categories: ``R-Models''(which comprise swatches of a generic galactic disk, sampled at various radii) and an ``Arm model'' (which is a box constructed to follow the motion of a galactic spiral arm). The details of the models selected are discussed below, and a table summarizing their important characteristics is provided as \autoref{tab:models}. For each model, we select and process 8 snapshots at various times within the typical range of $0.5 < t/t_\mathrm{orb} < 3$, where $t_\mathrm{orb} = 2\pi/\Omega$ is the orbital time (column 7 in \autoref{tab:models}), and $\Omega$ is the orbital frequency at the center of the simulation domain. This time range covers many star formation/feedback cycles and outflow/inflow events \citep{Vijayan+20}. Due to lack of data, one R-Model (R4) is a slight exception, with 6 snapshots drawn from $0.25 < t/t_\mathrm{orb} < 1.25$.\
\subsubsection{R-Models}
The R-Models are TIGRESS simulations designed to emulate the conditions in a generic Milky-way-like galactic disk at various radial distances from the galactic center. To realistically represent these environments, they adopt appropriately varied gas surface densities, as well as gravitation potentials for an old stellar disk and dark matter halo. In this work, we investigate the same 3 models previously analyzed in \cite{Lucia2022}, namely the R2 (2 kpc from the galactic center), R4 (4 kpc), and R8 (8kpc) models, with the R8 model corresponding broadly to the solar neighborhood. In this work, we return to these models with a more developed post-processing method (discussed in \S\ref{ssec:PPmeth}) and a method for MHD back-reaction (discussed in \S\ref{ssec:MHDmeth}). As a result, we are able to corroborate and expand upon the results from \cite{Lucia2022}.\

The R2 and R4 models have box size $L_x = L_y = 512$ pc and $L_z = 3584$ pc and resolution of $\Delta x = 4$ pc. The dimensions of the R8 model (box size and resolution) are twice as large. The $z$-axis corresponds to the extraplanar altitude, while the $x$- and $y$-axes are oriented in the radial and azimuthal directions of the galactic disk, respectively. For further details on these models, please refer to \cite{Lucia2022} and \cite{Kim2020a}. Overall, these models sample a wide range of galactic environments, spanning two orders of magnitude in gas surface density (an input quantity), two orders of magnitude in midplane total pressure (an emergent quantity), and almost three orders of magnitude in SFR surface density (also an emergent quantity). Amongst the quantities delineated, the emergent gas scale height ($H_\mathrm{gas}$) varies the least, being essentially identical between the R-models and only mildly larger in the R8 Arm model. This behavior stems from the fact that gas scale heights are directly proportional to the square of the velocity dispersion and inversely proportional to the local vertical ($z$-direction) gravitational acceleration. As the velocity dispersion and gravitational force increase together in higher surface density environments, scale heights tend to be fairly comparable over a relatively large range of environments \citep[e.g.][]{Bacchini2020}.\

\subsubsection{Arm Model}
The remaining model studied in this work is an environment in which CR transport has not previously been simulated with our framework: the spiral arm. We select the F20B10 model from \cite{Kim2020b}, henceforth referred to as the R8 Arm model. The TIGRESS arm model utilizes local spiral arm coordinates \citep{Roberts1969} using a shearing-periodic box with the $y$-axis aligned with the longitudinal axis of a tightly wound spiral (a pitch angle of $\sin i=1/8$)  and revolving at the arm pattern speed ($\Omega_p=\Omega_0/2$ where $\Omega_0=30\,{\rm km\,s^{-1}\,kpc^{-1}}$). In the local spiral arm coordinates, the galactic differential rotation (adopting a flat rotation curve) gives rise to the traditional background shear velocity and an additional inclined translational velocity due to the difference between spiral arm pattern speed and galactic rotation speed. A sinusoidal potential is introduced to model the spiral arm in the old stellar disk
\begin{equation}\label{eq:spiral_pot}
    \Phi_\mathrm{arm}(x) = \Phi_\mathrm{arm}(0)~\mathrm{cos}\left(\frac{2\pi x}{L_x}\right).
\end{equation}
Here, the amplitude of the spiral potential $\Phi_{\rm arm}(0)$ is parameterized by the ratio of the radial force due to the spiral arm to the mean radial gravitational force
\begin{equation}\label{eq:spiral_F}
\mathcal{F} \equiv \frac{m}{\mathrm{sin}~i}\left( \frac{\vert\Phi_\mathrm{arm}(0)\vert}{R_\mathrm{0}^2\Omega_\mathrm{0}^2}\right),
\end{equation}
where $m=2$ is adopted as the arm number. The models have a box size of $L_x = \pi $ kpc (corresponding to the arm-to-arm distance of a two-armed spiral with these characteristics) and $L_y = L_z = 2\pi$ kpc with a resolution of $\Delta x = 12.3$ pc, and are centered on a galactocentric radius of $R_0=8$ kpc. The full details of the implementation can be found in \cite{Kim2020b}. 
The R8 Arm model presented here has $\mathcal{F} = 0.2$ and initial uniform magnetic field parallel to the spiral arm $B_0=2.6\,\mu{\rm G}$ at the midplane.\

This arm model thus represents a novel environment for the study of CR propagation, with the potential for interesting variations on the vertical and lateral transport paradigms present in the disk models. Investigation of this regime will also allow a more comprehensive understanding of the CR distributions in star-forming galaxies, where spiral arms are often highly prominent. In several of our analyses, we will distinguish between ``arm'' and ``interarm'' regions, defined respectively as lying between (exterior to) $-0.5$ kpc $<$ x $<$ 1.0 kpc.

\begin{table*}
    
    \hspace{-2.75cm}\begin{tabular}{cccccccccccc}
        \hline
        \hline
         Model & $L_z$ & $\Delta x$ & $\rho_\mathrm{DM}$ & $\Sigma_\mathrm{star}$& $\Sigma_\mathrm{gas, ini}$ & $t_\mathrm{orb}$ & $\langle\Sigma_\mathrm{gas}\rangle$ & $\langle\Sigma_\mathrm{SFR}\rangle$ & $\langle n_\mathrm{mid}\rangle$ & $\langle P_\mathrm{mid}/k_\mathrm{B} \rangle$ & $\langle H_\mathrm{gas} \rangle$\\
         & [pc] & [pc] & [$M_\odot$ pc$^{-3}$] & [$M_\odot$ pc$^{-2}$] & [$M_\odot$ pc$^{-2}$] & [Myr] & [$M_\odot$ pc$^{-2}$] & [$M_\odot$ kpc$^{-2}$ yr$^{-1}$] & [cm $^{-1}$] & [K cm$^{-3}$] & [kpc] \\
         (1) & (2) & (3) & (4) & (5) & (6) & (7) & (8) & (9) & (10) & (11) & (12) \\
         \vspace{-8pt}\\
         \hline
         R2 & $\pm 1792$ & 4 & $8.0 \times 10^{-2}$ & 450 & 150 & 61 & 74 & 1.1 & 7.7 & $2.5 \times 10^{6}$ & 0.35\\
         R4 & $\pm 1792$ & 4 & $2.4 \times 10^{-2}$ & 208 & 50 & 110 & 30 & 0.13 & 1.4 & $4.1 \times 10^{5}$ & 0.34\\
         R8 & $\pm 3584$ & 8 & $6.4 \times 10^{-2}$ & 42 & 12 & 220 & 11 & $5.1 \times 10^{-3}$ & 0.9 & $1.9 \times 10^4$ & 0.33\\
         R8 Arm & $\pm 3141$ & 12.3 & $6.4 \times 10^{-2}$ & 42 & 13 & 205 & 10 & $4.2 \times 10^{-3}$ & 0.5 & $2.3\times10^4$ & 0.43 \\
         \hline
    \end{tabular}
    
    \caption{Physical parameters of the model environments sampled in this work. Columns: (1) Model name; (2) Simulation box vertical extent; (3) Spatial resolution; (4) dark matter volume density; (5) Old stellar disk surface density; (6) Initial gas surface density; (7) Orbital time; (8) Time-averaged gas surface density; (9) time averaged SFR surface densirt; (10) Time-averaged midplane gas number density; (11) Time-averaged midplane total gas pressure; (12) Time-averaged gas scale height. The time-averaged quantities are averaged over the interval 0.5  $< t/t_\mathrm{orb} <$  1.5.}
    \label{tab:models}
\end{table*}

\subsection{Implementation of CR Transport} \label{ssec:PPmeth}
To perform these simulations we utilize the two-moment algorithm for CR transport developed by \cite{Jiang2018} and incorporated within the MHD code \textit{Athena++} \citep{Stone2020}. 
The full set of ideal MHD and CR transport equations are as follows:

\begin{equation}
    \frac{\partial \rho}{\partial t} + \nabla \cdot (\rho \textbf{v}) = 0
\end{equation}

\begin{equation}\label{eq:mhdmom}
\begin{split}
    &\frac{\partial (\rho \textbf{v})}{\partial t} + \nabla \cdot \left( \rho \textbf{vv} + P_t \te{\textbf{I}}  + \frac{B^2}{8\pi} \te{\textbf{I}} - \frac{\textbf{BB}}{4\pi}\right) \\
    &= -\rho \nabla\Phi + \te{\sigma}_\mathrm{tot} \cdot [\textbf{F}_c - \textbf{v} \cdot (\te{\textbf{P}}_c + e_c \te{\textbf{I}})]
\end{split}
\end{equation}

\begin{equation}\label{eq:mhdeng}
\begin{split}
    &\frac{\partial e}{\partial t} + \nabla \cdot \left[ \left(e + P_t + \frac{B^2}{8\pi} \right) \textbf{v} - \frac{\textbf{B}(\textbf{B} \cdot \textbf{v})}{4\pi}\right] = -\rho \mathcal{L} \\
    &- \rho \textbf{v} \cdot \nabla \Phi +(\textbf{v} + \textbf{v}_s) \cdot \te{\sigma}_\mathrm{tot} \cdot [\textbf{F}_c - \textbf{v} \cdot (\te{\textbf{P}}_c + e_c \te{\textbf{I}})]
\end{split}
\end{equation}

\begin{equation}
    \frac{\partial\textbf{B}}{\partial t} - \nabla \times (\textbf{v} \times \textbf{B}) = 0
\end{equation}

\begin{equation} \label{eq:engtrans}
\begin{split}
  \frac{\partial e_c}{\partial t} + \nabla \cdot \textbf{F}_c = -(\textbf{v} +\textbf{v}_s&) \cdot \te{\sigma}_{\mathrm{tot}} \cdot [ \textbf{F}_c - \textbf{v}\cdot(\te{\textbf{P}}_c+e_c\te{\textbf{I}})] \\
    &+ S - \Lambda_\mathrm{coll} n_\mathrm{H} e_c  
\end{split}
\end{equation}
\begin{equation}\label{eq:momtrans}
\begin{split}
  \frac{1}{v_m^2}\frac{\partial \textbf{F}_c}{\partial t} + \nabla \cdot \te{\textbf{P}}_c = &- \te{\sigma}_{\mathrm{tot}} \cdot [ \textbf{F}_c - \textbf{v}\cdot(\te{\textbf{P}}_c+e_c\te{\textbf{I}})] \\
    &- \frac{\Lambda_\mathrm{coll} n_\mathrm{H}}{v_p^2} \textbf{F}_c  
\end{split}
\end{equation}
Here, $\rho$ is the gas density, \textbf{v} is the gas velocity, \textbf{B} is the magnetic field, and $e = (1/2) \rho v^2 + P_t / (\gamma - 1) + B^2 / (8\pi)$ is the gas energy density, with $P_t$ being the gas thermal pressure and $\gamma = 5/3$ being the gas adiabatic index. $e_c$ is the CR energy density, $\textbf{F}_c$ is the CR energy flux, and $\te{\textbf{P}}_c$ is the CR pressure tensor, while $S$ represents the CR source term (from SN injection) and the terms proportional to $\Lambda_\mathrm{coll}$ represent collisional losses. We assume approximately isotropic pressure, such that $\te{\textbf{P}}_c \equiv P_c\te{\textbf{I}}$,with $P_\mathrm{c} =(\gamma_c - 1)e_c = e_c/3$, where $\gamma_c = 4/3$ is the adiabatic index of the relativistic fluid, and $\te{\textbf{I}}$ is the identity tensor. $\Phi$ is the ``external'' gravitational potential from the old stellar disk and dark matter halo \citep[see][]{Kim2017}. Unlike the original TIGRESS simulations, self-gravity is not included in the simulations presented in this work, as the MHD variables are frozen during the post-processing step (see \S\ref{sssec:PP-methods}) and the duration of the ``MHD relaxation'' step is extremely short (see \S\ref{sssec:MHD-Relaxation}). \

In \autoref{eq:momtrans}, $v_\mathrm{m}$ is the maximum velocity at which CRs can propagate. In principle, $v_\mathrm{m}$ must be equal to the speed of light $c$ for relativistic CRs. However, here we adopt the “reduced speed of light” approach (see e.g. \citealt{Skinner2013}, for the corresponding two-moment radiation implementation) with $v_\mathrm{m} = 10^4$ km s$^{-1} \ll c$, as it has been demonstrated that the result is not sensitive to the exact value of $v_\mathrm{m}$ as long as $v_\mathrm{m}$ is much larger than any other speed in the simulation \citep{Jiang2018}. This enables larger numerical timesteps owing to a less restrictive CFL condition. \ 

$\rho \mathcal{L} = n_\mathrm{H}(n_\mathrm{H}\Lambda(T)-\Gamma$) is the net cooling function, where $n_\mathrm{H}$ is the hydrogen number density. The cooling coefficient $\Lambda(\mathrm{T})$ is computed using the fitting formula in \cite{Koyama2002} for T $< 2 \times 10^4$~K, and the tabulated values in \cite{Sutherland1993} with solar metallicity for T $> 2 \times 10^4$~K. For warm and cold gas we apply heating to represent the photoelectric effect on grains. The heating rate $\Gamma$ scales with the instantaneous far-ultraviolet luminosity per area $\Sigma_{\rm FUV}$ from star particles with a global attenuation factor $f_\tau$ determined by a plane-parallel approximation \citep[][see \citealt{Linzer2024} for in-depth investigations of the validity of this and other approximations]{Ostriker2010,Ostriker2022}; written explicitly, $\Gamma/\Gamma_0 = f_\tau \Sigma_{\rm FUV}/\Sigma_{\rm FUV,0}$. For reference solar-neighborhood values, we adopt a heating rate of $\Gamma_0= 2 \times 10^{-26}$ erg s$^{-1}$, and a mean FUV intensity of $\Sigma_\mathrm{FUV,0} =4\pi J_\mathrm{FUV,0} = 2.7 \times 10^{-3}$ erg s$^{-1}$ cm$^{-2}$ (see \citealt{Kim2020a} for further details).\ 

In Equations \ref{eq:mhdmom} and \ref{eq:momtrans}, the term $\te{\sigma}_\mathrm{tot} \cdot [\textbf{F}_c - \textbf{v} \cdot (\te{\textbf{P}}_c + e_c \te{\textbf{I}})]$ represents the rate of transfer of momentum density between the CRs and thermal gas. Likewise, in Equations \ref{eq:mhdeng} and \ref{eq:engtrans}, the term $\textbf{v} \cdot \te{\sigma}_{\mathrm{tot}} \cdot [ \textbf{F}_c - \textbf{v}\cdot(\te{\textbf{P}}_c+e_c\te{\textbf{I}})]$ represents the direct CR pressure work done on or by the gas, and the term $\textbf{v}_s \cdot \te{\sigma}_{\mathrm{tot}} \cdot [ \textbf{F}_c - \textbf{v}\cdot(\te{\textbf{P}}_c+e_c\te{\textbf{I}})]$ represents the rate of energy transferred to the gas via damping of gyro-resonant Alfv\'en waves. In this expression, $\textbf{v}_s$ is the CR streaming velocity, defined as
\begin{equation}
    \mathbf{v_s} \equiv - \mathbf{v}_\mathrm{A,i}  \frac{\mathbf{B} \cdot (\nabla \cdot \te{\mathbf{P}}_c)}{|\mathbf{B} \cdot (\nabla \cdot \te{\mathbf{P}}_c)|} = - \mathbf{v}_\mathrm{A,i} \frac{\hat{B} \cdot (\nabla P_c)}{|\hat{B} \cdot (\nabla P_c)|}
\end{equation}
which represents the component of the local ion Alfv\'en velocity, $\mathbf{v}_\mathrm{A,i} = \mathbf{B}/\sqrt{4\pi \rho_i}$, oriented along the local magnetic field and pointing down the CR pressure gradient. $\rho_i$ is the ion mass density (computed in accordance with the method described in \S 2.2.5 of \citealt{Lucia2021}).\

The diagonal tensor $\te{\sigma}_\mathrm{tot}$ is the wave-particle interaction coefficient, defined to allow for both scattering and streaming along the direction parallel to the magnetic field,
\begin{equation}\label{eq:sigpardef}
    \sigma_\mathrm{tot,\parallel}^{-1} = \sigma_\parallel^{-1} + \frac{v_\mathrm{A,i}}{|\hat{B}\cdot\nabla P_c|} (P_c + e_c)
\end{equation}
and only scattering in the directions perpendicular to the magnetic field,
\begin{equation}
    \sigma_\mathrm{tot,\perp} = \sigma_\perp
\end{equation}
The second term in \autoref{eq:sigpardef} is equivalent to adding $\mathbf{v}_{A,i}$ to $\mathbf{v}$ on the right-hand side of \autoref{eq:engtrans} and \autoref{eq:momtrans}.  

For the relativistic case, $\sigma_\mathrm{\parallel} = \nu_\parallel/c^2 $, where $\nu_\parallel$ is the scattering rate parallel to the magnetic field direction due to Alfv\'en waves that are resonant with the CR gyro-motion. In the simulation, $\sigma_\mathrm{\parallel}$ is computed in a self-consistent manner (see below), while we set $\sigma_\mathrm{\perp} = 10 \, \sigma_\mathrm{\parallel}$. The latter can be understood as scattering by unresolved fluctuations of the mean magnetic field. In \cite{Lucia2021}, the transport of CRs was explored in the absence of perpendicular scattering ($\sigma_\mathrm{\perp} \gg \sigma_\mathrm{\parallel}$) as well as the case $\sigma_\mathrm{\perp} = 10 \, \sigma_\mathrm{\parallel}$, and no substantial difference in the CR distribution were found. We refer to \S\ref{sec:lattrans} for further discussion on the impact of perpendicular scattering on CR transport in the simulations presented in this work.\

In this work, the scattering coefficient parallel to the magnetic field, $\sigma_\mathrm{\parallel}$, is derived in accordance with the self-confinement scenario for CR scattering. In this picture, CR streaming excites Alfv\'en waves, which then, in turn, pitch-angle scatter the CRs \citep[e.g.,][]{Kulsrud1969,Wentzel1974,Bai2019}. In steady state, the excitation of such Alfv\'en waves is balanced by some form of wave damping in the plasma. We consider two contributors to the overall wave damping: NLL and IN damping \citep{Kulsrud1969,Kulsrud2005}. The former occurs when thermal ions have a resonance with the beat wave formed by the interaction of two resonant Alfv\'en waves, while the latter arises from friction between ions and neutrals in a partially ionized gas, where neutrals are not tied to magnetic fields. Formal derivations of $\sigma_\mathrm{\parallel}$ from both of these sources, as well as detailed discussions of the nature of wave excitement and damping, can be found in \S2.2.2 of \cite{Lucia2022} and \S2.2.4 of \cite{Lucia2021}. For the sake of concision, we only report here the final formulae.\

In the well-ionized, low-density gas, NLL is the dominant source of wave damping, and $\sigma_\mathrm{\parallel}$ reduces to:
\begin{equation}\label{eq:signll}
    \sigma_\mathrm{\parallel,nll} = \sqrt{\frac{\pi}{16}\frac{|\hat{\mathbf{B}}\cdot\nabla P_c|}{v_\mathrm{A,i} P_c}\frac{\Omega_0 c}{0.3 v_\mathrm{t,i}v_p^2}\frac{m_p}{m_i}\frac{n_1}{n_i}}
\end{equation}
whereas in the primarily neutral, denser gas, IN damping dominates, and $\sigma_\mathrm{\parallel}$ reduces to:
\begin{equation}\label{eq:sigin}
        \sigma_\mathrm{\parallel,in} = \frac{\pi}{8}\frac{|\hat{\mathbf{B}}\cdot\nabla P_c|}{v_\mathrm{A,i} P_c}\frac{\Omega_0}{\langle \sigma v \rangle _\mathrm{in}}\frac{m_p(m_n + m_i)}{n_n m_n m_i}\frac{n_1}{n_i}
\end{equation}
In Eq. \ref{eq:signll}, $\Omega_0 = e|\mathbf{B}|/(m_\mathrm{p}c)$ is the cyclotron frequency for the elementary charge $e$ and proton mass $m_\mathrm{p}$. $v_\mathrm{t,i}$ is the ion thermal velocity (which we set equal to the gas sound speed $c_s$), $m_i$ is the ion mass, $n_i$ is the ion number density, and $v_p = \sqrt{1-(m_p c^2/E)^2}$ is the CR proton velocity, where $E \equiv E_k + m_p c^2$ is the total relativistic energy, with $E_k$ being the kinetic energy. In Eq. \ref{eq:sigin}, $m_n$ is the neutral mass, $n_n$ is the neutral number density (derived from the ionization fraction $x_i \equiv n_i/n_\mathrm{H}$, see \S2.2.5 of \citealt{Lucia2021} for more details). $\langle\sigma v\rangle_\mathrm{in} \approx 3 \times 10^9$ cm$^3$ s$^{-1}$ is the rate coefficient for collisions between H and H$^+$ (see \citealt{Draine2011} Table 2.1).\

The quantity
$n_1$ is defined as 
\begin{equation}\label{eq:n1}
    n_1 \equiv 4\pi p_1 \int^{\infty}_{p_1}p F(p) dp
\end{equation}
where $F(p)$ is the CR distribution function in momentum space, and is proportional to the total number density of CRs with momentum above the resonant momentum $p_1 = m_p \Omega_0/k$ for wavenumber $k$ (for a power law with slope 4.7, $n_1=0.6\, n(p>p_1)$, with $n$ the CR number density). For CRs with $E_k \simeq 1$ GeV, $n_1 = 1.1 \times 10^{-10}$ [$e_c(E_k \geq 1 \mathrm{GeV})/1\mathrm{eV}$] cm$^{-3}$ (see Appendix A1 of \citealt{Lucia2021} for the details of the calculation). We refer to \S\ref{sec:subgridmodel} for further elaboration on the scattering coefficient in this work.\

Finally, in Equations \ref{eq:engtrans} \& \ref{eq:momtrans}, the terms $\Lambda_\mathrm{coll} n_\mathrm{H} e_c$ and $\Lambda_\mathrm{coll} n_\mathrm{H} \mathbf{F}_c/v_p^2$ represent, respectively, the rates of CR energy density and CR flux lost due to collisional interactions with the ambient gas. GeV CRs interact with the ambient gas through ionization of the neutral atomic/molecular gas and hadronic collisions which lead to decays of pions into $\gamma$-rays \citep[e.g.,][]{Padovani2020}. In the above, $\Lambda_\mathrm{coll}$ is the CR collisional coefficient. In \cite{Lucia2021}, we estimate that $\Lambda_\mathrm{coll}$ is equal to $4 \times 10^{-16}$ cm$^3$ s$^{-1}$ for CRs with $E_k$ = 1 GeV (see also \citealt{Padovani2020}).\

CR deposition is performed by injecting energy in the regions near star cluster particles, to model CR acceleration in connection to supernova events. For each discrete star cluster particle, the energy injection rate is calculated as\
\begin{equation}
    \dot{E}_\mathrm{c} = \epsilon_c E_\mathrm{SN}\dot{N}_\mathrm{SN}
\label{eq:eninj}
\end{equation}
where $\epsilon_\mathrm{c}$ is the fraction of supernova energy that goes into the production of CRs, which we take to be 10\% \citep[e.g.,][]{Morlino2012,Ackermann2014}. $E_\mathrm{SN} = 10^{51}$ ergs is the energy released by a single core-collapse supernova (CCSN) event, and $\dot{N}_\mathrm{SN}$ is the number of CCSN per unit time determined from the \textsc{\small STARBURST99} code, based on the age and mass of the cluster. In keeping with the convention of \cite{Kim2017} and the TIGRESS models broadly, non-core-collapse supernovae are not incorporated (although considering the timescales and injection distributions involved, this choice is very unlikely to produce any change in the CR distribution).\ 

\subsection{Application to the TIGRESS Simulations}\label{ssec:MHDmeth}

\subsubsection{Post-Processing}\label{sssec:PP-methods}
Utilizing the CR transport algorithms discussed in \S \ref{ssec:MHDmeth}, we first post-process the selected snapshots. During this phase, the MHD variables are frozen in time; the CR energy density and flux are evolved via \autoref{eq:engtrans} and \autoref{eq:momtrans} under the influence of the source terms set by the background gas and magnetic field, together with CR injection from star clusters,until reaching convergence. We define the convergence condition as occurring once the total CR energy density, summed throughout the box, reaches a steady state, i.e. $(e_\mathrm{c,tot}(t) - e_\mathrm{c,tot}(t-0.1~\mathrm{Myr})/e_\mathrm{c,tot}(t) < 10^{-6})$ with $e_\mathrm{c,tot} = \int_\mathrm{Vol} {e}_\mathrm{c} dx^3$.
The time taken for post-processing to conclude varies significantly between models, as a function of box size, grid resolution, and environmental parameters; however, due to the frozen nature of the MHD values, the convergence time of post-processing is not meaningful in itself.\

Different from \cite{Lucia2022}, we omit the adiabatic work term from the energy density equation ($\textbf{v} \cdot \te{\sigma}_{\mathrm{tot}} \cdot [ \textbf{F}_c - \textbf{v}\cdot(\te{\textbf{P}}_c+e_c\te{\textbf{I}})]$ in \autoref{eq:engtrans}) during the post-processing step. As explained in \citealt{Lucia2024} (see \S 3.1), this avoids unphysical enhancement of the CR energy when there is no CR backreaction on the gas to reorient magnetic field lines and the direction of the velocity in response to large CR pressure gradients at interfaces between cold/warm and hot gas.

\subsubsection{MHD Relaxation}\label{sssec:MHD-Relaxation}
While the post-processing methodology provides a representative view of the CR distribution, a fully self-consistent approach requires a treatment of the backreaction of the CRs on the gas. To do so, we adopt the MHD-relaxation method of \cite{Lucia2024}. The converged results of post-processing each snapshot (performed under frozen MHD) are taken as an initial condition and the simulations are restarted, now with the MHD variables (thermal gas distribution, magnetic field, etc.) permitted to evolve alongside the CRs. The adiabatic work term in \autoref{eq:engtrans}, describing transfer of kinetic energy between the CRs and the gas, is also enabled for this step, fully coupling the gas and CRs. While these simulations properly compute simultaneous evolution of the MHD and CRs, they do not include all of the physical effects incorporated in the original TIGRESS simulations. Namely, self-gravity is not enabled nor is new star formation followed. Most critically, injection of energy and momentum from radiation and SN feedback (into either the gas or CRs) is not performed during this ``MHD relaxation'' step. As a result, no new hot gas is produced during the relaxation step, which would significantly deplete the midplane of this ISM phase if the extant high-velocity hot gas is allowed sufficient time to advect out of the disk. To avoid this, the MHD relaxation step is performed for only for a brief time, between 0.5 and 2 Myr, as necessary to achieve convergence in the relative orientation of velocity/magnetic fields and CR pressure gradients (see below).

The main goal of the MHD relaxation step is to allow CRs to rearrange the magnetic field and gas velocity topology near phase boundaries. As discussed in \cite{Lucia2024} (which previously applied the MHD relaxation step to the R8 model), in the post-processing step, the orientation of the velocity and magnetic field lines near these boundaries confines CRs within the cold/warm, dense gas regions. However, by applying the backreaction step, the excess CR pressure in these regions exerts forces which reorient the magnetic and gas velocity topology, allowing the dense gas and the CRs trapped within it to propagate into adjacent hotter regions.
The main result of this process is a more uniform CR distribution. In the interest of allowing this effects of the CR backreaction while minimizing the impact of methodological drawbacks, we define our criteria to stop the simulation based on the relative orientation of the CR gradients with respect to the velocity and magnetic field directions. The details of this process are provided in \S3 of \cite{Lucia2024}, but the basic concept of the method is to MHD relax the system for sufficient time for the distributions of $\hat{B}\cdot\nabla P_\mathrm{c}/|\nabla P_\mathrm{c}|$ and $\hat{v}\cdot\nabla P_\mathrm{c}/|\nabla P_\mathrm{c}|$ to reach a steady state.\

However, as we shall show in \autoref{ssec:pressuredist} (See \autoref{fig:PP_MHD_comp}), the CR backreaction, while indeed influential to the velocity and magnetic field topology locally, makes little difference to the horizontally averaged vertical profiles; the slight smoothing produced in the CR profiles represents a deviation well within the range of typical temporal variations. As such, while important for controlling the detailed structure of the CR distribution (especially laterally and across phase interfaces), the MHD relaxation step is not critical for an analysis of the spatially averaged quantities.\ 
 
The majority of the work to follow concerns bulk CR transport properties and vertical profiles, upon which the MHD relaxation step produces no significant changes. For this reason, we focus our analysis primarily on the results of the post-processing simulations. This choice is further motivated by the aforementioned lack of energy injection in the current MHD relaxation, which leads to some reduction in the hot-gas velocity. Nevertheless, for the results to come, we validate our post-processing findings against the post-MHD-relaxation outcomes where such comparisons are possible and appropriate. The only exception is found in \S\ref{sec:subgridmodel}, where we focus on the CR scattering coefficients across different gas phases, rather than averaging over the total gas. In this case, we use the outcomes of the MHD relaxation simulations, as spatial variations are paramount.

The only model for which we do not apply the MHD relaxation step is the Arm model, due to its more complex potential structure and boundary conditions, which would require additional implementation in the code. We include Arm model results only for analyses applied to post-processing simulation outputs, which constitutes the bulk of the forthcoming discussion. The Arm model is naturally excluded from any discussions that make use of the MHD relaxation results. 

\section{Cosmic Ray Transport in Diverse Environments}\label{sec:Transport}
\begin{figure*}
    \centering
    \includegraphics[width=1\textwidth]{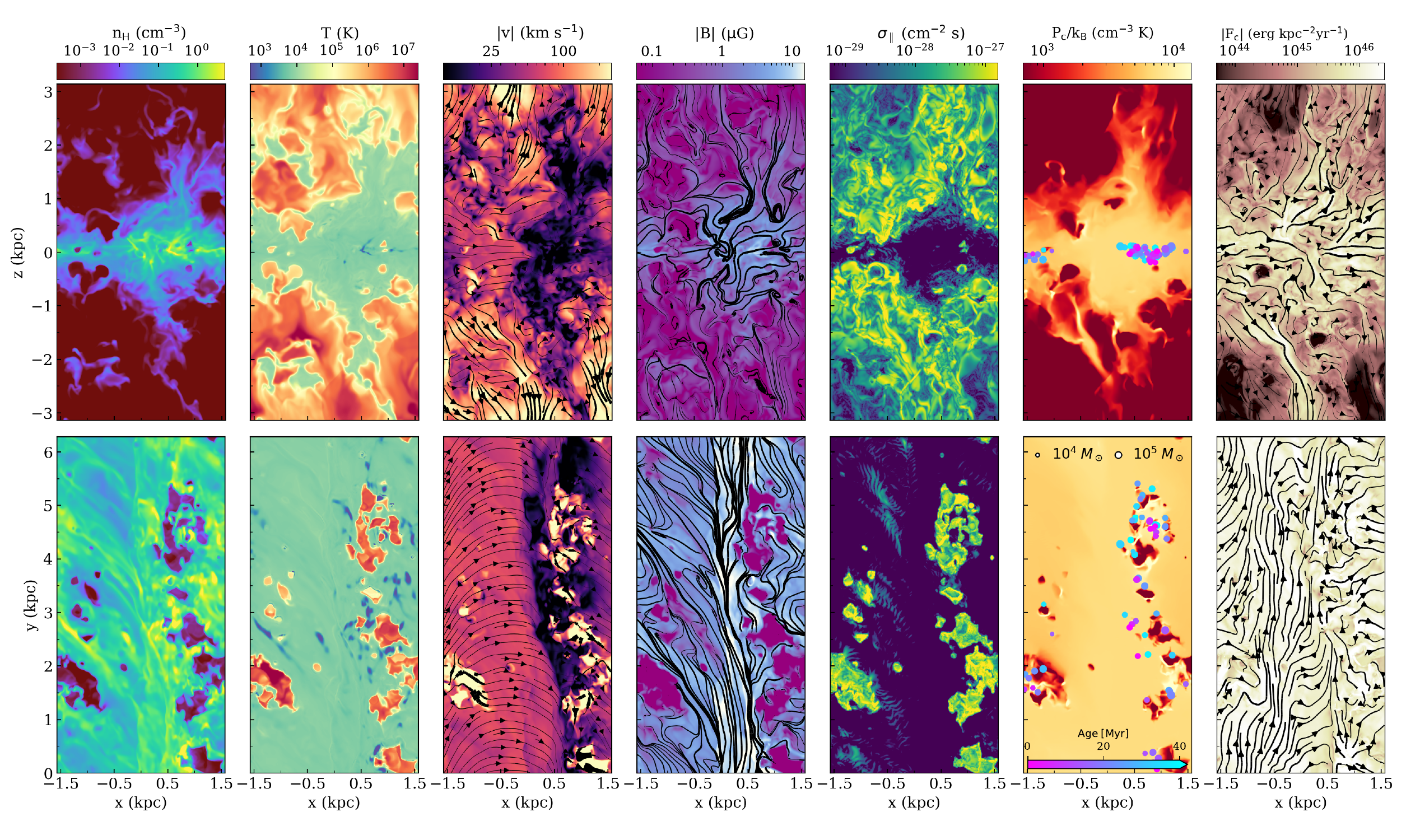}
    \caption{Slices taken from an exemplar snapshot of the R8 Arm model. The top row shows vertical slices through the center of the simulation box, while the lower panels show horizontal slices through the midplane. In these coordinates, $z$ represents the extraplanar altitude, while the $y$ ($x$) axis is parallel (perpendicular) to the local spiral arm segment. From left to right, the columns show hydrogen number density ($n_\mathrm{H}$), gas temperature ($T$), gas speed ($|\mathbf{v}|$), magnetic field strength ($|\mathbf{B}|$), scattering coefficient ($\sigma_{\parallel}$), cosmic-ray pressure ($P_c$), and cosmic-ray flux ($F_c$). Note that cosmic-ray flux panel displays the magnitude of component of the flux in the plane of each panel (i.e. the upper row displays the magnitude of the $x-z$ components and the bottom row displays the $x-y$ components). The location of star clusters are represented by overlayed points, color-coded by cluster age, in the $P_c$ panel. For vector quantities, only the magnitude of the vector components in the plane of the slice is shown, with variable width streamlines to indicate direction and magnitude. In the magnetic field and CR flux panels, the streamline widths are scaled logarithmically with vector magnitude, while the velocity panel utilizes a linear scaling to enhance contrast over a small dynamic range. We note that, to better highlight transport across the arm, the mean $y$-component of circular rotation velocity (in the frame of the arm) has been subtracted, as has the mean $y$-component of the average CR flux. Periodicity of the simulation along the arm's longitudinal axis means these components result in no net transport of gas or CRs with respect to the box.
    }
    \label{fig:slice}
\end{figure*}

In this section we provide an overview and intercomparsion of the results on CR and gas structure in the simulations across 
galactic environments. \autoref{fig:slice} displays slices of a selected snapshot of the R8 Arm model, at the end of post-processing. The topmost row displays vertical slices through the center of the simulation box ($y=0$), while the lower row are horizontal slices through the midplane ($z=0$); the columns show (from left to right) hydrogen number density ($n_\mathrm{H}$), gas temperature ($T$), gas speed ($|\mathbf{v}|$), magnetic field strength ($|\mathbf{B}|$), parallel scattering coefficient ($\sigma_{\parallel}$), cosmic-ray pressure ($P_\mathrm{c}$), and cosmic-ray flux ($F_\mathrm{c}$). The location of star clusters with age younger than 40~Myr are represented by overlayed points in the cosmic-ray pressure panel, color-coded by cluster age.
The gaseous arm is best visible in the density and magnetic field strength panels as the peaks located near $x=0-0.5$ kpc. The corresponding dip in gas speed is an indication of galactic spiral shocks \citep[e.g.,][]{Roberts1969,Kimwt2002}, but the shock front is highly corrugated and fragmented due to the multiphase and turbulent nature of the ISM \citep{Kim2020b}, which overwhelms the (magneto)hydrodynamical instabilities of spiral shocks \citep[e.g.,][]{Wada2004,Kimwt2006,Kimwt2014,Kimy2015}. CR injection predominantly occurs at the immediately downstream edge of the arm, where the young star clusters are located. 
The arm feature also displays a noticeable vertical extent, resulting in the presence of cold/warm, dense gas at higher altitudes above the disk than in the interarm regions. 

The CR scattering coefficient is seen to correlate with the gas phase: $\sigma_\parallel$ is low in the dense, mostly neutral gas (where Alfv\'en waves are efficiently damped by IN frictions) and high in the diffuse, mostly ionized gas (where Alfv\'en wave damping via NLL is less effective).
In keeping with this low scattering coefficients (and in conjunction with  the magnetic field structure, discussed below), the CR pressure distribution is remarkably uniform throughout the arm's vertical and planar extent as well as within the midplanar region of the interarm. Exterior to these cool, dense regions, CR scattering is significantly enhanced, and CR pressure gradients become noticeable, especially along the vertical direction.
Consistent with previous works \citep{Lucia2021, Lucia2022}, the R-models similarly display low scattering coefficients and resultant uniform CR pressures in the cooler, higher-density gas comprising most of the ISM.

\subsection{Magnetic and velocity fields}\label{ssec:Bv}

The arm is characterized by a strong and coherent magnetic field running along its longitudinal axis. Further coherent magnetic field structures are observed to run along the ``spurs'' jutting out from the dense arm into the interarm region. As a result, the midplane is characterized by a largely toroidal magnetic field, with enhanced strength in the arm. Exterior to the midplane, the magnetic field weakens somewhat and reorients to a near-vertical orientation due to feedback driven vertical outflows. This is consistent with the well established results of multi-wavelength polarimetry observations, which show spiral galaxies tend to have coherent, plane-parallel magnetic fields within the disk and weaker, more vertical fields at high altitude \citep[e.g.][]{Borlaff2023,Beck2013,Planck2016,SOFIA2023}. CR propagation is highly anisotropic and, both through streaming and diffusion, happens primarily along magnetic field lines. As such, the extraplanar field reorientation can provide a path for CR transport from midplane injection regions to high altitudes in addition to the advection with the hot winds. The effect of the midplane magnetic structure on lateral transport is discussed in \S\ref{sec:lattrans}.
We note that just as there can be a net inward flow of gas along spiral arms \citep[e.g.][]{Kimwt2002,Kim2020b,Shetty2007}, there could be a net flow of CRs along the magnetic field that are aligned with arms, which in real galaxies would preferentially be radially outward (because the mean CR pressure decreases outward). \

In \autoref{fig:slice}, the mean \textit{y}-component of the circular velocity in the spiral arm frame ($R_0(\Omega-\Omega_p)=120~\mathrm{km/s}$) has been subtracted. As the simulation box is periodic in \textit{y}, this motion results only in cyclic motion along the arm, with no particular net CR transport. By removing this component, the remaining velocity field is more easily visualized. In the midplane, velocities are largely planar and in addition to the background differential rotation display significant contributions from the rotation of the arm, sweeping upstream gas from the interarm region into the arm, and then transporting gas into the downstream interarm region. This bulk motion is punctuated by high-velocity supernova bubbles embedded in the midplane. Outside of the midplane and arm regions, velocities are significantly higher in magnitude and characterized by near-vertical hot chimney outflows. While the R-models use the co-rotating frame of the gas, thereby lacking the bulk planar motion seen in the arm model (instead being dominated by expanding superbubbles), they still display the same general velocity structure: moderate velocities in warm/cold gas (which fills the majority of the volume near the midplane), and high velocities in hot gas that vents through near-vertical chimneys. Advective transport modes thus also provide potentially significant avenues for CR transport away from the disk region, as previously discussed in \cite{Lucia2022} and \cite{Lucia2024}.\

\begin{figure*}
    \centering
    \subfloat{\includegraphics[width=0.5\textwidth]{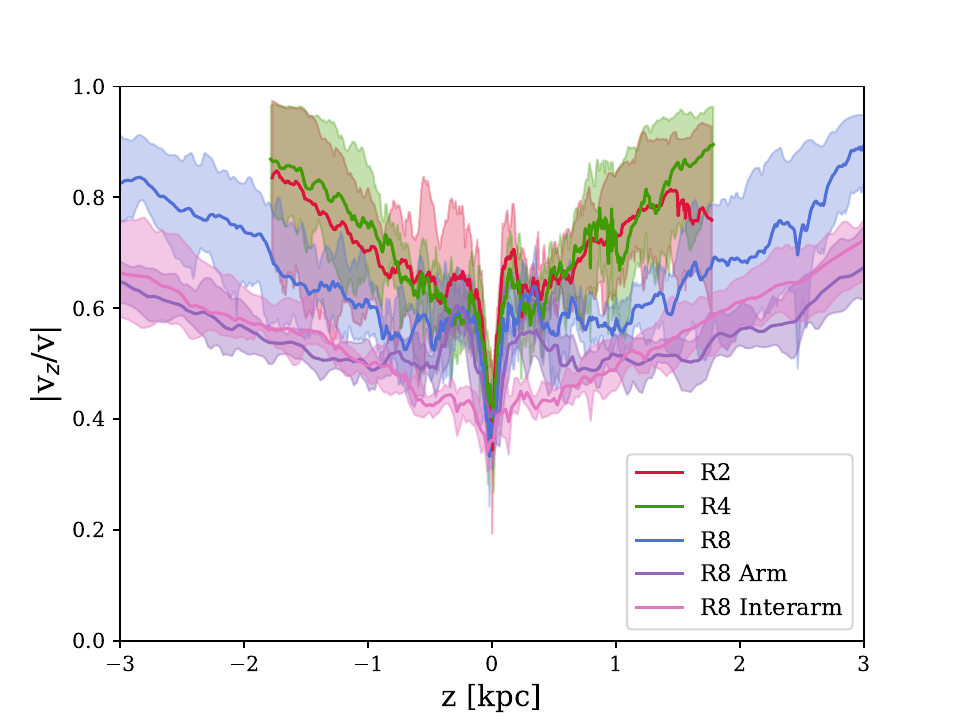}}
    \subfloat{\includegraphics[width=0.5\textwidth]{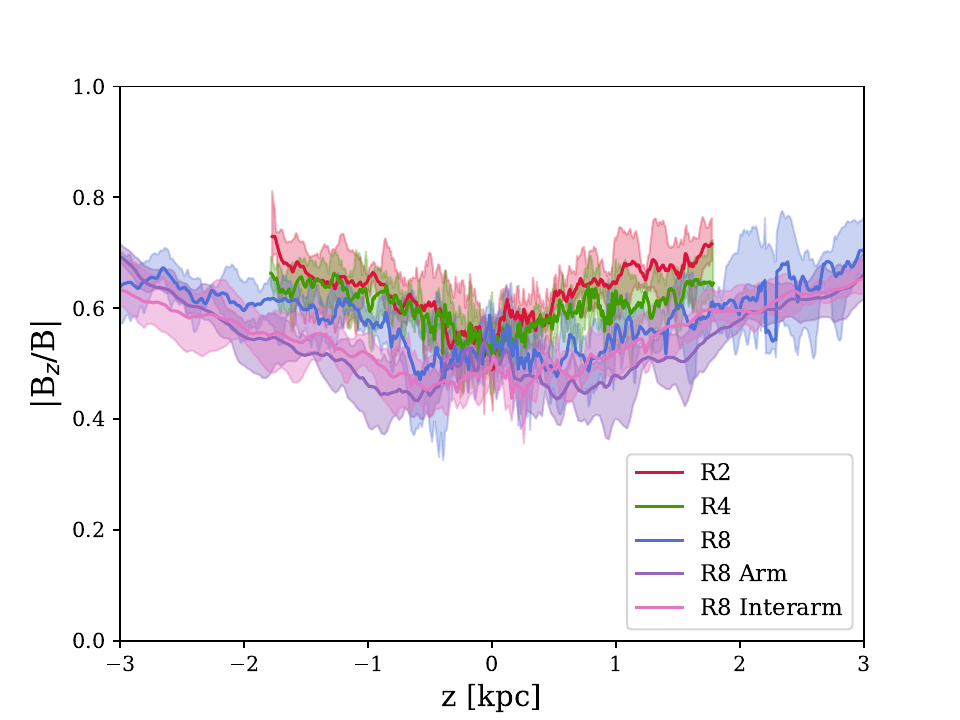}}\hfill
    \subfloat{\includegraphics[width=0.5\textwidth]{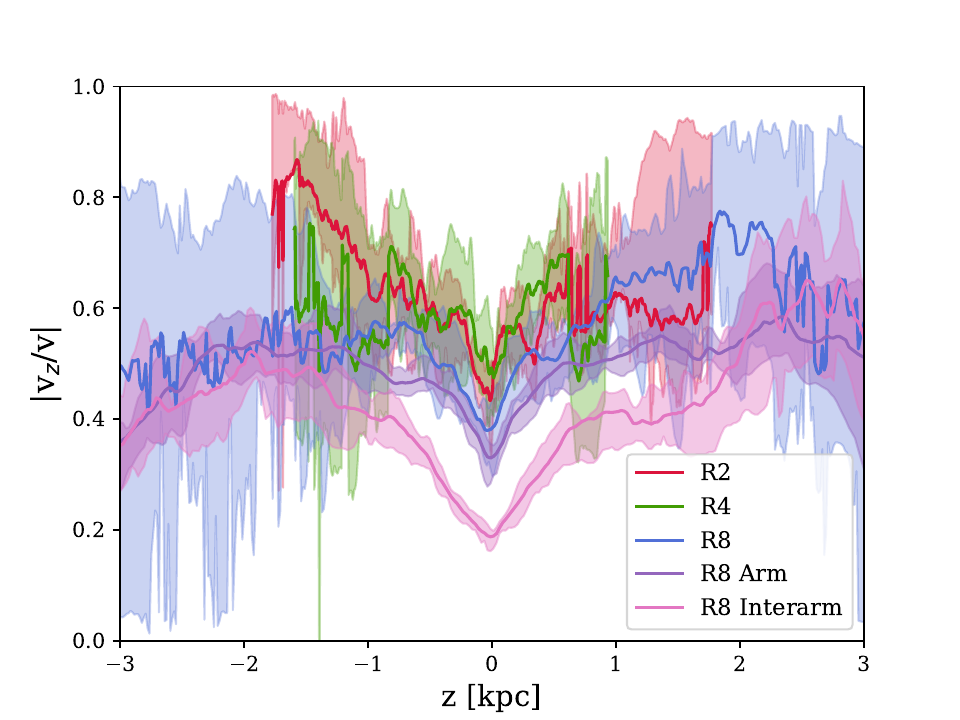}}
    \subfloat{\includegraphics[width=0.5\textwidth]{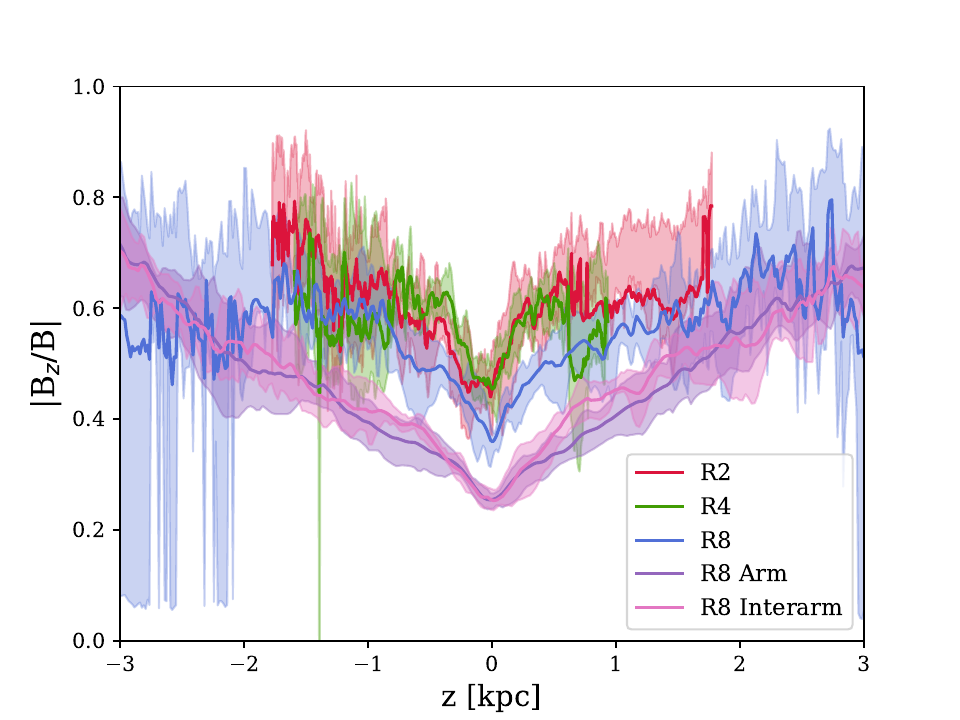}}
    
    \caption{Horizontally and temporally averaged vertical profiles of the ``relative verticality'' of the gas velocity ($\vert v_\mathrm{z}/v \vert$; left) and magnetic field ($\vert B_\mathrm{z}/B \vert$; right). The top row shows the results for hot gas ($>2\times10^4$ K), while the bottom row shows warm/cold gas ($<2\times10^4$ K). Lines with different colors represent different models, with the arm models differentiated into the arm and interarm regions: R2 (red), R4 (green), R8 (blue), R8 Arm (purple), R8 Interarm (pink). The arm (interarm) region is defined as lying between (exterior to) $-0.5$ kpc $<$ x $<$ 1.0 kpc. The shaded regions represent the 16th - 84th percentiles of temporal variation. For the Arm model, we subtract the mean velocity of the gas relative to the arm before computing $v$.}
    \label{fig:verticality}
\end{figure*}

In order to further characterize the transition of the velocity and magnetic field orientation between the midplane and the extraplanar region, we create horizontally and temporally averaged vertical profiles of the \textit{z} components of these fields ($v_\mathrm{z}$ and $B_\mathrm{z}$), normalized by their local magnitude ($v$ and $B$). This quantity provides a measure of the ``relative verticality'' of both fields. The results are shown in \autoref{fig:verticality}, where they have been split between hot ($>2\times10^4$K) gas on the top row, and warm/cold ($<2\times10^4$K) gas on the bottom row. In practice, we first create horizontally-averaged vertical profiles of $\vert v_\mathrm{z}/v \vert$ and $\vert B_\mathrm{z}/B \vert$ for each snapshot, and then average over all snapshots. In so doing, we are able to marginalize over spatial and temporal variations to produce representative distributions for each model. Note that, for the Arm model, the regions interior and exterior to the spiral arm are shown separately. We operationally define the spiral arm as lying in the region of $-500$ pc $<$ x $<$ 1000 pc, as the arm location and extent vary minimally between snapshots. For the Arm model, the gas circular motion in the arm frame is removed from the velocity magnitude used in normalization.\

In confirmation of the visually observable trends in \autoref{fig:slice}, the vertical components of both the velocity and magnetic field structure are significantly lower near the midplane ($|B_z/B| = 0.45$ corresponds to a horizontal field component twice the magnitude of its vertical counterpart). The warm/cold gas (lower row) displays significantly more horizontally-dominated motion compared to the hot material; this is consistent with the primarily warm/cold gas makeup of the midplane material. Midplane hot gas, by contrast, is only found in injection regions, where it tends to form vertical ``chimneys''. The predominantly hot extraplanar regions show stronger verticality as a result of such hot outflows, with the velocity field reaching verticality values above 0.8 for all R-models. Further, we find that the vertical velocity components that are present near the midplane are characterized by disorganized, symmetric fluctuations, rather than coherent motion. Nevertheless, it should be noted that for both gas phases, verticality of the velocity and magnetic fields increases with increasing altitude, suggesting strong collimating effects exist within each phase. This result implies that the total high-altitude verticality is a function of both the transition of dominant filling factor (i.e. more vertically oriented hot gas becomes the primary population with increasing altitude) and of more general physical effects which act to vertically orient the velocity and magnetic fields at high altitude. \ 

The interarm region of the R8 Arm model displays a deeper minimum in $|v_z/v|$ near the midplane when compared to the arm region or non-arm models. This deeper trough in the velocity panels of \autoref{fig:verticality} is a combination of both the somewhat lower vertical velocity (numerator), due to the lack of star formation-driven feedback to launch vertical outflows, and a larger total velocity (denominator), due to shear over a larger horizontal distance. Additionally, the deeper apparent minimum of magnetic field verticality in the Arm model stems from the stronger, coherent planar component of the magnetic field introduced by the arm and associated spurs. Lastly, We find that this verticality is independent of the computation of $|v_z/v|$ or $v_z/|v|$, indicating bulk collimation of the high-altitude vertical flows.

\subsection{CR Pressure Distributions}\label{ssec:pressuredist}
\begin{figure*}
    \centering
    \subfloat{\includegraphics[width=0.5\textwidth]{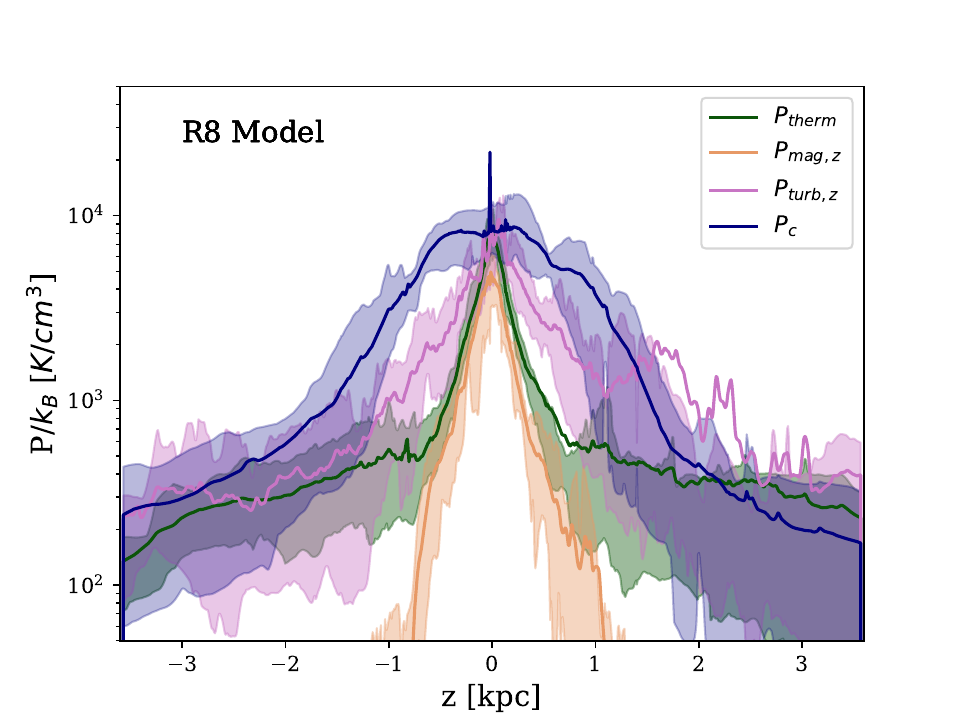}}
    \subfloat{\includegraphics[width=0.5\textwidth]{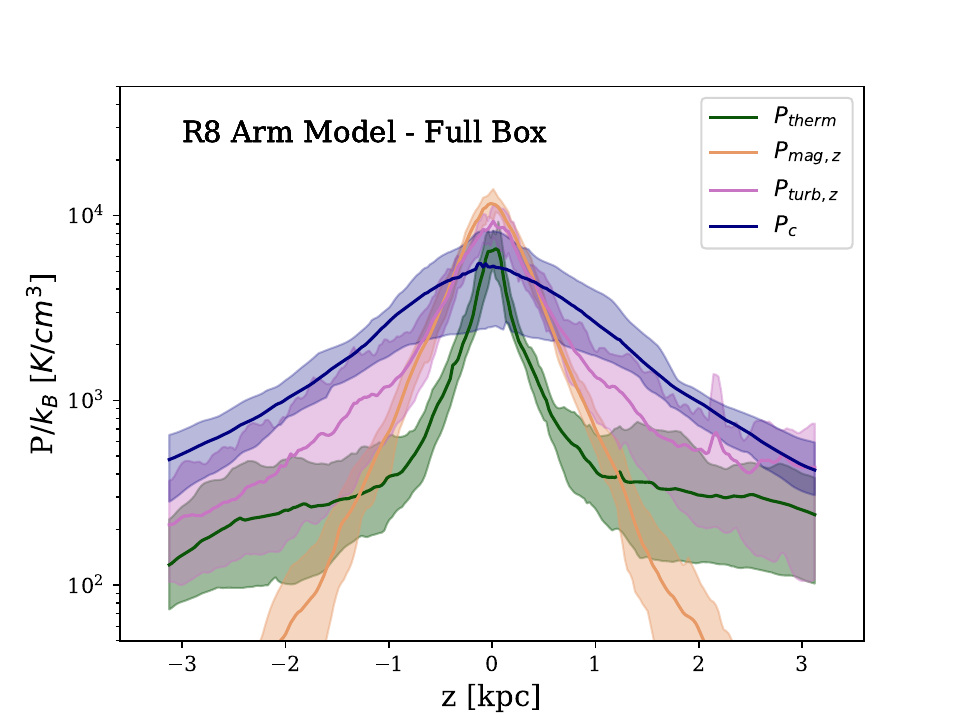}}\hfill
    \subfloat{\includegraphics[width=0.5\textwidth]{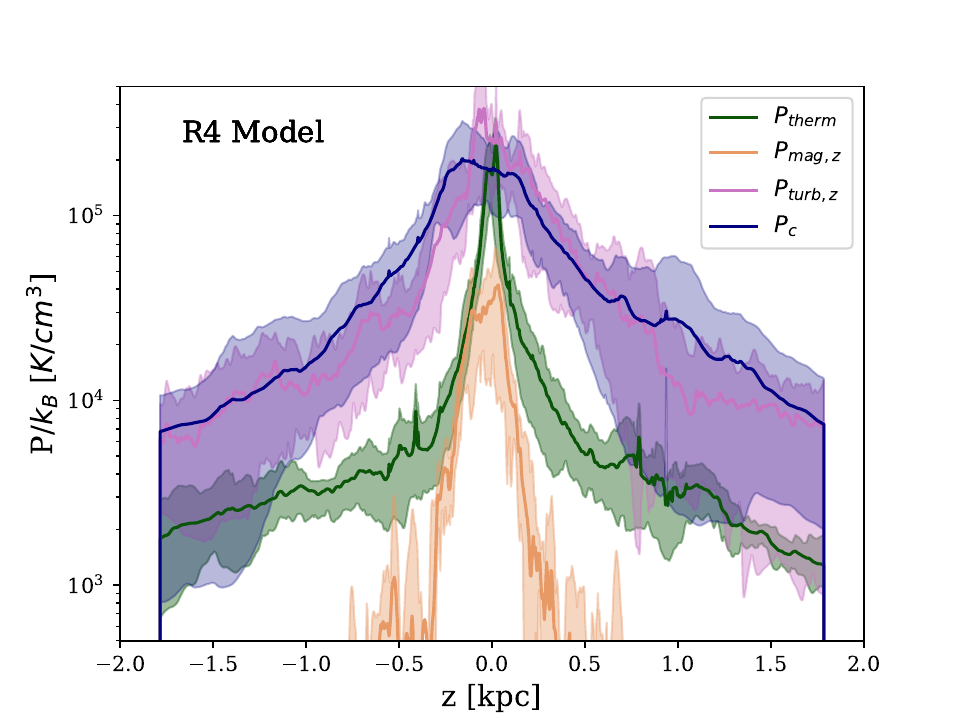}}
    \subfloat{\includegraphics[width=0.5\textwidth]{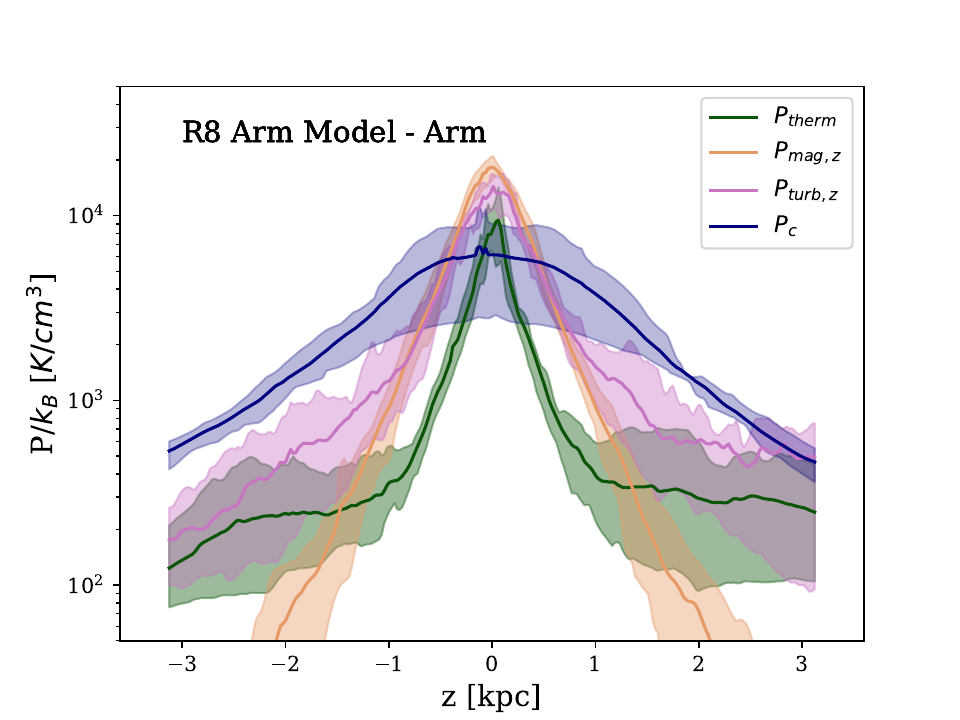}}\hfill
    \subfloat{\includegraphics[width=0.5\textwidth]{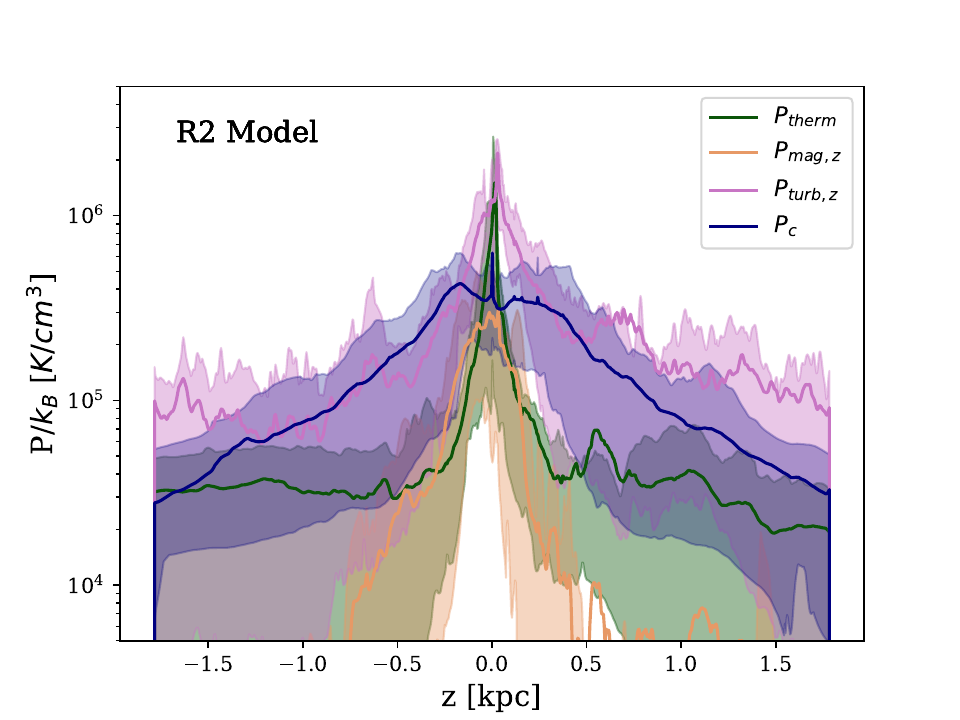}}
   \subfloat{\includegraphics[width=0.5\textwidth]{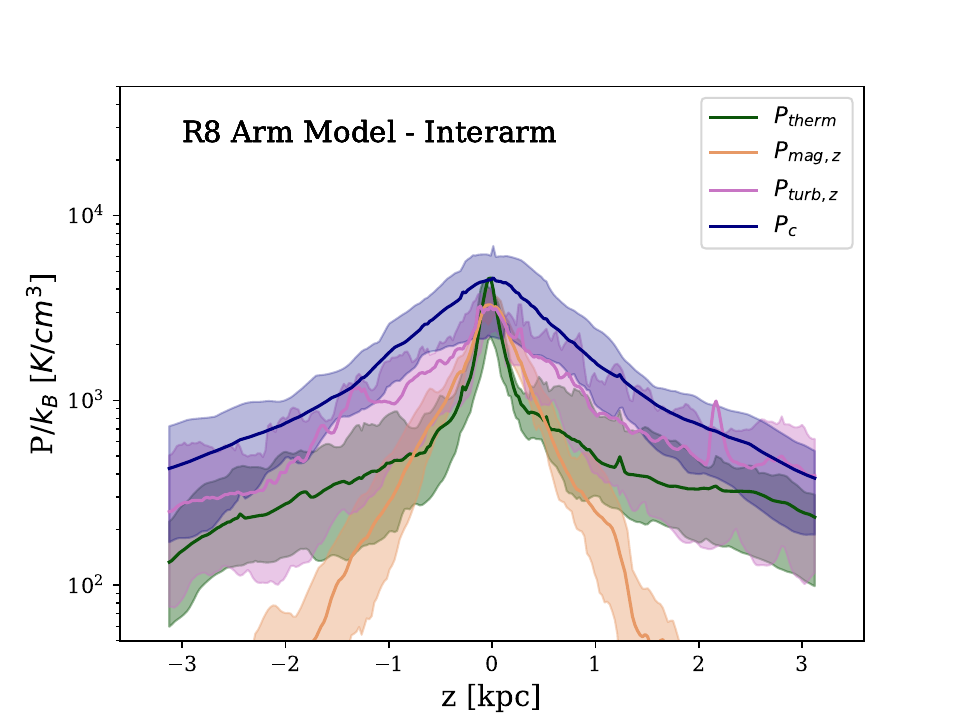}}\hfill
    \caption{Horizontally and temporally averaged pressure profiles for each model: R8 (top left), R4 (middle left), R2 (bottom left) and R8 Arm (right panels). Each panel shows the profiles of thermal pressure ($P_\mathrm{therm}$; green line), vertical magnetic stress ($P_\mathrm{mag,z}$; yellow line), vertical turbulent pressure ($P_\mathrm{turb,z}$; pink line), and CR pressure ($P_\mathrm{c}$; blue line) as a function of $z$. The shaded regions represent the 16th - 84th percentiles of temporal variation. The R8 Arm model is presented in its entirety (top right), as well as split into the arm (middle right) and interarm (bottom right) region. 
    }
    \label{fig:allpress}
\end{figure*}

In this section, we analyze the vertical distribution of CR pressure across the different models. \autoref{fig:allpress} shows, for each model, horizontally and temporally averaged vertical profiles of CR pressure ($P_\mathrm{c}$), thermal pressure ($P_\mathrm{therm}$), vertical turbulent pressure ($P_\mathrm{turb,z}$), and vertical magnetic stress ($P_\mathrm{mag, z}$). The vertical turbulent pressure is defined as $P_\mathrm{turb,z} = \rho v_\mathrm{z}^2$, while the vertical magnetic stress is defined as $P_\mathrm{mag,z} =  (B_\mathrm{x}^2+B_\mathrm{y}^2-B_\mathrm{z}^2)/8\pi$.
These profiles are taken at the end of the post-processing step. For the Arm model, the vertical pressure profiles are shown for the entire simulation regime, as well as separately for the regions interior and exterior to the spiral arm.

For all models, the midplane CR pressure is in rough equipartition with the turbulent and thermal pressures, with overlapping and comparably sized temporal variation ranges. The rough equipartition among the various pressure terms is in agreement with observations in the local ISM \citep[e.g.,][]{Boulares&Cox90,Beck01}.
By comparing the CR pressure profiles for the three R-models in \autoref{fig:allpress} to those in Figure~4 of \citet{Lucia2022}, it can be observed that the midplane CR pressures in this study are consistently a factor of $2-3$ lower than those in \citet{Lucia2022}. The difference stems from the updated transport algorithm for the post-processing step, which no longer enhances the injected CR energy content beyond the adopted 10\% of SN energy  (see \S \ref{ssec:PPmeth} and \S 3.1 in \citealt{Lucia2024}).
While in overall equipartition, we do find that the ratio of CR pressure ($P_\mathrm{c}$) to MHD pressures ($P_\mathrm{MHD}$) at the midplane decreases slightly with increasing SFR for the R-models (i.e. $P_\mathrm{c}(z=0)/P_\mathrm{MHD}(z=0)$ steadily decreases from R8 to R2), consistent with the earlier findings of \citet{Lucia2022}. However, factors other than star formation may affect the ratio of cosmic ray pressure to other pressures.  For example, the R8 and R8 Arm models have $\Sigma_\mathrm{SFR}$ differing by only 20\%, but the ratio $P_c/P_\mathrm{MHD}$ at the midplane is a factor two lower in the arm model.

While the MHD pressure profiles are tightly peaked near the midplane and decrease rapidly outside of the disk (see \citealt{Kim2017,Vijayan+20,Kim2020a} for a detailed analysis), the CR pressure profiles show a near-midplane region of roughly constant pressure, transitioning to an exponential decay (visually linear on a log-linear axes of \autoref{fig:allpress}) between 0.3 and 1 kpc above the midplane, depending on the model. This transition between uniform pressure and exponential decay can be understood as the transition between a near-midplane region with low CR scattering coefficients (high diffusion) and an extraplanar region with higher scattering coefficients (low diffusion). Referencing \autoref{fig:slice}, one can observe that the portion of the interarm region characterized by low scattering coefficients is more vertically limited than the corresponding zone in the arm region of the same box; this is directly in keeping with the narrower isobaric region in the interarm panel of \autoref{fig:allpress} when compared to the arm panel.

\begin{figure}
    \centering
\includegraphics[width=0.5\textwidth]{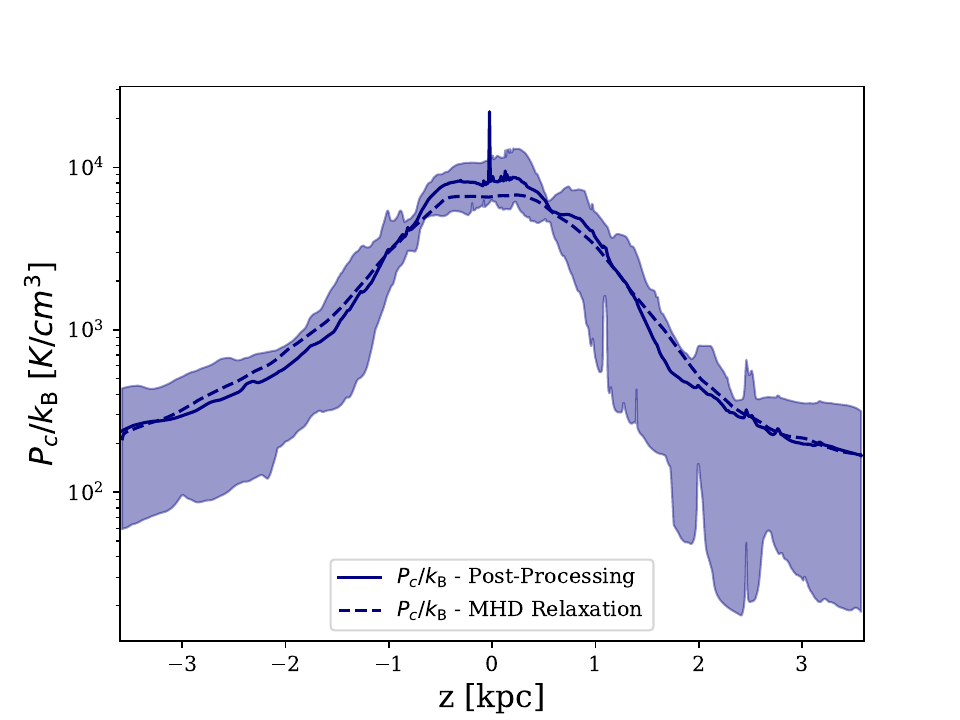}
    \caption{Horizontally and temporally averaged profiles of the CR pressure in the R8 model, chosen as an example. Profiles are shown after the post-processing step and after the MHD relaxation step. The shaded region again represents the 16th - 84th percentiles of temporal variation. Note the minimal variation in these vertical profiles with and without the MHD step.}
    \label{fig:PP_MHD_comp}
\end{figure}

 A comparison of the vertical CR pressure profiles before and after the MHD relaxation step for the R8 model is shown \autoref{fig:PP_MHD_comp}. For this model, the midplane CR pressure (averaged over $\pm300$pc) decreases very slightly (less than 20\%), from $8200 ~k_B~\mathrm{K~cm}^{-3}$ to $6700 ~k_B~\mathrm{K~cm}^{-3}$, before and after the MHD relaxation step.  We find similarly small changes for the other models.  In \autoref{tab:veff}, we report measured midplane values of the CR pressure for all models.  
 
 As alluded to in \autoref{sssec:MHD-Relaxation}, the minor variations in vertically profiled quantities produced by the MHD relaxation step are well within the range of typical temporal variations. As such, while important for controlling the the detailed structure of the CR variations (especially by reorienting magnetic and velocity fields), the MHD relaxation step is not critical for analysis of spatially-averaged quantities.

 In the interest of easing direct comparison to observations, we offer a conversion between our numerical results for CR pressure and a spectral flux at $E_\mathrm{k} = 1$~GeV. We adopt the spectral shape of CR protons proposed by \cite{Padovani+18} for the solar neighborhood, which is based on a fit to the Voyager \citep[e.g.,][]{Cummings+16} and AMS \citep[e.g.,][]{Aguilar+15} data:
 \begin{equation}\label{eq:spectrum}
     j(E_\mathrm{k}) = A~e_c \frac{E_\mathrm{k}^\delta}{(E_\mathrm{k} + E_\mathrm{t})^{2.7+\delta}} ~\mathrm{GeV}^{-1}~\mathrm{cm}^{-2}~\mathrm{s}^{-1}~\mathrm{sr}^{-1}
 \end{equation}
with $E_\mathrm{t} = 650$~MeV. Here, the kinetic energy of a given particle,  related to the total energy by $E_k=E - m_\mathrm{p}c^2$. We have assumed that $e_\mathrm{c}$ is the total energy density of CRs with $E_\mathrm{k} > E_\mathrm{t}$. The high-energy slope of the spectrum is well determined at $-2.7$, but the low-energy slope ($\delta$) is poorly constrained; we adopt a value of $-0.35$ for this work, which lies between the two values proposed by \cite{Padovani+18}. We obtain the normalization factor A as 
 \begin{equation}
     A = \left( \int_{E_t}^{\infty}\frac{EE_\mathrm{k}^\delta}{v_\mathrm{p}(E_\mathrm{k} + E_\mathrm{t})^{2.7+\delta}}dE_\mathrm{k} \right)^{-1} ~\mathrm{GeV}^{0.7}~\mathrm{cm}~\mathrm{s^{-1}}~\mathrm{sr^{-1}}\\
 \end{equation}
 where the CR proton velocity $v_\mathrm{p}\equiv c\sqrt{1-\left({m_\mathrm{p}c^2}/{E}\right)^2}$.\

Propagating these formulae for our adopted $\delta$ yields 
$A=1.57\times10^{9}~\mathrm{GeV}^{0.7}~\mathrm{cm}~\mathrm{s^{-1}}~\mathrm{sr^{-1}}$ and 
$j (E_\mathrm{k} = 1 \, \mathrm{GeV}) = 0.48 ~e_c  $ GeV$^{-1}$ cm$^{-2}$ s$^{-1}$ sr$^{-1}$, for $e_c$ in eV cm$^{-3}$. The average value of $j (E_\mathrm{k} = 1 \, \mathrm{GeV})$ within the galactic disk ($\vert z \vert < 300$ pc) is $1.03$ GeV$^{-1}$ cm$^{-2}$ s$^{-1}$ sr$^{-1}$ at the end of the post-processing step and 0.83 GeV$^{-1}$ cm$^{-2}$ s$^{-1}$  sr$^{-1}$ at the end of the MHD relaxation step. These values are a
factor of $\sim2.2-2.8$ larger than the value 
 $j (E_\mathrm{k} = 1 \, \mathrm{GeV}) = 0.37  $ GeV$^{-1}$ cm$^{-2}$  s$^{-1}$ sr$^{-1}$, estimated using the empirical formula provided in \cite{Padovani+18} with $\delta = -0.35$. Note that \cite{Padovani+18} investigates values of $\delta$ of -0.8 and 0.1 rather than $\delta = -0.35$; however, $j(E_\mathrm{k} = 1 \, \mathrm{GeV})$ is not particularly sensitive to this choice, varying only by $\sim 20\%$ compared to the $\delta = -0.35$ result. The higher CR energy density in our simulations, as previously noted in \citet{Nora2025} and \citet{Lucia2025}, could stem from a higher SFR in our simulations than the present level in the solar neighborhood, or suggest that local SNe inject less than 10\% of their energy in the form of $>1$ GeV CRs. An enhancement of the gas outflow velocity due to acceleration from large-scale gradients in the CR pressure (which we cannot capture within the present simulation framework) would also lead to a reduction in the CR energy density for a given input flux (see \autoref{sec:1Dmodel}).

\subsection{CR Pressure Scale Height and Scattering Rates}\label{ssec:scaleheight}
\begin{figure}
    \centering
    \includegraphics[width=0.5\textwidth]{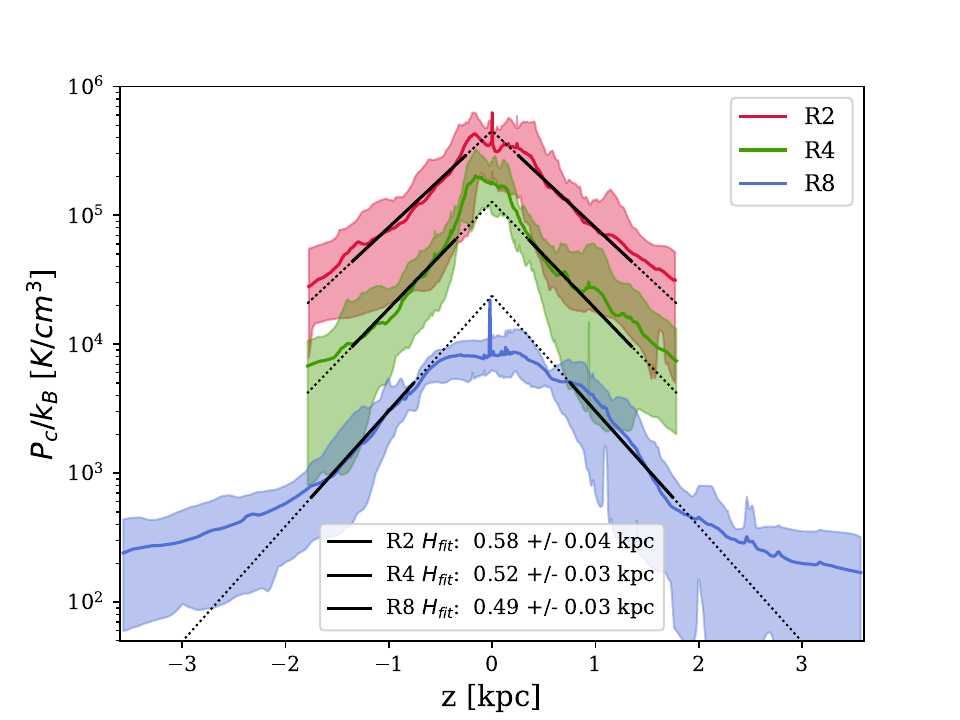}
    \includegraphics[width=0.5\textwidth]{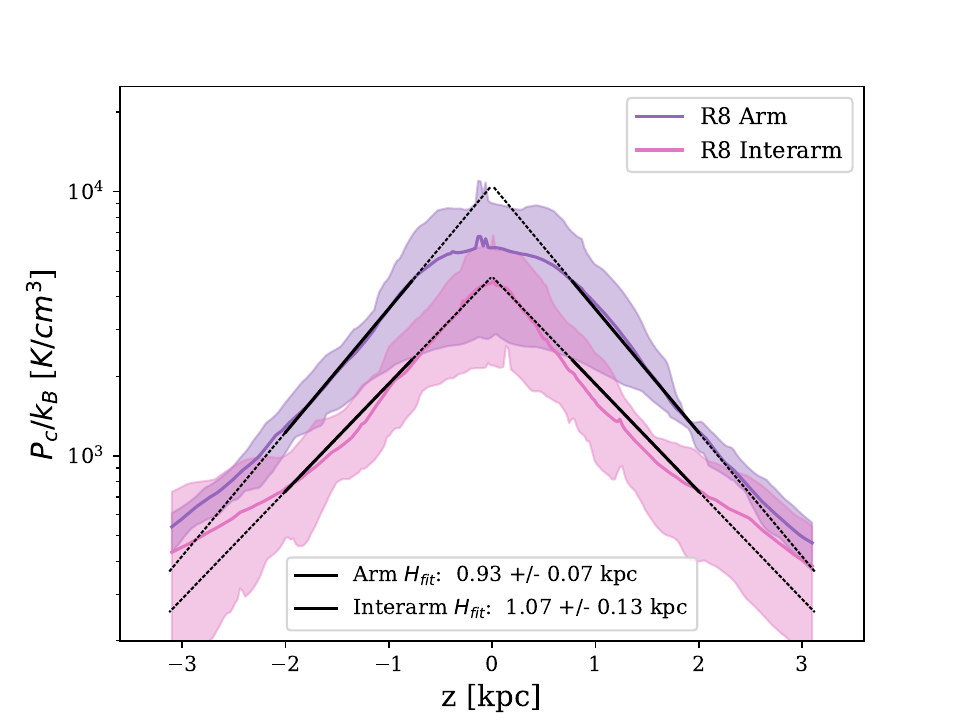}
    \caption{Comparative CR pressure profiles for the various models. The top panel shows the R-models, while the bottom panels shows the R8 Arm model, split into the arm and interarm regions. The shaded regions represent the 16th - 84th percentiles of temporal variation. Exponential fits to the extraplanar distributions of each model are shown in dotted black, with solid black indicating the fitting domain.}
    \label{fig:pcfits}
\end{figure}

In \autoref{fig:pcfits}, we compare the time-averaged vertical profiles of CR pressure across the three R-models and separately for the arm and interarm regions of the Arm model. We derive values for the CR scale height (outside of the midplane region) by fitting to the functional form: 
\begin{equation}
    P_\mathrm{c} = P_\mathrm{0}e^{-|z/H_\mathrm{c}|}
\end{equation}
where $P_{0}$ is a normalization factor and $H_\mathrm{c}$ is the constant CR pressure scale height.\
The resultant scale heights are overplotted in \autoref{fig:pcfits}, and prove to be accurately representative of the pressure structure. The fitting domains are shown in solid black on this figure and chosen to sample roughly 1.5 kpc vertical regions immediately above and below each model's respective disk. However, the fit results prove to be fairly insensitive to the choice of extraplanar fitting domain. 

The scale heights for the R-models are similar, lying between 0.6 kpc and 0.5 kpc, with a scatter of the order of the fit uncertainties. Similar fits to the post-MHD relaxation CR distributions yield scale heights in statistical agreement with the corresponding results at the end of post-processing, with equivalent goodness of fit. 
The scale heights for the Arm model are slightly higher than those of the R-models, around 1 kpc. Exact CR scale heights for each model are provided in the legend of \autoref{fig:pcfits} and \autoref{tab:veff}.\ It is striking that there is so little variation in the measured CR scale heights, considering the large factors by which many (but not all) other properties vary across the models.  
The R-models differ by almost three orders of magnitude in SFR surface density and span roughly 2 orders of magnitude in midplane CR density. Nevertheless, their CR scale heights are nearly identical. For the R8 Arm model, the SFR surface density averaged over the domain is very similar to that in the R8 model, and the midplane CR density is also quite comparable to the sans-arm R8 model, yet the vertical CR scale height differs by a factor of $\lesssim 2$. 
The apparent independence of the vertical CR scale height from any of these parameters presents a potential advantage for modeling efforts (see \S \ref{sec:subgridmodel}), but raises important questions about what factors govern these scale heights. \ 

\begin{figure}
    \centering
    \includegraphics[width=0.5\textwidth]{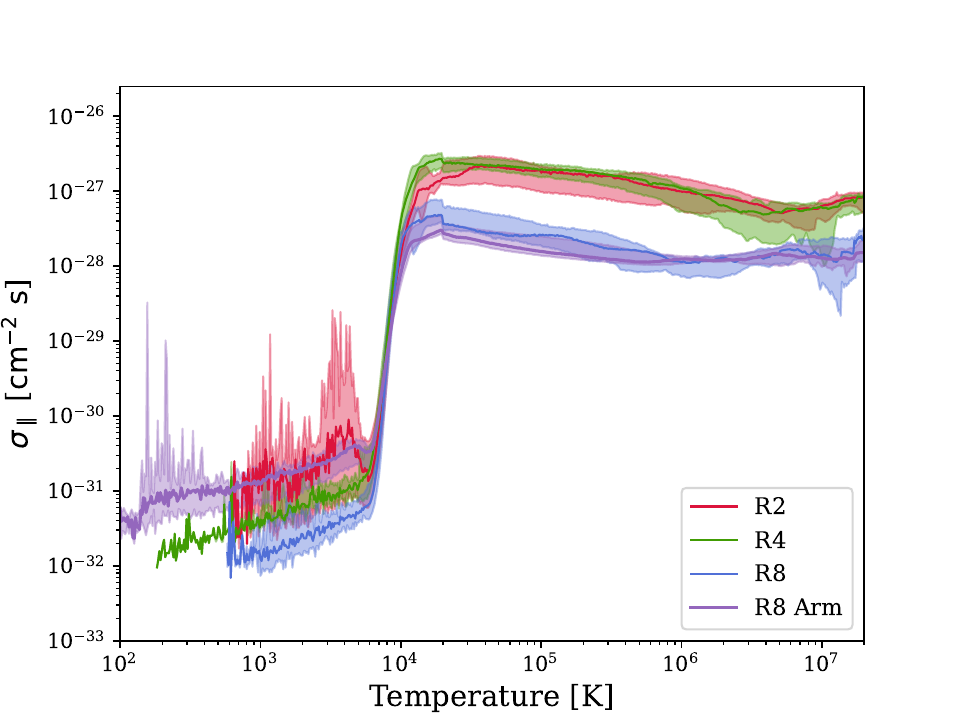}
    \caption{Temporal medians of the median CR parallel scattering coefficient ($\sigma_\parallel$) profiles as a function of temperature. The shaded regions represent the 16th - 84th percentiles of temporal variation. The arm and interarm regions of the R8 Arm model displayed no meaningful variation, and are thus not differentiated for this figure.}
    \label{fig:sigmavstemp}
\end{figure}

Since CRs are confined by scattering, a natural avenue of investigation to understand scale heights is to explore how CR scattering coefficients vary across models. \autoref{fig:sigmavstemp} shows the time-averaged profiles of the parallel CR scattering coefficient as a function of temperature for each model. 
We emphasize that the scattering rates shown here are for GeV CRs, since it is the number density ($n_1$ from \autoref{eq:n1}) at this energy that is used in evaluating \autoref{eq:signll} and \autoref{eq:sigin}; at higher energy $n_1$ is smaller and the scattering rates  are lower. 
Because the arm and interarm regions display nearly the same profiles, the Arm model is presented without differentiation of these regions. All four models display a sharp transition around $10^4$ K, each seeing the scattering coefficient increase by $4-6$ orders of magnitude. This is a result of the rapid change in ionization fraction around this temperature, prompting a transition between the NLL and IN damping regimes. \ 

In the low-temperature regime, where IN damping is dominant, the value of the scattering coefficient in the R-models increases monotonically from R8 to R2, consistent with the similarly increasing CR pressure in these models, in accordance with Equations ~\ref{eq:sigin}-\ref{eq:n1}. The R8 Arm model, despite possessing comparable CR pressures to the R8 model, displays a higher IN scattering rate due to having a lower typical density at a given temperature. Nevertheless, the IN scattering coefficient is extremely low across all models, helping explain the flat midplane CR distribution via efficient diffusion.\

Above the $10^4$~K break, the gas is primarily ionized, and the scattering coefficient is governed by NLL damping. As before, the scattering coefficient increases from R8 to R2 due to the growing CR pressure. However, we note that R2 and R4 exhibit similar scattering coefficients. As discussed in \cite{Lucia2022}, both advection and streaming are more efficient in R2 than in R4 (see also \S\ref{sec:1Dmodel}), particularly in low-density gas at high latitudes. This more effective transport mechanism results in similar CR pressure gradients along the magnetic field direction for R2 and R4 ($\sigma_\parallel \propto {|\hat{\mathbf{B}}\cdot\nabla P_\mathrm{c}|}^{0.5}$, see Eq. \ref{eq:signll}).

To test the possibility that CR scale heights are controlled solely by scattering, we consider the regime where diffusion dominates CR propagation. In this case, the CR energy flux equation (\autoref{eq:momtrans}) in the vertical direction would reduce to:
\begin{equation}\label{eq:Hcdiff}
    F_\mathrm{c,z} \approx  \dfrac{1}{\sigma_\parallel} \dfrac{d P_\mathrm{c}}{dz},
\end{equation} 
which would imply the CR pressure scale height would be
\begin{equation}
    H_\mathrm{c}^{\mathrm{diff}} \approx -\frac{P_\mathrm{c}}{\sigma_\parallel |F_\mathrm{c,z}|}
\end{equation}
with $|dP_\mathrm{c}/dz| \equiv P_\mathrm{c} / H_\mathrm{c}^{\mathrm{diff}}$.\ 

As can be seen in \autoref{fig:sigmavstemp}, in the NLL regime of the extraplanar region, $\sigma_\parallel$ increases by roughly an order of magnitude between the R8 models and the R2/R4 models and is roughly equivalent between the R8 and R8 Arm models. The ratio $|F_\mathrm{c,z}|/P_\mathrm{c,z}$ experiences more inter-snapshot variability, but overall also increases by $0.5-1$ dex between the R8 model and the R2/R4 models, and is once again consistent between the R8 and R8 Arm models. Thus, a “diffusion dominant'' model would imply the CR scale height should decrease by more than an order of magnitude between R8 and R2/R4 and that R8 and R8 Arm should have approximately equal scale heights. This conclusion is seen to hold whether $|F_\mathrm{c,z}|/P_\mathrm{c,z}$ is computed as a vertical profile of the local ratio or by dividing the average CR input flux (defined as half the total input energy divided by box footprint) by the average disk pressure.\

Using \autoref{eq:Hcdiff}, we would estimate that $H_c^{\mathrm{diff}}\sim 40-80$~pc in the R8 model and $H_c^{\mathrm{diff}}\sim 1-5$~pc in the R2/R4 models.  All of these are well below the true scale heights of $500-600$~pc in all three R-models (see \autoref{fig:pcfits}). Further, the estimates of $H_c^{\mathrm{diff}}$ for the R8 Arm model roughly align with the estimates for the standard R8 model, though their actual scale heights differ by almost a factor of 2. Both the magnitude and inter-model variation of these simple predictions stand at odds with the actual scale heights, implying diffusion is not responsible for setting the CR scale height in the extraplanar region. We additionally note that these conclusions do not qualitatively change post-MHD relaxation. 

In short, the $10^4$~K transition between IN and NLL damping mechanisms is critical for setting the boundary between very low scattering (extremely high diffusion) in the midplane and higher scattering (but still appreciable diffusion) in the extraplanar region. However, for the GeV CRs considered in this work, the level of diffusion in the extraplanar region is far too low to be able to explain the measured CR pressure gradients there.  
An alternative model is thus required.\

\section{1D Dynamical Model of CR transport}\label{sec:1Dmodel}

The previous section demonstrates that a pure-diffusion model is unable to explain the vertical profiles measured in our simulations of GeV CRs. In this section, we develop an alternative model in which CRs are primarily transported dynamically, by a combination of gas advection and streaming along the magnetic field lines (as constrained by frequent Alfv\'en wave scattering). 

\subsection{Expectation for Dynamically-Dominated  CR Transport}\label{ssec:dynamtrans}
Before developing a 1D dynamical model for CR transport, we first verify that it is a reasonable approximation to assume that transport of GeV CRs in the extraplanar region is primarily governed by dynamical mechanisms, i.e. streaming plus advection.\
In this limit, 
\autoref{eq:engtrans} and \autoref{eq:momtrans} under the steady-state approximation ($\partial/\partial t \approx 0$)
become
\begin{equation} \label{eq:engtrans1d}
\begin{split}
  d_zF_c = (v_\mathrm{z} + v_\mathrm{s,z}&)~d_zP_c +S - \Lambda_\mathrm{coll} n_\mathrm{H} e_c  
\end{split}
\end{equation}
\begin{equation}\label{eq:momtrans1d}
\begin{split}
  d_zP_c = &- \sigma_\parallel [ F_c - (v_\mathrm{z} + v_\mathrm{s,z})(P_c+e_c)] 
\end{split}
\end{equation}

We note that we have omitted the collisional loss term in the flux equation (\autoref{eq:momtrans1d}) because it is negligible compared to the other terms.

In the regime where diffusion is subdominant (very low $1/\sigma_\parallel$) and CR transport is governed by advection and streaming, \autoref{eq:momtrans1d} reduces to:
\begin{equation}\label{eq:fhighsig}
    F_c = (v_\mathrm{z} + v_\mathrm{s,z})(P_c+e_c)
\end{equation}
Additionally, because the low gas density and lack of young clusters in the extraplanar region render collisional losses and supernova injection, respectively, unimportant, \autoref{eq:engtrans1d} 
simplifies to: $ d_zF_c =(v_\mathrm{z} + v_\mathrm{s,z})(d_z P_c)$. Combining these expressions (and using $P_c=e_c/3$) yields
\begin{equation}
    4~d_z[(v_\mathrm{z} + v_\mathrm{s,z})P_c] = (v_\mathrm{z} + v_\mathrm{s,z})d_z P_c,
\end{equation}
which can be rearranged to yield:
\begin{equation}\label{eq:hcr_vs_hvel}
    d_z \mathrm{ln}P_c = -\frac{4}{3} d_z \mathrm{ln}(v_\mathrm{z} + v_\mathrm{s,z}).
\end{equation}
Thus, to the extent that dynamical transport mechanisms are predominantly responsible for controlling the CR distribution in the low-density, source-free, high-scattering extraplanar region, the vertical CR pressure profile should follow the form:
\begin{equation}\label{eq:analyticPc}
    \frac{P_\mathrm{c}}{P_0} = \left(\frac{v_\mathrm{tot}}{v_0}\right)^{-4/3}
\end{equation}
where $v_\mathrm{tot} \equiv v_\mathrm{z} + v_\mathrm{s,z}$ is the total dynamical transport speed, and $P_0$ and $v_0$ are normalization constants. This relation also implies $H_\mathrm{c} = (4/3) H_\mathrm{a}$, where $H_\mathrm{c}$ is the CR scale height, and $H_\mathrm{a}\equiv |d_z\ln(v_\mathrm{tot})|^{-1}$ is the acceleration length scale (the scale height of the total dynamical velocity $v_\mathrm{tot}$).\

\begin{figure*} 
    \centering
    \subfloat{\includegraphics[width=0.5\textwidth]{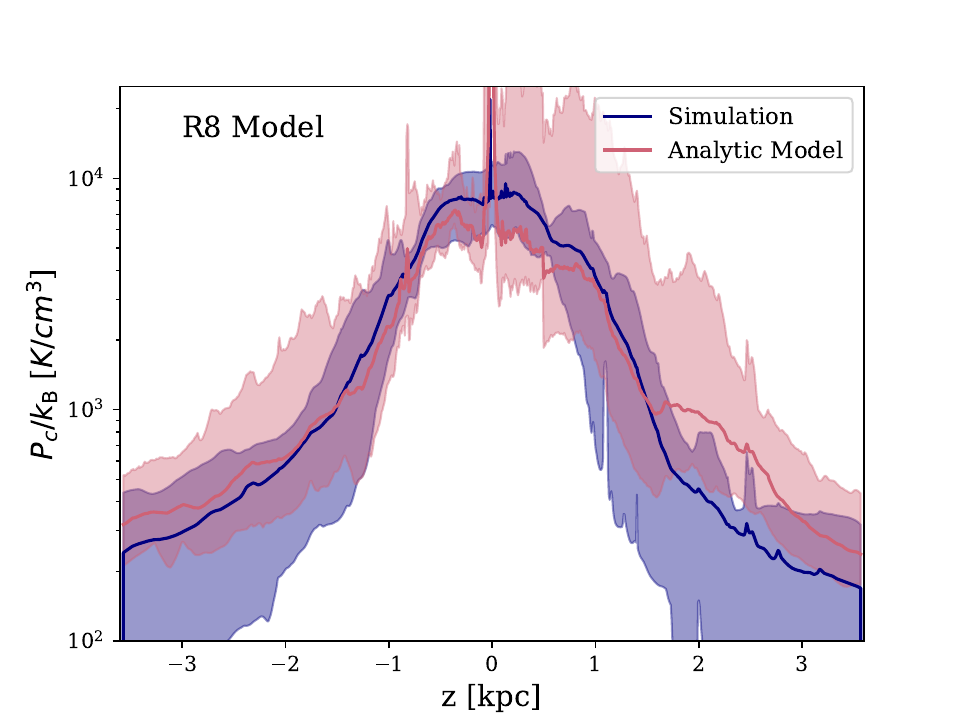}}
    \subfloat{\includegraphics[width=0.5\textwidth]{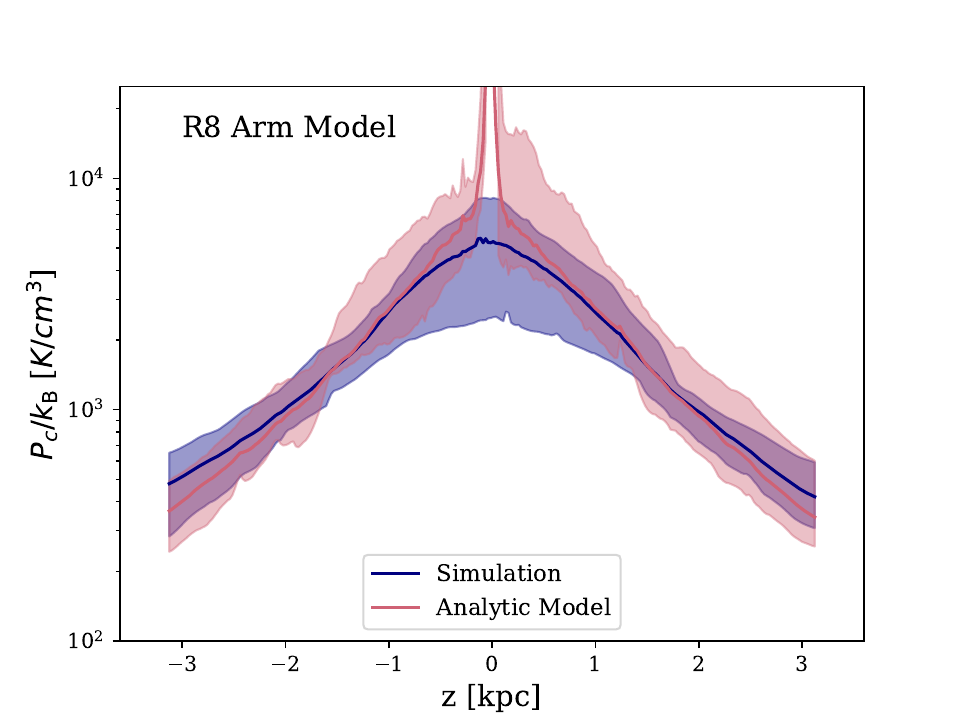}}\hfill
    \subfloat{\includegraphics[width=0.5\textwidth]{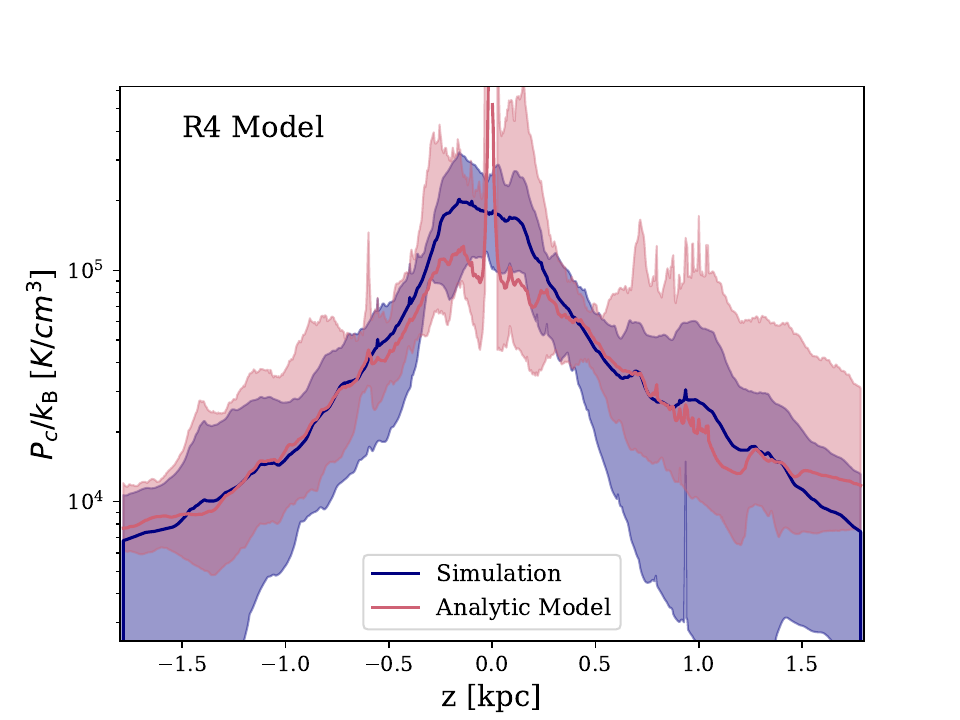}}
    \subfloat{\includegraphics[width=0.5\textwidth]{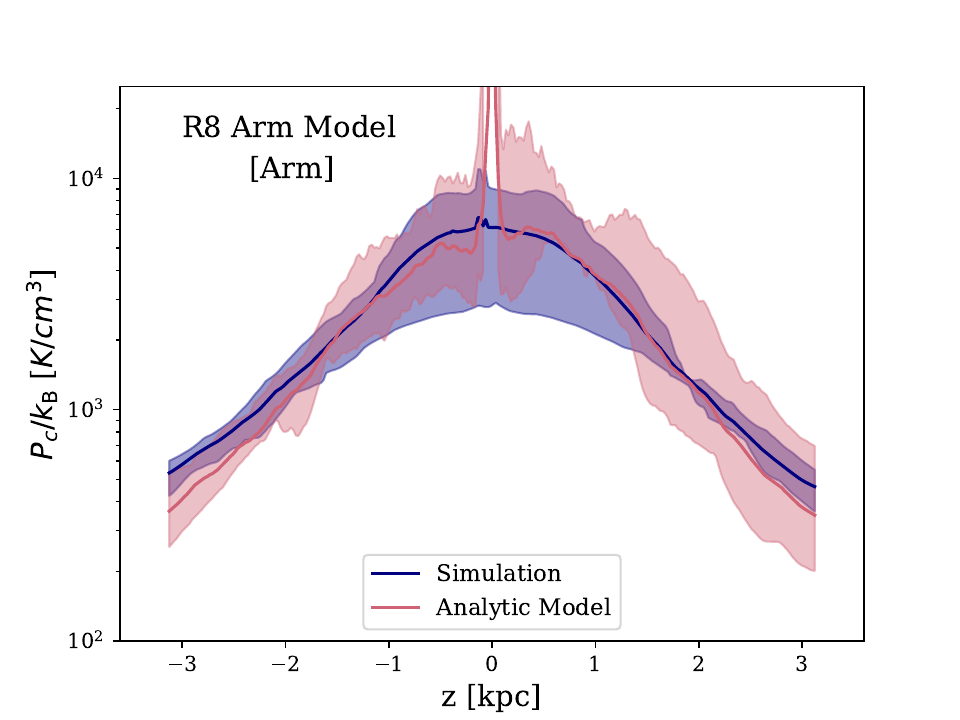}}\hfill
    \subfloat{\includegraphics[width=0.5\textwidth]{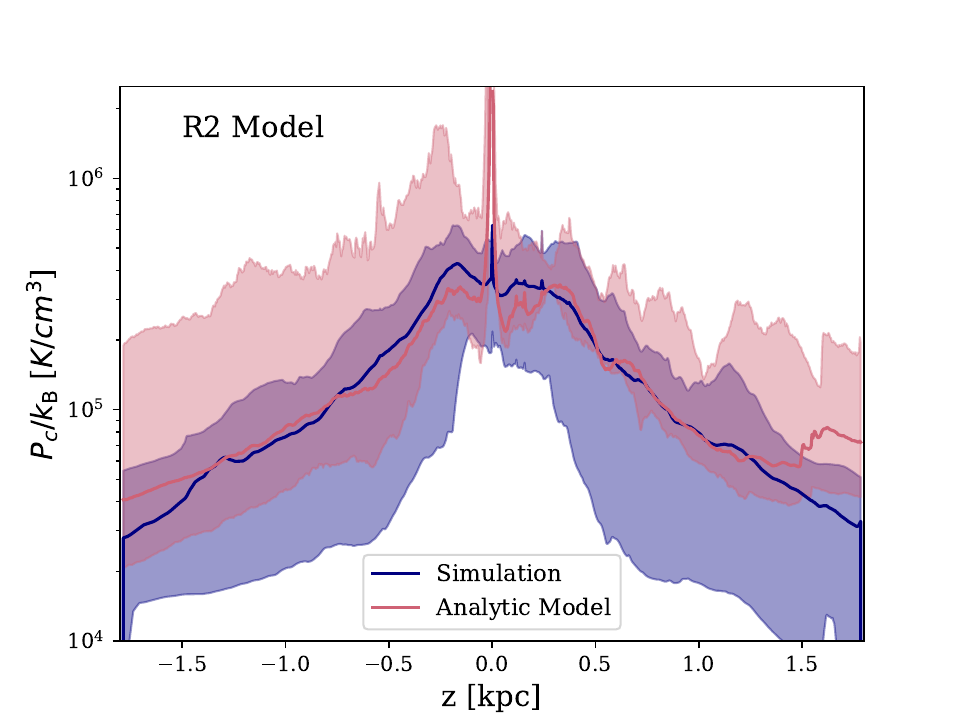}}
    \subfloat{\includegraphics[width=0.5\textwidth]{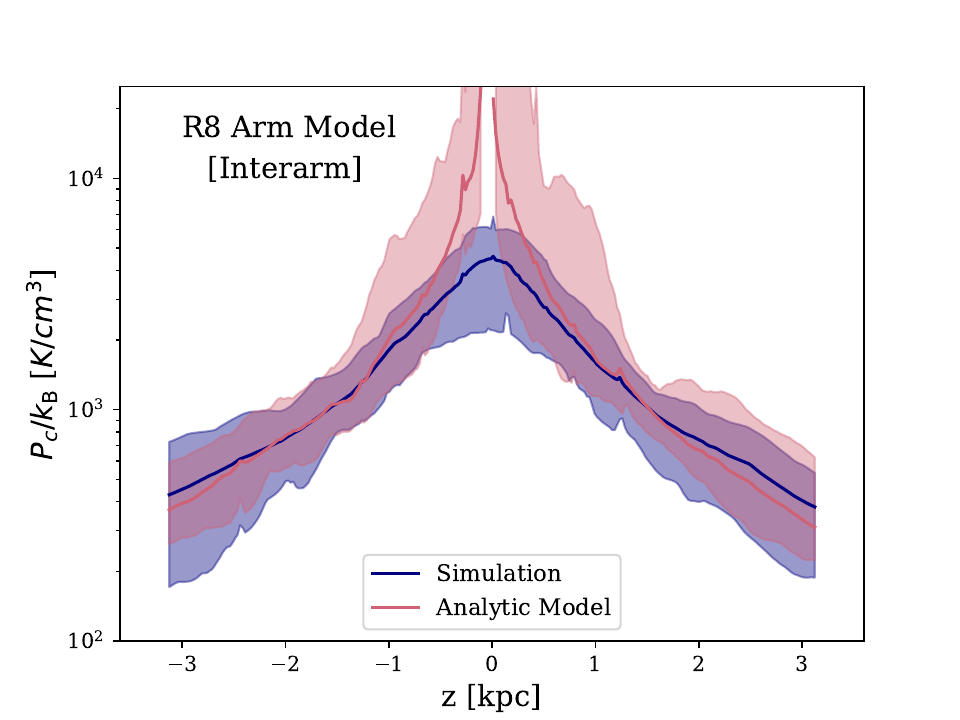}}\hfill
   
    \caption{Comparisons between the normalized vertical CR pressure profiles obtained from the simulations (blue lines) and those predicted by the analytic model assuming dynamically controlled vertical CR transport (red lines; \autoref{eq:analyticPc}). The pressure profile is the temporal mean of unweighted horizontal-mean profiles, while the analytic model profile (computed from \autoref{eq:analyticPc}) uses the temporal mean of CR pressure-weighted horizontal-mean velocity profiles, to better capture the spatial regions most responsible for CR transport. The shaded regions represent the 16th - 84th percentiles of temporal variation. Each panel refers to a different model. The R8 Arm model is shown both in its entirety and split into an arm and interarm region. The analytic models are normalized to the CR pressure at $\vert z\vert = 0.75$ kpc for R2 and R4, and $\vert z\vert = 1.5$ kpc in the R8 and R8 Arm models.} 
    \label{fig:pc_vs_v}
\end{figure*}

\autoref{fig:pc_vs_v} displays, for each environment, the horizontally and temporally averaged profiles of CR pressure measured in the simulations, in comparison to the corresponding analytic predictions using $v_\mathrm{tot}$ in case of dynamically controlled CR transport (\autoref{eq:analyticPc}). The analytic models are normalized to the CR pressures at arbitrary altitudes for visual clarity ($\vert z\vert = 0.75$ kpc for R2 and R4, and $\vert z\vert = 1.5$ kpc in the R8 and R8 Arm models). This normalization is performed separately above and below the midplane, as each side of the bipolar outflows are independent. In the extraplanar region, the predicted and measured profiles display clear and robust agreement for all models. \

This result confirms that dynamical transport mechanisms dominate vertical CR transport. 
It also helps us interpret the minimal variation in CR pressure scale heights between the three R-models: while the galactic environments probed by these simulations vary extensively, their vertical velocity structures are quite similar.  This is perhaps unsurprising, given that the characteristic turbulent velocity increases only slightly with star formation rate, and that the density scale height is similar in all models.\

The slightly larger CR scale height for the R8 Arm model can be explained as follows. The spiral arm concentrates (and mildly enhances) star formation. This concentration leads to a locally higher SFR surface density in the arm compared to the R8 model, but retains a comparably deep galactic potential. The result is more efficient outflow launching and a longer transport velocity scale height (the stronger and more ordered magnetic field in the Arm model increases the Alfv\'en velocity, further accentuating the effect). Thus, the R8 Arm model displays a somewhat larger CR scale height.\

In summary, we find that the vertical scale height of the CR pressure is primarily governed by underlying MHD parameters, namely the scale height of the gas velocity (set by feedback injection and the gravitational potential) and ion Alfv\'en velocity (controlled by the gas density distribution and magnetic field characteristics).

Our results are bolstered by the independent findings of \citealt{Reichenherzer2022}. The simulations presented in this work consider CRs with substantially larger energies, thus lying in the regime of extrinsic turbulent confinement rather than self-confinement. This limits any potential for direct comparison to our results, but nonetheless offers an informative point of comparison. This work suggests that the observed steepening of the high-energy ($E_\mathrm{k} > 100~\mathrm{GeV}$) CR spectral slope in the inner galaxy is inconsistent with a diffusion-dominated model when extending beyond quasi-linear theory. Extrapolation of their findings to the energies considered in our work (see their Eq. 24 and 25) suggests GeV CRs should indeed be dominated by dynamical transport across all our simulated environments, in keeping with our findings. This agreement between both different computational approaches and widely different CR transport regimes offers further independent and observationally-motivated support to our findings.

As previously discussed, the velocity profiles in the MHD-relaxed simulations are somewhat unreliable in the current implementation. Repetition of this particular investigation is therefore not fully possible for the simulations after MHD back-reaction. However, given that the scattering coefficients demonstrate no significant changes following the MHD relaxation step, we strongly expect CR transport to continue to be dominated by advection and streaming. As a result, while the validity of this model is expected to continue to hold under the influence of a fully self-consistent MHD back-reaction, further validation in this regime must await future work.\

\subsection{Prediction for Midplane CR Pressure}\label{ssec:1DmodelPdisk}
Having confirmed that diffusion and dynamical transport dominate the disk and extraplanar transport, respectively, we now proceed to develop a simplified 1D model for vertical CR transport, in which all quantities will represent horizontal averages. We start by solving the vertical component of \autoref{eq:momtrans} in the mostly neutral disk (or arm) region. In this region, the low scattering rate renders $\partial P_\parallel \approx 0$. We note that diffusive transport operates almost exclusively along the magnetic field lines, and thus high diffusivity does not automatically imply efficient diffusive transport throughout the disk's vertical extent. For magnetic geometries with field lines entirely confined to the plane, CRs would be unable to access higher altitudes, regardless of the field-aligned diffusion. Detailed analysis of field lines (M. Winter et al, in prep) shows, however, that while the magnetic geometry in the disk is fairly horizontal on average (see \autoref{fig:verticality}), individual field lines oscillate vertically, with sufficiently large vertical excursions as to allow field-confined CRs to rapidly reach the vertical edges of the disk.\

Given the low scattering rate within the midplane region, integration of \autoref{eq:engtrans} over the whole region yields:
\begin{equation}\label{eq:ft1}
    F_\mathrm{out} = F_\mathrm{in} - \dfrac{3}{2}N_\mathrm{disk} \Lambda_{\mathrm{coll}}P_{\mathrm{disk}}
\end{equation}
where $F_\mathrm{out}$ is the vertical CR flux at the interface between the disk and extraplanar regions, $F_\mathrm{in}$ is the input flux to each side of the disk, 
$P_{\mathrm{disk}}$ is the mean CR pressure in the disk, $\Lambda_\mathrm{coll}$ is the CR collisional coefficient defined in \S\ref{ssec:PPmeth}, and $N_{\mathrm{disk}}$ is the hydrogen column density, integrated over the vertical extent of the disk. Since the majority of the gas mass in the simulation box resides in the neutral gas phase within the disk, to high accuracy, $N_{\mathrm{disk}} \approx N$, with $N$ the column density integrated over the entire simulation box.\ 

\autoref{eq:ft1} relates the flux emerging from the midplane, $F_\mathrm{out}$, to the injected flux and integrated losses.  At the interface between disk and extraplanar regions, both the flux and the CR pressure must match on the two sides. On the extraplanar side, from \autoref{eq:fhighsig}
we are motivated to write $F_\mathrm{out} = 4 v_{\mathrm{eff}} P_c$, where we have 
replaced $v_{\mathrm{tot}}$ with $v_{\mathrm{eff}}$, the effective CR transport speed.

By combining 
\begin{equation}
    F_\mathrm{out} = 4 v_{\mathrm{eff}} P_\mathrm{c,disk}
\end{equation}
with 
\autoref{eq:ft1} 
we obtain:
\begin{equation}\label{eq:Pdisk}
    P_{\mathrm{disk}} = \frac{F_\mathrm{in}}{4 v_\mathrm{eff} + \dfrac{3}{2} N \Lambda_{\mathrm{coll}}}.  
\end{equation}

We thus expect the midplane CR pressure/energy density to vary directly with the injected CR flux and inversely with both the effective transport velocity and the rate of collisional losses to the gas. With $N$ being the gas column density,
and the input flux directly coming from the supernova injection rate (which itself is a function of the SFR), \autoref{eq:Pdisk} provides a straightforward method for predicting midplane CR pressure from basic environmental parameters -- provided $v_\mathrm{eff}$ can be reliably determined.\

For the simulations presented in this work, $P_\mathrm{c,disk}$, $F_\mathrm{in}$, and $N$ are known, allowing a determination of $v_\mathrm{eff}$ for calibration purposes. Being interested principally in a single characteristic vertical transport speed, we perform a temporal and horizontal average of all quantities. As the CR pressure varies minimally within the disk, $P_{\mathrm{disk}}$ is computed as the mean pressure within 300~pc of the midplane. $F_\mathrm{in}$ is computed as $0.5 \dot{E}_\mathrm{c}/(L_{x} L_{y})$, with $\dot{E}_\mathrm{c}$ the energy injection rate (\autoref{eq:eninj}). This method is consistent with our horizontal-average approach and -- given the efficiency with which CRs are transported in the midplane and the resultant uniformity of pressure -- provides a good representation of the actual CR flux injection. $N$ is computed as $\Sigma_\mathrm{gas} / (1.4 m_\mathrm{p})$, with $\Sigma_\mathrm{gas}$ from \autoref{tab:models}. The results of this analysis are presented in \autoref{tab:veff}. The resulting effective velocities are relatively low, with their mean values increasing from R8 ($v_\mathrm{eff} \simeq 15$ km~s$^{-1}$) to R4 ($v_\mathrm{eff} \simeq 35$ km~s$^{-1}$) to R2 ($v_\mathrm{eff} \simeq 57$ km~s$^{-1}$). The mean effective velocity of the Arm R8 model ($v_\mathrm{eff} \simeq 22$ km~s$^{-1}$) is slightly larger than that of the R8 model.

\begin{figure*}
    \centering
    \includegraphics[width=0.8\textwidth]{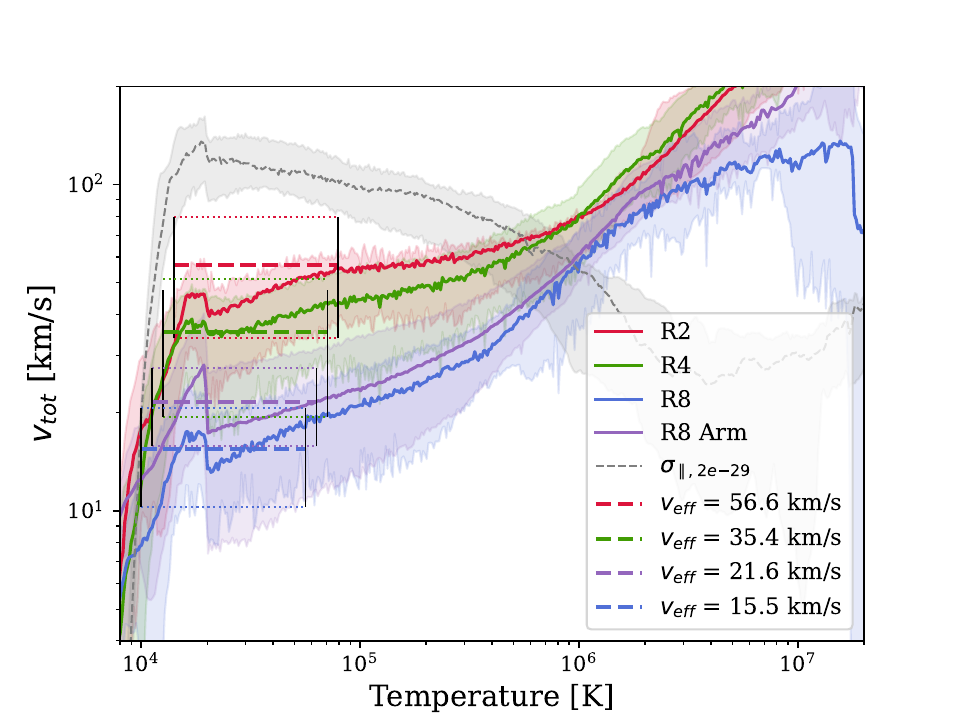}
    \caption{CR pressure-weighted mean vertical dynamical velocity ($v_\mathrm{tot} = v_\mathrm{z} + v_{s,z}$) as a function of temperature for all models (solid lines). Different colors correspond to different models: R2 (red), R4 (green), R8 (blue), R8 Arm (purple). The arm and interarm regions of the R8 Arm model display no informative differences in this regime and are therefore combined for clarity. Horizontal dashed lines display the mean value of $v_\mathrm{eff}$ (defined via \autoref{eq:Pdisk}) corresponding to each model, with $1\sigma$ temporal variations shown with correspondingly colored dotted lines. These horizontal lines, as well as the black vertical lines that bound the variation ranges, are offset for legibility, rather than to imply any differences in their applicable temperature domain. The gray dashed curve represents a characteristic profile of the scattering coefficient, normalized for visual comparison. The good agreement between $v_\mathrm{eff}$ and $v_\mathrm{tot}$ in the regime where scattering transitions from low to high values motivates \autoref{eq:final_1d_model}.}
    \label{fig:vel_vs_t}
\end{figure*}

Recalling our previous conclusion that CR transport out of the disk is primarily driven by dynamical mechanisms, we now aim to investigate whether the inferred effective transport velocities are indeed comparable with the typical dynamical velocities in the extraplanar region. In \autoref{fig:vel_vs_t}, the solid curves show the CR pressure-weighted  mean vertical dynamical velocity ($v_\mathrm{tot} \equiv v_z + v_\mathrm{s,z}$) as a function of gas temperature, while the dashed lines represent the computed $v_\mathrm{eff}$ for the models of the corresponding color. The R8 Arm model is not differentiated into an arm and interarm region on the grounds of clarity: while the Alfv\'en velocity in the arm region is higher, a serendipitously lower value of the gas velocity offsets this change, and both regions display essentially identical velocity curves. The dotted lines hemmed by black vertical lines denote the $1\sigma$ temporal variations of $v_\mathrm{eff}$. A normalized, representative curve of the CR scattering coefficient from the R8 model is included for reference.\

\begin{table*}
    \begin{tabular}{ccccccc}
        \hline
        \hline
         Model & $H_\mathrm{c}$ & $P_\mathrm{c,disk}/k_B$ & $F_\mathrm{in}$ &  N &$v_\mathrm{eff}$ & $v_\mathrm{tot}$ \\
         & [kpc] & [K $\mathrm{cm}^{-3}$] & [erg kpc$^{-2}$ yr$^{-1}$] & [cm$^{-2}$] & [km s$^{-1}$] &  [km s$^{-1}$] \\
         (1) & (2) & (3) & (4) & (5) & (6) & (7)\\
         \vspace{-8pt}\\
         \hline
         R2 & 0.58 $\pm$ 0.04 & 3.8 $\pm$ 1.5 $\times$ 10$^5$ & 4.7 $\pm$ 3.0 $\times$ 10$^{47}$& 6.6 $\pm$ 2.0$\times$ 10$^{21}$&56.6 $\pm$ 23& 46.0 $\pm$ 13\\
         R4 & 0.52 $\pm$ 0.03 & 1.8 $\pm$ 0.8 $\times$ 10$^5$ & 1.1 $\pm$ 0.5 $\times$ 10$^{47}$ & 3.1 $\pm$ 0.6$\times$ 10$^{21}$ & 35.4 $\pm$ 16& 37.7 $\pm$ 12\\
         R8 & 0.49 $\pm$ 0.03 & 8.1 $\pm$ 2.3 $\times$ 10$^3$& 2.4 $\pm$ 1.1 $\times$ 10$^{45}$ & 8.2 $\pm$ 0.4$\times$ 10$^{20}$&15.5 $\pm$ 5.2& 16.4 $\pm$ 5.6\\
         Arm R8 & 1.02 $\pm$ 0.04 & 5.3 $\pm$ 2.5 $\times$ 10$^3$ & 1.9 $\pm$ 0.7 $\times$ 10$^{45}$ & 8.3 $\pm$ 0.5$\times$ 10$^{20}$ &21.6 $\pm$ 5.9& 20.4 $\pm$ 10\\
         \vspace{-10pt}\\
         \hline

    \end{tabular}
    \caption{Tabulated parameters of relevance to our dynamical model of CR transport, along with their inter-snapshot variation. $H_\mathrm{c}$ is the CR scale length, $P_\mathrm{c,disk}/k_B$ is the average CR pressure within the central 300 pc of the disk, $F_\mathrm{in}$ is the total CR flux injected by SNe, and N is the average gas column density.  The velocity $v_\mathrm{eff}$ encodes the effective CR transport speed, as calculated via \autoref{eq:Pdisk}, while $v_\mathrm{tot}=v_\mathrm{z} + v_\mathrm{s,z}$ is the measured sum of vertical streaming and vertical advective motion, averaged over the temperature range of 15,000K to 65,000K (see \autoref{fig:vel_vs_t}).}
    \label{tab:veff}
\end{table*}

As can be clearly seen, the effective CR transport velocity corresponds quite consistently to the typical values of the dynamical velocity in the temperature regime where the CR scattering coefficient is maximized. Physically, this corresponds to a statement that the effective CR transport speed is set by the dynamical transport velocities at the interface between low and high scattering regimes, which occurs around 20,000~K.\

In light of this, we can replace $v_\mathrm{eff}$ in \autoref{eq:Pdisk} with $v_\mathrm{tot, WIM}$, the typical velocity in the extraplanar warm ionized medium (WIM; $T \approx 20,000$~K): 
\begin{equation}\label{eq:final_1d_model}
    P_{\mathrm{disk}} = \frac{F_\mathrm{in}}{4 v_\mathrm{tot, WIM} + \dfrac{3}{2} N \Lambda_{\mathrm{coll}}}
\end{equation}
This formula, validated over a wide regime of galactic environmental conditions, therefore allows the accurate prediction of the CR pressure in galactic disks from purely MHD properties.\

We note that \autoref{eq:final_1d_model} was derived for GeV CRs, and since GeV CRs dominate the energy spectrum, it provides a good approximation of the total CR pressure in the disk. However, \autoref{eq:final_1d_model} is not applicable for computing the pressure of CRs in kinetic energy ranges $\gg 1$~GeV. As the scattering coefficient decreases with increasing kinetic energy, diffusive transport becomes increasingly important at higher energies, and the effective velocity exceeds the typical dynamical velocity in the WIM \citep[][]{Lucia2025}.

\subsection{Cosmic Ray Feedback Yield}\label{ssec:yield}
By surveying a wide range of galactic environments with our simulations, we are able to derive empirical relations between the star formation rate and the resultant yield of CR pressure support. We present our results in the language of the pressure regulated, feedback modulated (PRFM) theory of star formation \citep{Ostriker2022}, to facilitate direct comparisons to other pressures in the ISM that are sourced from star formation feedback. In this framework, galactic gas disks in equilibrium must satisfy vertical force balance between ISM weight and various sources of support (e.g. thermal pressure, turbulent pressure, magnetic stress). As star formation is responsible for injecting the energy that maintains individual pressures (by balancing heating with cooling and turbulent driving with dissipation), in equilibrium its
level must adjust to provide the needed pressure. 
A decrease in supporting pressures would result in an increase in star formation, thereby injecting more energy and momentum into the ISM via feedback and reestablishing equilibrium. The effect of star formation feedback in producing each type of pressure can be easily parameterized via  ``Feedback Yield'' parameters, defined as $\Upsilon_i \equiv P_i/\Sigma_\mathrm{SFR}$.\\

In this work we provide fits for the effective CR feedback yield as a function of SFR surface density ($\Sigma_\mathrm{SFR}$). We find the following:
\begin{equation}\label{eq:Ups_SFR_scaling}
    \Upsilon_\mathrm{c} = (284\pm33)~\mathrm{km~s^{-1}} \left(\frac{\Sigma_\mathrm{SFR}}{0.01~\mathrm{M_\odot~ pc^{-2}~Myr^{-1}}}\right)^{-0.23\pm0.05}
\end{equation}

The feedback yield can also be equivalently parameterized as a function of the observable ISM weight estimator $P_\mathrm{DE} = \pi G\Sigma_\mathrm{gas}^2/2+ \Sigma_\mathrm{gas}\sigma_\mathrm{eff}\sqrt{2G\rho_\mathrm{sd}}$, where $\sigma_\mathrm{eff}$ is the effective ISM dispersion velocity, $\rho_\mathrm{sd}$ is the sum of stellar and dark matter densities, and $\Sigma_\mathrm{gas}$ is the gas surface density. This formulation yields:
\begin{equation}\label{eq:Ups_P_scaling}
    \Upsilon_\mathrm{c} = (426\pm42) ~\mathrm{km~s^{-1}} \left(\frac{P_\mathrm{DE}/k_B}{10^4~\mathrm{K ~cm^{-3}}}\right)^{-0.29\pm0.04}
\end{equation}
Compared to other pressure components \citep[see][Eq. 24-25]{Ostriker2022}, feedback is similarly efficient in driving CR pressure as in producing turbulent and magnetic pressure, but the yield for CR pressure exceeds that found for thermal pressure. The CR pressure amounts to $\sim 40$\% of the combined pressure from turbulent, magnetic, and thermal terms, with similar dependence on galactic conditions.\

In terms of environmental dependence, CRs again fall in the middle, being significantly more sensitive to $\Sigma_\mathrm{SFR}$ (or $P_\mathrm{DE}$) than turbulent support, but noticeably less sensitive than thermal pressure support. Interestingly, the exponent of $\sim -0.2$ is approximately the same as the environmental dependence of the total feedback yield.

We note that in equilibrium it is the \textit{difference} of CR pressure between the midplane and the top of the gas disk that contributes to counter-balancing the ISM weight. While CR pressure near the disk midplane is as large as other pressure components (or equivalently, the CR feedback yield is comparable to other feedback yields), the contribution from the CR pressure difference in balancing the ISM weight is greatly suppressed if the scale height of the CR pressure is significantly larger than that of the gas. \autoref{fig:allpress} shows that the CR pressure profiles are indeed quite flat near the midplane (especially, for R8 models) and have larger scale heights than the gas in the extraplanar region. The flat CR pressure profile in the midplane region is due to the very low scattering rate in the neutral gas that makes up the majority of the ISM mass. Our expectation is therefore that CR pressure support does not significantly contribute to vertical dynamical equilibrium and hence to star formation regulation. However, a direct test of this, as well as a quantitative assessment of the role of CRs in extraplanar ISM dyanmics, requires dynamically coupled CR-MHD simulations. We defer these investigations to our future work. 

In \autoref{eq:Pdisk} and \autoref{eq:final_1d_model}, the ratio of the loss term to the transport term in the denominator is a factor $\sim0.05-0.20$ for GeV CRs, for the TIGRESS simulation environments considered in the current work.  Since the transport term is dominant, this means that the scaling with $\Sigma_\mathrm{SFR}$ or $P_\mathrm{DE}$ in \autoref{eq:Ups_SFR_scaling} or  \autoref{eq:Ups_P_scaling} represents the inverse of the scaling of $v_\mathrm{eff} \approx v_\mathrm{tot,WIM}$ with these environmental properties.  At different CR energies, $\Lambda_\mathrm{coll}$ would vary, while the transport velocity would be essentially unchanged.  It is therefore useful to directly report fits for velocity,
\begin{equation}
 v_\mathrm{eff}=  21\pm 3 \ \mathrm{km\ s^{-1} }\left(\frac{\Sigma_\mathrm{SFR}}{0.01~\mathrm{M_\odot~ pc^{-2}~Myr^{-1}}}\right)^{0.20\pm0.05}  
\end{equation}
and 
\begin{equation}
 v_\mathrm{eff} = 15\pm2\ \mathrm{km\ s^{-1} }\left(\frac{P_\mathrm{DE}/k_B}{10^4~\mathrm{K ~cm^{-3}}}\right)^{0.26\pm0.05}.
\end{equation}
As might be anticipated, the scaling with SFR and ISM pressure is very similar to that reported for warm+cold gas outflow velocities $v_\mathrm{out}$ in \citet[][see ``cool'' results in Table 5 and Figure 11a there]{Kim2020a}, which vary as $v_\mathrm{out}\propto \Sigma_\mathrm{SFR}^{0.23}$ and $v_\mathrm{out}\propto {\cal W}^{0.27}$ for gas weight ${\cal W} \approx P_\mathrm{DE}$.  

Further, it should be noted that these fits apply to CRs with energies of approximately 1 GeV. As such, the total CR pressure would be somewhat lower than implied by $\Upsilon_\mathrm{c}\Sigma_\mathrm{SFR}$, as increasingly significant diffusive contributions will increase $v_\mathrm{eff}$ above $v_\mathrm{tot,WIM}$ at higher CR energies. To compute the total correction factor $f_\mathrm{corr} \equiv P_\mathrm{c,tot}/P_\mathrm{c,GeV}$ one must consider the change in $v_\mathrm{eff}$ at all energies by integrating over the CR spectrum.  Omitting the (negligible for protons) loss term in \autoref{eq:Pdisk}, the correction factor would be 
\begin{equation}\label{eq:fcorr}
    f_\mathrm{corr}=\frac{\int E(p) ~p^{-\gamma_\mathrm{inj}} ~v_\mathrm{eff}(p)^{-1} ~p^2dp}{\int E(p) ~p^{-\gamma_\mathrm{inj}} ~v_\mathrm{eff}(\mathrm{1~GeV})^{-1}~p^2dp}
\end{equation}
where $\gamma_\mathrm{inj}=4.3$ is the power-law slope of the injected CR spectrum.  
Theoretically, we expect $v_\mathrm{eff}(p)$ to be given by  Eq. 21 of \citealt{Lucia2025}, 
\begin{equation}\label{eq:veff}
    v_\mathrm{eff}(p) = v_\mathrm{tot,WIM} \left(\frac{5}{8} + \frac{3}{8}\sqrt{1+\frac{16}{9}\frac{\kappa_\parallel(p)}{H_a v_\mathrm{tot,WIM}}}\right),
\end{equation}
where $H_\mathrm{a}$ is the scale length for vertical acceleration of the gas and $\kappa_\parallel = 1/\sigma_\parallel$ $\propto~p^{(\gamma-3)/2}\propto p^{(2/3)\gamma_\mathrm{inj} -2}$ is the energy-dependent diffusion coefficient in the regime dominated by nonlinear Landau damping. For 
$\kappa_\parallel(p)/(H_a v_\mathrm{tot,WIM})\ll 1$, $v_\mathrm{eff} \rightarrow v_\mathrm{tot,WIM}$. At energies $\sim$ 1 GeV, $v_\mathrm{eff}/v_\mathrm{tot,WIM} \sim 1$ (as we have seen),  but the ratio becomes significantly higher at higher CR energies since $\kappa \propto p^{0.9}$ . The correction is somewhat environmentally dependent, so we provide predicted corrections for each model. To do so, values for  $H_\mathrm{a}$ and $v_\mathrm{tot,WIM}$ are drawn from fits to the simulations, and $\kappa_\parallel$ is normalized by the average value found in the WIM (see \autoref{fig:sigmavstemp}). We use $\gamma = 4/3~\gamma_\mathrm{inj} -1$ (see \citealt{Lucia2025}), though the results are not particularly sensitive to this choice. We predict $f_\mathrm{corr} = 0.82,~0.82,~0.68,~0.69$ for R2, R4, R8, and R8 Arm, respectively. These corrections are modest, but nonetheless notable; we observe, as expected, that the models corresponding to the inner galaxy (with higher scattering and stronger outflows) have smaller corrections for diffusive transport. The accuracy of our predictions can be validated by future multi-environment multi-energy simulation.

\section{Lateral CR Transport}\label{sec:lattrans}

While the analysis presented in \S\ref{sec:Transport} and \S\ref{sec:1Dmodel} focuses on the mechanisms driving large-scale vertical CR transport, this section shifts attention to the mechanisms governing small-scale lateral ($xy$) CR transport. Of particular interest is the lateral transport within the disk region, as our simplified transport model for deriving the midplane CR pressure, based on environmental properties (\autoref{eq:final_1d_model}), tacitly assumes a uniform CR pressure across the disk. While this is empirically observed in our simulations, it warrants a degree of explanation, as CR injection is performed by a relatively small number of discrete cluster particles and is thus highly stochastic. A uniform CR pressure distribution therefore requires highly efficient transport throughout the disk. Although CR diffusion is highly effective in the cold gas of the disk, it is still predominantly constrained along magnetic field lines, meaning that the uniform CR distribution cannot be justified by the efficiency of field-aligned diffusion alone.\

\begin{figure*}
    \centering
    \includegraphics[width=1.0\textwidth]{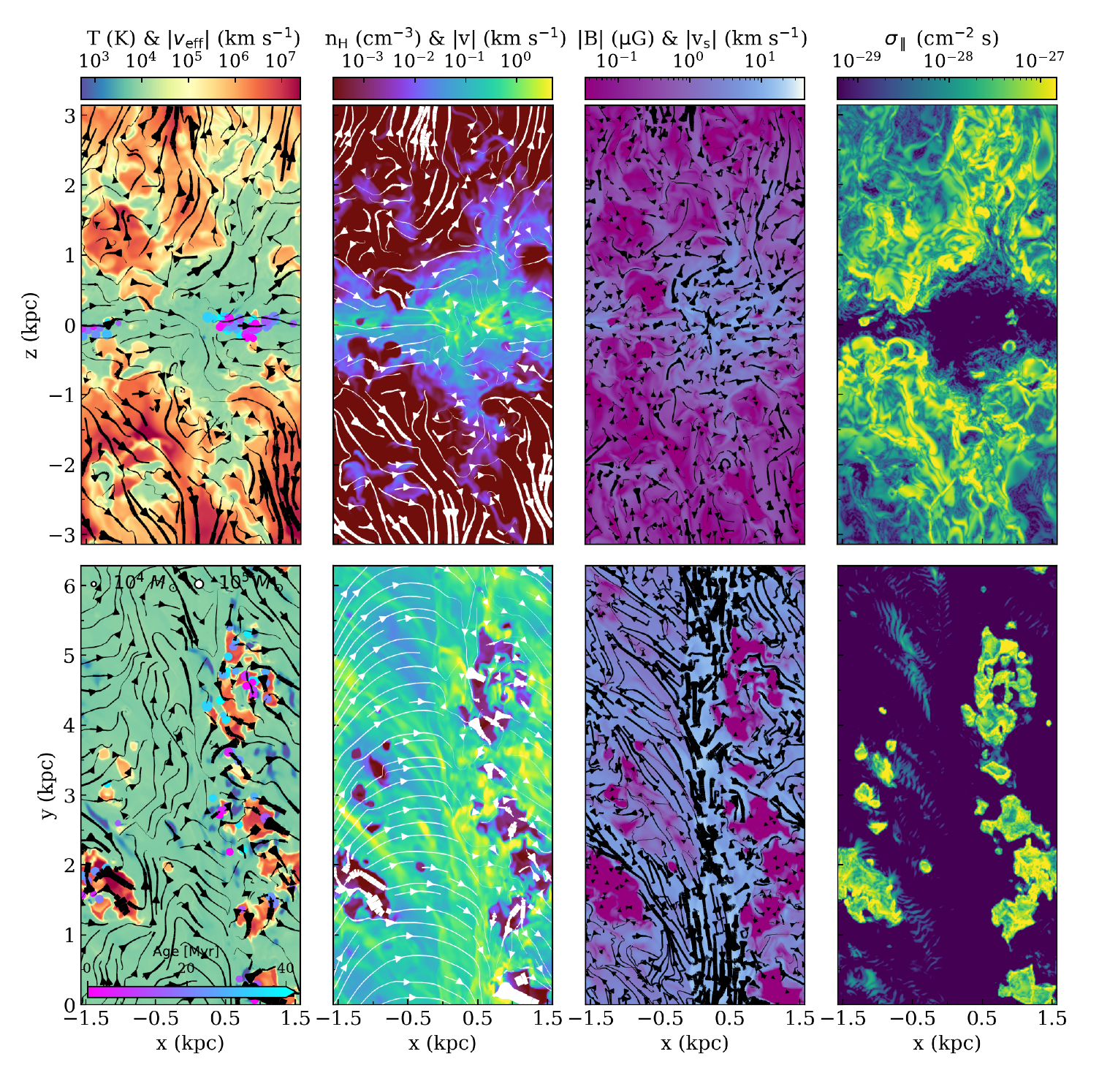}
    \caption{Slices from a snapshot of the R8 Arm model. The top/bottom row represents vertical/horizontal slices through the center of the simulation box. The color maps represent scalar quantities: from left to right, these are gas temperature (T), hydrogen number density ($n_\mathrm{H}$), magnetic field strength (B), and scattering coefficient ($\sigma_{\parallel}$). The superimposed streamlines show vectored quantities, with line width denoting magnitude. From left to right, these are the effective flux-transport speed ($\mathbf{v}_\mathrm{eff} \equiv F_\mathrm{c}/4P_\mathrm{c}$), the gas velocity ($\mathbf{v}$), and the streaming velocity ($\mathbf{v}_s$).   For clarity, the $y$-component of the arm rotation has been subtracted from both $\mathbf{v}_\mathrm{eff}$ and $\mathbf{v}$.}
    \label{fig:lateral}
\end{figure*}

To address this question, in \autoref{fig:lateral}, we show horizontal and vertical slices of a variety of parameters critical to the moderation of lateral transport, taken from a snapshot of the R8 Arm simulation. Scalar quantities are represented by the color mesh: they include, from left to right, gas temperature (T), hydrogen number density ($n_\mathrm{H}$), magnetic field strength ($|\mathbf{B}|$), and scattering coefficient ($\sigma_{\parallel}$). The superimposed streamlines show vector quantities, with line width denoting magnitude. These vector quantities are, from left to right, the effective flux-transport speed, $\mathbf{v_\mathrm{eff}} \equiv \mathbf{F_\mathrm{c}}/(4P_\mathrm{c})$, which represents an effective CR transport velocity implied by the local CR flux and pressure, the gas velocity ($\mathbf{v}$), and the streaming velocity ($\mathbf{v_s}$).
Note that $\mathbf{v_\mathrm{eff}}$ is the locally computed vector field denoting effective CR transport velocities, whereas $v_\mathrm{eff}$, referenced extensively in preceding sections, is the temporally and spatially averaged effective transport speed in the vertical direction.
For clarity, the mean $y$-component of the background flow ($R_0(\Omega-\Omega_p)$) has been subtracted from both $\mathbf{v}_\mathrm{eff}$ and $\mathbf{v}$; this operation allows easier study of more physically interesting motions. It is worth emphasizing that the streaming velocity, equal in magnitude to $v_{A,i}$, is physically meaningful for CR transport only in regions that have high enough scattering rate that diffusion is negligible.  Thus, in any region that has sufficiently small $\sigma_\parallel$, indicative of high diffusion, CRs will not actually stream at $v_s$.  \

The top row of \autoref{fig:lateral}, which shows vertical slices of the aforementioned quantities, displays many of the characteristic trends discussed in \S \ref{sec:Transport} and \S\ref{sec:1Dmodel}. Specifically, the CR flux, and therefore $\mathbf{v}_\mathrm{eff}$ in the extraplanar region is vertically oriented, with higher velocity in the hot gas, and lower velocity in the warm gas. Further, by comparison to the 2nd panel, it is evident that  $\mathbf{v}_\mathrm{eff}$ at high altitudes is closely aligned with both the strength and direction of the advective speed. Likewise, comparison to the 3rd panel shows that the streaming motion also makes substantial contributions to the total transport, especially where the advective motion is weaker. 
For the ionized extraplanar gas, we also find that $|\Delta \mathbf{v}|/|\mathbf{v}_\mathrm{eff}| \equiv |\mathbf{v_\mathrm{eff}} - \mathbf{v} - \mathbf{v_s}|/|\mathbf{v}_\mathrm{eff}|$ is small (median value of 0.1, with 89\% of the gas below unity), meaning that the gas is dominated by a combination of advection and streaming. 

Moving to the bottom row of \autoref{fig:lateral}, which shows horizontal slices through the midplane, we observe that within the hot SN bubbles where CRs are injected, the effective CR velocity follows the direction of the gas advection velocity, being oriented outward from the bubbles as they expand into the cooler surrounding medium. The 2nd panel clearly shows that the rotation of the arm first sweeps upstream gas into the arm and then sweeps it downstream from the trailing edge injection regions. Therefore, once CRs propagate from their injection sites into the arm, the gas flow (in the arm frame) advects them downstream into the interarm region. Immediately downstream from the spiral arm, the streamlines of $\mathbf{v}_\mathrm{eff}$ are similar to those for $\mathbf{v}$. 

As observed for \autoref{fig:slice}, the arm features a strong, aligned magnetic field along its longitudinal axis; aligned field lines are also evident in the ``spurs'' connecting the arm to the interarm region. The alignment between field lines and the arm and spurs is evident in orientation of $\mathbf{v}_\mathrm{s}$ in the 3rd column of the bottom row of \autoref{fig:lateral}. 
Due to the low fractional ionization and high magnetic field strength in the dense gas of the midplane, the magnitude of the streaming speed can be quite large.  
We note, however, that, in some locations, the streaming speed direction along these coherent field lines appears somewhat random. This is due to the streaming direction being defined relative to the CR gradient: in the disk, where CR pressure gradients are near zero, small gradient fluctuations can lead to direction reversal. This is especially visible in the arm itself, where the small gradients result in frequent reversals. In contrast, the bulk streaming flow remains overall well defined in the ``spurs'', where it is observed to flow in net towards the interarm region. In the injection regions, the magnetic field is weak and highly disordered, enabling CRs to follow the chaotic field lines to the injection region interface (transported to these interfaces by efficient advection). Once outside of these injection regions, the presence of strong and ordered magnetic structures in the arm's ``spurs'' allows streaming to carry CRs from the arm along field lines into the interarm region. Because the scattering rate is low in both the arm and interarm neutral gas in the midplane (rightmost panel in bottom row of \autoref{fig:lateral}), $|v_\mathrm{eff}|$ is not limited to the sum of $v_{A,i}$ and $v$, and indeed we find that the median value of $|\Delta \mathbf{v}|/|\mathbf{v}_\mathrm{eff}|$ exceeds unity. 

From the 2nd and 3rd columns of the bottom row of \autoref{fig:lateral}, it is noteworthy that the direction of streaming (and diffusion) of CRs upstream from arms is in many places near-perpendicular to the direction of the advection velocity, since magnetic field lines tend to align with the dense gas filaments that make up interarm spurs, with gas motions perpendicular to the spurs. 
Since CRs can move in either direction along magnetic field lines, and the gas velocity is nearly transverse to field lines in the region upstream from the arm, the resulting directions of $\mathbf{v}_\mathrm{eff}$ shown in the lower left panel of \autoref{fig:lateral} are nearly random. The directionally incoherent lateral CR flows evident here explain why CR energy density in the disk midplane is so uniform (\autoref{fig:slice}). 

We have confirmed that diffusion perpendicular to the magnetic field lines makes a negligible contribution to lateral CR transport. To check this, we post-processed the R8 Arm snapshots by setting the perpendicular scattering coefficient to an extremely high value, effectively rendering perpendicular diffusion negligible. These post-processed snapshots show only minor differences in the planar CR distribution compared to the default snapshots, with no alteration of our qualitative findings.\

The R-models lack the bulk advective motion of the Arm model and the strong, highly ordered magnetic field of the arm itself, making the visual picture less obvious. Nevertheless, the R-models still display superbubble regions with low-magnitude, randomly oriented magnetic fields, and strong advection velocities. The CRs are efficiently advected to the edges of the injection regions by these flows, whereupon they ``join'' the stronger and more ordered magnetic fields of the disk's cold/warm medium, where they are transported by streaming and parallel diffusion. In this sense the R-model results mirror those of the Arm model; however, the lack of bulk advective motion makes this process more ``patchy'' and thus results in a somewhat less uniform midplane CR distribution (despite uniformity within the cold/warm medium).\

It must be noted that the question of lateral transport is far more sensitive to small-scale variation than the vertical transport, owing to the far greater preponderance of sharp midplane ISM phase boundaries and the inherently 2D nature of horizontal transport. Our current analysis shows that the ``MHD back-reaction'' at phase interfaces tends to enhance transport across these interfaces, creating more tenable paths for CRs to reach pressure equilibrium between the hot injection regions and warm disk. However, strong shocks driven by active supernova events might tend to vitiate this to some extent.  For this reason, it will be important to further explore lateral transport in future simulations where CR transport, MHD dynamics, and SN events (depositing both thermal and CR energy) are simultaneously evolved over long periods.    

Overall, we find that advection is able to efficiently transport CRs out of the high-scattering, poorly magnetized SN injection regions to the interface with the CNM/WNM. From there, low-scattering and more ordered fields allow 
parallel transport to equilibrate CR pressure throughout the rest of the midplane. The net result is the efficient ability to equalize the CR energy density through the disk midplane, without the need to rely on strong perpendicular diffusion from highly tangled field lines. From there, as we have seen in previous sections, the CRs can leak into the high-scattering extraplanar regions, where dynamical mechanisms govern their vertical transport. \\

\section{Subgrid Model for Local CR Transport}\label{sec:subgridmodel}
In keeping with previous results \citep{Lucia2021,Lucia2022,Lucia2024}, we find that the efficiency of CR scattering can vary drastically depending on local gas properties and environmental conditions. Specifically, the CR scattering coefficient changes by more than 4 orders of magnitude within any given simulation,  depending on the local gas temperature (and density), and by up to an order of magnitude between equivalent temperature regimes in different galactic environments. The primary reason for this variation is the much stronger damping in primarily neutral warm/cool gas that in hotter ionized gas.  Furthermore, in this work, we have extensively discussed the critical role played by the increase in scattering in the fountain region (which coincides with the disk/halo interface in real galaxies) in controlling large-scale vertical CR transport and the resulting spatial distribution: CR pressure transitions from being nearly uniform in the dense, mostly neutral disk (where scattering coefficients are low) to exponentially decreasing in the low-density, mostly ionized extraplanar region (where scattering coefficients are relatively high). The sharp increase in the scattering coefficient effectively confines CRs within the disk, with the velocity at which CRs escape the disk primarily determined by the advection and streaming speeds in the temperature regimes of maximum scattering ($T \approx 10^4$~K). 

In light of the above findings, accurate simulations of CRs in the multiphase galactic ISM require a more realistic treatment of CR scattering than is afforded by models which adopt a single value for the CR scattering coefficient. Nevertheless, detailed transport simulations of the kind described in this work are not computationally feasible for all applications. In this section, we aim to present a practical parameterization of the CR scattering coefficient that captures the most critical variations and is valid across a wide range of regimes. The ultimate goal is to provide a straightforward subgrid model that offers a more accurate representation of CR scattering without requiring significant additional computational expense when compared to single-value models.\

Inserting typical ISM values for gas and CRs 
in  \autoref{eq:signll} and \autoref{eq:sigin} for the CR scattering rates 
yields
\begin{equation}\label{eq:signll2}
    \sigma_\mathrm{\parallel,nll} \sim 5.8 \times 10^{-28} \mathrm{\frac{s}{cm^2}} ~|\hat{\mathbf{B}}\cdot\nabla P_c|_{-4}^{1/2} ~ c_\mathrm{s,1}^{-1/2}(x_i n_\mathrm{H,-3})^{-1/4}
\end{equation}
in the well-ionized, low-density regime of NLL damping primacy, and 
\begin{equation}\label{eq:sigin2}
        \sigma_\mathrm{\parallel,in} \sim 3.4 \times 10^{-31} \mathrm{\frac{s}{cm^2}} ~|\hat{\mathbf{B}}\cdot\nabla P_c|_{-4} ~ x_\mathrm{i,-2}^{-1/2} n_\mathrm{H,0}^{-3/2}
\end{equation}
in the primarily neutral, denser gas where IN damping is dominant.
Here, $|\hat{\mathbf{B}}\cdot\nabla P_c|_{-4} = |\hat{\mathbf{B}}\cdot\nabla P_c|/(10^{-4}\mathrm{eV~cm}^{-3}\mathrm{pc}^{-1})$,  $c_\mathrm{s,1} = c_\mathrm{s}/(10~\mathrm{km~s}^{-1})$, and $n_\mathrm{H,-3} = n_\mathrm{H}/(10^{-3} \mathrm{cm}^{-3})$, $x_\mathrm{i,-2} = x_\mathrm{i}/10^{-2}$, and $n_\mathrm{H,0} = n_\mathrm{H}/(1 ~\mathrm{cm}^{-3})$.

The chief challenge in directly applying these formulae in MHD simulations lies in the term $|\hat{\mathbf{B}}\cdot\nabla P_c|$, which depends on both local variations of the CR distribution and magnetic field topology. Accurate computation of this term requires high spatial resolution and realistic representation of ISM physics and CR transport. Local-disk simulations of the type described in this work are well-suited for this purpose.\ 

In the following, we use the outcomes of our simulations to simplify Equations \ref{eq:signll2} and \ref{eq:sigin2}. Specifically, we employ the results from the MHD-relaxed simulations, as they provide a more realistic magnetic field topology and corresponding CR pressure gradients along the magnetic field lines. Further, as this section does not focus on gas velocities, the principal drawbacks that prevented the use of post-MHD relaxation results in previous section are no longer of concern.\

\begin{figure*}
    \subfloat{\includegraphics[width=0.5\textwidth]{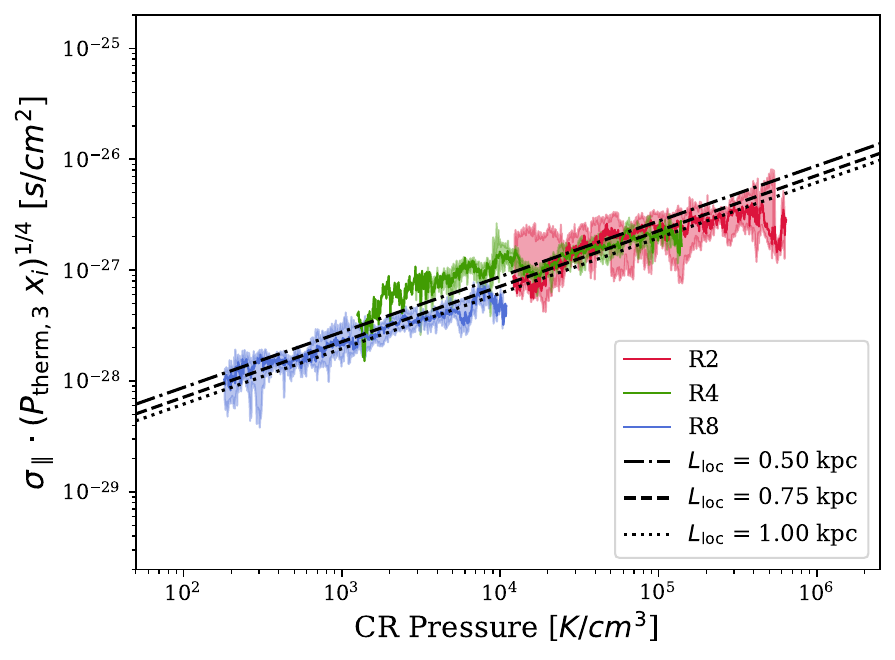}}
        \subfloat{\includegraphics[width=0.5\textwidth]{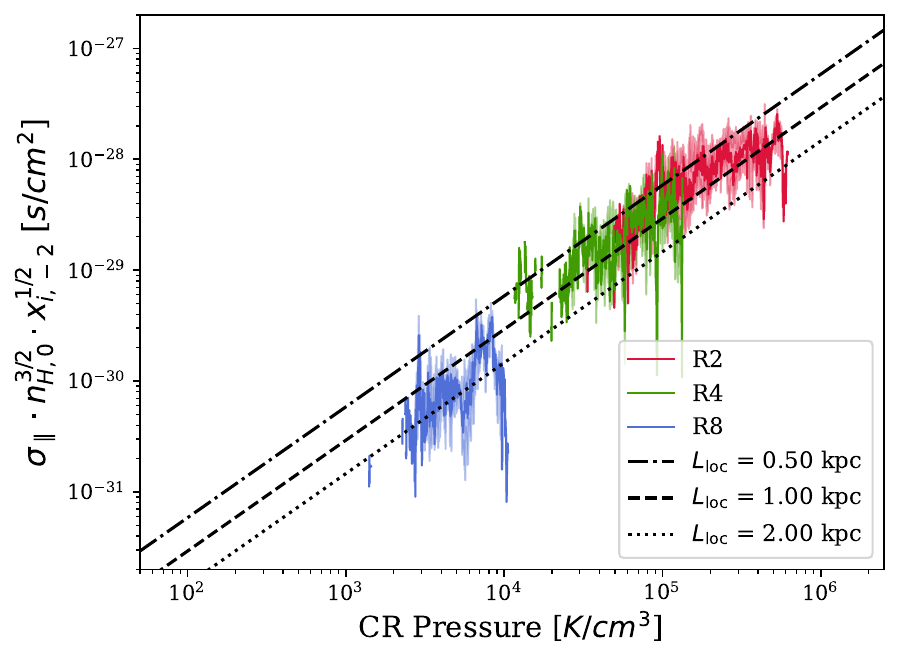}}\hfill
    \caption{Spatial and temporal median values of $\sigma_\parallel$ (normalized by a specific combination of gas parameters motivated by the form of Equations \ref{eq:signll2} and \ref{eq:sigin2}) as a function of local CR pressure. The left panel shows results for gas in the NLL regime and right panel for gas in the IN regime. The black lines show predictions from our analytic prescriptions for each regime, for a range of input scale heights.}
    \label{fig:sigvspc}
\end{figure*}

We approximate $|\hat{\mathbf{B}}\cdot\nabla P_c|$ in Equations \ref{eq:signll2} and \ref{eq:sigin2} as $P_c/L_\mathrm{loc}$ where $L_\mathrm{loc}$ is the local scale length of CR pressure in the direction of the magnetic field. In principle, this value can (and does) vary spatially, and as such, we then investigate whether there exists a representative value (or a simple parameterization) of $L_\mathrm{loc}$ in terms of other local quantities that can be substituted into Equations \ref{eq:signll2} and \ref{eq:sigin2}. The results are shown in \autoref{fig:sigvspc}, displaying spatially and temporally averaged profiles of $\sigma_\parallel$ (normalized by a specific combination of gas parameters motivated by the form of Equations \ref{eq:signll2} and  \ref{eq:sigin2}) as a function of local CR pressure.\

The left panel is limited to the warm/hot ionized gas -- where NLL is the dominant damping mechanism -- which we operationally define as gas with ionization fraction $x_\mathrm{i} > 0.5$ and temperature $8000 < \mathrm{T} < 5\times10^4$~K. It is immediately obvious that the normalized scattering coefficients for all three R-models fall upon a power law $\propto P_c^{0.5}$ (indicated by the back dashed lines), consistent with the dependence of $\sigma_\parallel$ on $P_\mathrm{c}$ in \autoref{eq:signll2}. This behavior is also observed for gas with $\mathrm{T} > 5\times10^4$~K (albeit with greater scatter); however, we do not include this hotter gas in the analysis, as in that temperature regime transport is dominated by advection, and variations in the scattering coefficient are not particularly impactful.\  

Notably, all three models appear to share nearly the same constant normalization for this power-law relation. As the gas dependencies in \autoref{eq:signll2} have been rolled into the normalization of the $y$-axis, this constant power-law normalization implies not only that $L_\mathrm{loc}$ is approximately constant within each model, but also across them. If $L_\mathrm{loc}$ varied significantly in different pressure regimes (either within a single model or between them), the different segments of power-law slope would be misaligned. 

Although there are slight offsets among the three models, the characteristic length scale $L_\mathrm{loc}$ is clearly quite similar. 
The black lines show that this characteristic value lies within $0.5 \lesssim L_\mathrm{loc}\lesssim 1$ kpc, corresponding to the typical values of the vertical scale height of CRs $H_\mathrm{c}$ (see \autoref{fig:pcfits} and \autoref{tab:veff}).\

The right panel of \autoref{fig:sigvspc} displays the relation between the normalized scattering coefficient and the CR pressure for the cold/warm neutral ISM, defined as gas with $x_\mathrm{i} < 0.5$ and $\mathrm{T} < 8000$~K. Because this phase occupies only a small portion of the volume, the number of cells containing this gas is limited, resulting in significantly noisier relations compared to the left panel. Nevertheless, these relations follow a shared power law (as indicated by the dashed black lines), consistent with the linear dependence of $\sigma_\parallel$ on $P_c$ in the IN damping regime (see \autoref{eq:sigin2}). Unsurprisingly, given that the neutral ISM is characterized by very low CR pressure gradients (and thus long scale lengths), the $\sigma_\parallel - P_\mathrm{c}$ relation is reproduced by larger values of $L_\mathrm{loc}$  compared to the ionized ISM, with $0.5 < L_\mathrm{loc}  < 2$ kpc. Given that the absolute magnitude of the scattering coefficient is universally low in the neutral ISM, for subgrid modeling purposes, the key properties of CR transport and distribution in this regime (i.e. relatively free streaming and uniform pressure distributions) can be successfully captured even if the value adopted for $L_\mathrm{loc}$ differs by a factor of a few from the true value.

In essence, we have shown that replacing $|\hat{\mathbf{B}}\cdot\nabla P_\mathrm{c}|$ with $P_\mathrm{c}/L_\mathrm{loc}$ in Equations \ref{eq:signll2} and \ref{eq:sigin2} provides a good approximation of $\sigma_\parallel$. Further, we have found that the local CR pressure scale length, $L_\mathrm{loc}$, 
is nearly constant across a wide range of physical regimes, and its value is consistent with the bulk CR pressure scale height $H_\mathrm{c}$. In light of these conclusions, we revise Equations \ref{eq:signll2} and \ref{eq:sigin2} as follows:

\begin{equation}\label{eq:signllfinal}
    \sigma_{\parallel, \mathrm{nll}} \sim 1.96 \times 10^{-28}\mathrm{\frac{s}{cm^2}}~ \sqrt{\frac{P_\mathrm{c,3}}{H_\mathrm{c,3}}}({P_\mathrm{th,3}   ~x_i})^{-1/4}
\end{equation}
and 
\begin{equation}\label{eq:siginfinal}
    \sigma_{\parallel, \mathrm{in}} \sim 2.93 \times 10^{-31}\mathrm{\frac{s}{cm^2}}~ \frac{P_\mathrm{c,3}}{H_\mathrm{c,3}}(   x_{i,-2}~n_\mathrm{H,0}^3)^{-1/2}.
\end{equation}
Here, $H_{c,3}=H_c/\mathrm{kpc}$, $P_\mathrm{c,3}=P_c/(10^3 \mathrm{K \ cm^{-3}})$ and 
$P_\mathrm{th,3}=P_\mathrm{th}/(10^3 \mathrm{K \ cm^{-3}})$. \ 

It is worth noting also that because CR transport is dominated by advection and streaming in the extraplanar region, $H_c$ is very similar to the acceleration scale there (see \S\ref{sec:1Dmodel}). As such, it is possible to accurately parameterize the local CR scattering coefficient purely as a function of MHD gas characteristics and the local CR energy density. This offers the ability for simulators to capture the primary qualitative features of CR transport without the need for a full self-consistent simulation framework. \

In conclusion, Equations \ref{eq:signllfinal} and  \ref{eq:siginfinal} provide a compelling and practical alternative to single-value scattering treatments. By use of these two formulae as a subgrid prescription (with a sharp transition between the NLL and IN regime at $\sim$8,000K), the leading-order impacts of the multiphase ISM on CR transport can be included, without significant increases in computational expense.

\section{Conclusions}\label{sec:Conclusions}
In this work, we investigate the dominant mechanisms that control galactic-scale transport of CRs with $\sim$GeV energy in the dynamic, multiphase ISM. To do so, we apply the CR transport scheme previously developed in \citet[][]{Lucia2021}, and \citet[][]{Lucia2024} to a diverse range of TIGRESS MHD simulations. The simulations presented in this work consider a number of environments covering conditions prevailing in massive disk galaxies similar to the Milky Way at a range of radii, including a case with prominent spiral structure.  
The environments studied span several orders of magnitude in pressure, density, and SFR.
Our simulations differ from many others in that scattering rates are set by local conditions, based on  a balance between wave excitation and damping, which leads to very large differences between scattering rates in gas at the midplane vs. coronal regions. Additionally, high spatial resolution in our local shearing boxes leads to realistic distributions of the gas velocities and magnetic fields that are crucial in controlling CR transport.\  

Our principal findings are as follows:
\begin{enumerate}
    \item In all environments, we observe approximate (within a factor of a few) energy equipartition between the CRs and the thermal and turbulent pressures (\autoref{fig:allpress}), in agreement with observation. Further, we find that the ratio of CR to MHD pressures decreases slightly with increasing SFR, validating the earlier findings of \citet[][]{Lucia2022} with a more self-consistent methodology, over a expanded range of environments. See \S \ref{ssec:pressuredist} for further details.
    
    \item The CR pressure $P_\mathrm{c}$ is found to be near-uniform within the primarily neutral gas near the disk midplane, declining exponentially in 
    the extraplanar region, where CR scattering increases by several orders of magnitude (\autoref{fig:slice}). \autoref{fig:sigmavstemp} demonstrates this transition to be the result of the transition at $T\sim 10^4$K in the primary wave damping mechanism. In the cooler, more neutral gas of the disk midplane, ion-neutral collisions strongly suppress small scale Alfv\'en waves which are necessary for CR confinement; in the hotter, more ionized gas of the extraplanar region, the  nonlinear Landau effect produces less efficient wave damping. See \S \ref{ssec:PPmeth} for details of CR scattering and damping, and \S\ref{sec:Transport} for elaboration on CR transport characteristics. These findings reinforce the paramount importance of accurately modeling the complex structure of the multiphase ISM for realistic CR transport.
    
    \item Measured CR scale heights in the extraplanar regions are remarkably similar across the entire diverse range of environments probed -- in all cases falling between 0.5 and 1 kpc (see \autoref{fig:pcfits}). We show that this behavior cannot be explained by CR diffusion and demonstrate instead that a dynamical transport mechanism in extraplanar regions is able to explain the observed CR 
    scale heights. We further show that a predicted analytic relation (\autoref{eq:analyticPc}) between the CR pressure and dynamical velocities (gas and ion Afv\'en speeds) indeed holds in our simulations, and that the predicted relationship between CR scale length and acceleration scale, $H_\mathrm{c} = 4/3 H_\mathrm{a}$, agrees well with measurements in the simulations (see \autoref{fig:pc_vs_v}).

    \item Combining our findings of uniform CR pressure in the disk with our results regarding dynamically moderated extraplanar transport, we develop a simple analytic prediction for CR pressure in the disk (\S\ref{ssec:1DmodelPdisk}). 
    In this model (see \autoref{eq:final_1d_model}), the disk pressure of $\sim$ GeV CRs depends on the CR flux injected by supernovae, collisional losses dependent on the gas column density, and an effective speed at which CRs are transported out of the disk. This effective CR transport speed is demonstrated to agree with the total dynamical transport speed $v_\mathrm{tot}$ as measured in the WIM, which forms the boundary layer between disk and hot extraplanar region. 
    
    \item Leveraging the diverse breadth of environments simulated in this study, we are able to provide fits to the CR feedback yield $\Upsilon_\textrm{c}\equiv P_\textrm{c}/\Sigma_{\rm SFR}$ as a function of environment (either as SFR surface density, $\Sigma_\mathrm{SFR}$, or the observable ISM weight estimator, $P_\mathrm{DE} = \pi G\Sigma_\mathrm{gas}^2/2+ \Sigma_\mathrm{gas}\sigma_\mathrm{eff}\sqrt{2G\rho_\mathrm{sd}}$). Although the CR feedback yield is comparable to turbulent and magnetic feedback yields (consistent with overall energy equipartition), the uniform CR pressure within the gas scale height makes its contribution to the vertical support negligible. See \S\ref{ssec:yield} for the results of these fits.
    
    \item We further investigate lateral transport within the disk. We find that advection is responsible for carrying CRs out of high-scattering, poorly-magnetized SN injection regions to the CNM/WNM, where low scattering enables CRs to uniformly fill the neutral gas via streaming and diffusion parallel to the magnetic field, without the need for perpendicular diffusion. In galaxies with strong spiral structure, magnetic field lines (compressed in interarm spurs) are often nearly perpendicular to large-scale gas flows, leading to incoherent and patchy CR flows in the midplane region.
    \item We utilize results from our simulations to develop simple prescriptions for CR scattering rates in multiphase gas. \autoref{eq:signllfinal} and 
    \autoref{eq:siginfinal} provide 
    analytic formulae for the CR scattering coefficient in NLL and IN regimes, solely as functions of local MHD values and the CR pressure. This model offers a prescription that captures the key influences of ISM variation on CR transport, and can be applied in simulations (as an alternative to constant values) without significant computational expense. 
\end{enumerate}
In short, we find that while efficient field-aligned diffusion ensures CR homogeneity within the cooler, more neutral gas of the galactic disk, extraplanar regions have much higher ionization fractions and, therefore, higher scattering.  The transport of GeV CRs out of the disk and through the extraplanar region is thus principally controlled by the high-altitude advection and Alfv\'en speeds, i.e. by dynamical rather than diffusive mechanisms. Both the CR energy density within the disk and the profile of CRs in the extraplanar regions can be predicted with simple analytic models. We emphasize that the present conclusions regarding transport of CRs applies to protons with energy close to 1 GeV.  In separate work, we consider transport of both protons and electrons over a wider range of energies \citep{Lucia2025,Nora2025}.

\section*{Acknowledgments}
We thank Nora Linzer, Mila Winter, and Nick Choustikov for useful and insightful discussions. We are grateful to the referee for a useful and constructive report.
This work was supported in part by grant 510940 from the Simons Foundation to E. C. Ostriker, and in part by grant AST-2407119 from the National Science Foundation. L.~A.\ was supported in part by the INAF Astrophysical fellowship initiative.
The work of C.-G.K. was in part supported by NASA ATP grant No. 80NSSC22K0717.
Data analysis was greatly benefited by the use of the open-access software packages \textsc{YT} \citep[][]{Turk2025}, \textsc{NumPy} \citep[][]{Harris2020}, and \textsc{Matplotlib} \citep[][]{Hunter2007}.

\bibliographystyle{aasjournalv7}
\bibliography{bibmain}

\begin{thebibliography}{}
\expandafter\ifx\csname natexlab\endcsname\relax\def\natexlab#1{#1}\fi
\providecommand{\url}[1]{\href{#1}{#1}}
\providecommand{\dodoi}[1]{doi:~\href{http://doi.org/#1}{\nolinkurl{#1}}}
\providecommand{\doeprint}[1]{\href{http://ascl.net/#1}{\nolinkurl{http://ascl.net/#1}}}
\providecommand{\doarXiv}[1]{\href{https://arxiv.org/abs/#1}{\nolinkurl{https://arxiv.org/abs/#1}}}

\bibitem[{M. {Ackermann} {et~al.}(2014){Ackermann}, {Ajello}, {Albert}, {Baldini}, {Ballet}, {Barbiellini}, {Bastieri}, {Bellazzini}, {Bissaldi}, {Blandford}, {Bloom}, {Bottacini}, {Brandt}, {Bregeon}, {Bruel}, {Buehler}, {Buson}, {Caliandro}, {Cameron}, {Caragiulo}, {Caraveo}, {Cavazzuti}, {Charles}, {Chekhtman}, {Cheung}, {Chiang}, {Chiaro}, {Ciprini}, {Claus}, {Cohen-Tanugi}, {Conrad}, {Corbel}, {D'Ammando}, {de Angelis}, {den Hartog}, {de Palma}, {Dermer}, {Desiante}, {Digel}, {Di Venere}, {do Couto e Silva}, {Donato}, {Drell}, {Drlica-Wagner}, {Favuzzi}, {Ferrara}, {Focke}, {Franckowiak}, {Fuhrmann}, {Fukazawa}, {Fusco}, {Gargano}, {Gasparrini}, {Germani}, {Giglietto}, {Giordano}, {Giroletti}, {Glanzman}, {Godfrey}, {Grenier}, {Grove}, {Guiriec}, {Hadasch}, {Harding}, {Hayashida}, {Hays}, {Hewitt}, {Hill}, {Hou}, {Jean}, {Jogler}, {J{\'o}hannesson}, {Johnson}, {Johnson}, {Kerr}, {Kn{\"o}dlseder}, {Kuss}, {Larsson}, {Latronico}, {Lemoine-Goumard}, {Longo}, {Loparco}, {Lott}, {Lovellette}, {Lubrano},
  {Manfreda}, {Martin}, {Massaro}, {Mayer}, {Mazziotta}, {McEnery}, {Michelson}, {Mitthumsiri}, {Mizuno}, {Monzani}, {Morselli}, {Moskalenko}, {Murgia}, {Nemmen}, {Nuss}, {Ohsugi}, {Omodei}, {Orienti}, {Orlando}, {Ormes}, {Paneque}, {Panetta}, {Perkins}, {Pesce-Rollins}, {Piron}, {Pivato}, {Porter}, {Rain{\`o}}, {Rando}, {Razzano}, {Razzaque}, {Reimer}, {Reimer}, {Reposeur}, {Saz Parkinson}, {Schaal}, {Schulz}, {Sgr{\`o}}, {Siskind}, {Spandre}, {Spinelli}, {Stawarz}, {Suson}, {Takahashi}, {Tanaka}, {Thayer}, {Thayer}, {Thompson}, {Tibaldo}, {Tinivella}, {Torres}, {Tosti}, {Troja}, {Uchiyama}, {Vianello}, {Winer}, {Wolff}, {Wood}, {Wood}, {Wood}, {Charbonnel}, {Corbet}, {De Gennaro Aquino}, {Edlin}, {Mason}, {Schwarz}, {Shore}, {Starrfield}, {Teyssier}, \& {Fermi-LAT Collaboration}}]{Ackermann2014}
{Ackermann}, M., {Ajello}, M., {Albert}, A., {et~al.} 2014, \bibinfo{title}{{Fermi establishes classical novae as a distinct class of gamma-ray sources},} Science, 345, 554, \dodoi{10.1126/science.1253947}

\bibitem[{M. {Aguilar} {et~al.}(2015){Aguilar}, {Aisa}, {Alpat}, {Alvino}, {Ambrosi}, {Andeen}, {Arruda}, {Attig}, {Azzarello}, {Bachlechner}, {Barao}, {Barrau}, {Barrin}, {Bartoloni}, {Basara}, {Battarbee}, {Battiston}, {Bazo}, {Becker}, {Behlmann}, {Beischer}, {Berdugo}, {Bertucci}, {Bindi}, {Bizzaglia}, {Bizzarri}, {Boella}, {de Boer}, {Bollweg}, {Bonnivard}, {Borgia}, {Borsini}, {Boschini}, {Bourquin}, {Burger}, {Cadoux}, {Cai}, {Capell}, {Caroff}, {Casaus}, {Castellini}, {Cernuda}, {Cerreta}, {Cervelli}, {Chae}, {Chang}, {Chen}, {Chen}, {Chen}, {Chen}, {Cheng}, {Chou}, {Choumilov}, {Choutko}, {Chung}, {Clark}, {Clavero}, {Coignet}, {Consolandi}, {Contin}, {Corti}, {Gil}, {Coste}, {Creus}, {Crispoltoni}, {Cui}, {Dai}, {Delgado}, {Della Torre}, {Demirk{\"o}z}, {Derome}, {Di Falco}, {Di Masso}, {Dimiccoli}, {D{\'\i}az}, {von Doetinchem}, {Donnini}, {Duranti}, {D'Urso}, {Egorov}, {Eline}, {Eppling}, {Eronen}, {Fan}, {Farnesini}, {Feng}, {Fiandrini}, {Fiasson}, {Finch}, {Fisher}, {Formato}, {Galaktionov},
  {Gallucci}, {Garc{\'\i}a}, {Garc{\'\i}a-L{\'o}pez}, {Gargiulo}, {Gast}, {Gebauer}, {Gervasi}, {Ghelfi}, {Giovacchini}, {Goglov}, {Gong}, {Goy}, {Grabski}, {Grandi}, {Graziani}, {Guand alini}, {Guerri}, {Guo}, {Haas}, {Habiby}, {Haino}, {Han}, {He}, {Heil}, {Hoffman}, {Hsieh}, {Huang}, {Huh}, {Incagli}, {Ionica}, {Jang}, {Jinchi}, {Kanishev}, {Kim}, {Kim}, {Kirn}, {Korkmaz}, {Kossakowski}, {Kounina}, {Kounine}, {Koutsenko}, {Krafczyk}, {La Vacca}, {Laudi}, {Laurenti}, {Lazzizzera}, {Lebedev}, {Lee}, {Lee}, {Leluc}, {Li}, {Li}, {Li}, {Li}, {Li}, {Li}, {Li}, {Li}, {Li}, {Li}, {Lim}, {Lin}, {Lipari}, {Lippert}, {Liu}, {Liu}, {Liu}, {Lolli}, {Lomtadze}, {Lu}, {Lu}, {Lu}, {Luebelsmeyer}, {Luo}, {Luo}, {Lv}, {Majka}, {Ma{\~n}{\'a}}, {Mar{\'\i}n}, {Martin}, {Mart{\'\i}nez}, {Masi}, {Maurin}, {Menchaca-Rocha}, {Meng}, {Mo}, {Morescalchi}, {Mott}, {M{\"u}ller}, {Nelson}, {Ni}, {Nikonov}, {Nozzoli}, {Nunes}, {Obermeier}, {Oliva}, {Orcinha}, {Palmonari}, {Palomares}, {Paniccia}, {Papi}, {Pauluzzi}, {Pedreschi},
  {Pensotti}, {Pereira}, {Picot-Clemente}, {Pilo}, {Piluso}, {Pizzolotto}, {Plyaskin}, {Pohl}, {Poireau}, {Putze}, {Quadrani}, {Qi}, {Qin}, {Qu}, {R{\"a}ih{\"a}}, {Rancoita}, {Rapin}, {Ricol}, {Rodr{\'\i}guez}, {Rosier-Lees}, {Rozhkov}, {Rozza}, {Sagdeev}, {Sandweiss}, {Saouter}, {Schael}, {Schmidt}, {von Dratzig}, {Schwering}, {Scolieri}, {Seo}, {Shan}, {Shan}, {Shi}, {Shi}, {Shi}, {Siedenburg}, {Son}, {Song}, {Spada}, {Spinella}, {Sun}, {Sun}, {Tacconi}, {Tang}, {Tang}, {Tang}, {Tao}, {Tescaro}, {Ting}, {Ting}, {Tomassetti}, {Torsti}, {T{\"u}rko{\v{g}}lu}, {Urban}, {Vagelli}, {Valente}, {Vannini}, {Valtonen}, {Vaurynovich}, {Vecchi}, {Velasco}, {Vialle}, {Vitale}, {Vitillo}, {Wang}, {Wang}, {Wang}, {Wang}, {Wang}, {Wang}, {Weng}, {Whitman}, {Wienkenh{\"o}ver}, {Willenbrock}, {Wu}, {Wu}, {Xia}, {Xie}, {Xie}, {Xiong}, {Xu}, {Xu}, {Yan}, {Yang}, {Yang}, {Yang}, {Ye}, {Yi}, {Yu}, {Yu}, {Zeissler}, {Zhang}, {Zhang}, {Zhang}, {Zhang}, {Zhang}, {Zhang}, {Zhang}, {Zheng}, {Zhuang}, {Zhukov}, {Zichichi},
  {Zimmermann}, {Zuccon}, \& {AMS Collaboration}}]{Aguilar+15}
{Aguilar}, M., {Aisa}, D., {Alpat}, B., {et~al.} 2015, \bibinfo{title}{{Precision Measurement of the Helium Flux in Primary Cosmic Rays of Rigidities 1.9 GV to 3 TV with the Alpha Magnetic Spectrometer on the International Space Station},} \prl, 115, 211101, \dodoi{10.1103/PhysRevLett.115.211101}

\bibitem[{E. {Amato} \& P. {Blasi}(2018){Amato} \& {Blasi}}]{Amato2018}
{Amato}, E., \& {Blasi}, P. 2018, \bibinfo{title}{{Cosmic ray transport in the Galaxy: A review},} Advances in Space Research, 62, 2731, \dodoi{10.1016/j.asr.2017.04.019}

\bibitem[{L. {Armillotta} {et~al.}(2021){Armillotta}, {Ostriker}, \& {Jiang}}]{Lucia2021}
{Armillotta}, L., {Ostriker}, E.~C., \& {Jiang}, Y.-F. 2021, \bibinfo{title}{{Cosmic-Ray Transport in Simulations of Star-forming Galactic Disks},} \apj, 922, 11, \dodoi{10.3847/1538-4357/ac1db2}

\bibitem[{L. {Armillotta} {et~al.}(2022){Armillotta}, {Ostriker}, \& {Jiang}}]{Lucia2022}
{Armillotta}, L., {Ostriker}, E.~C., \& {Jiang}, Y.-F. 2022, \bibinfo{title}{{Cosmic-Ray Transport in Varying Galactic Environments},} \apj, 929, 170, \dodoi{10.3847/1538-4357/ac5fa9}

\bibitem[{L. {Armillotta} {et~al.}(2024){Armillotta}, {Ostriker}, {Kim}, \& {Jiang}}]{Lucia2024}
{Armillotta}, L., {Ostriker}, E.~C., {Kim}, C.-G., \& {Jiang}, Y.-F. 2024, \bibinfo{title}{{Cosmic-Ray Acceleration of Galactic Outflows in Multiphase Gas},} \apj, 964, 99, \dodoi{10.3847/1538-4357/ad1e5c}

\bibitem[{L. {Armillotta} {et~al.}(2025){Armillotta}, {Ostriker}, \& {Linzer}}]{Lucia2025}
{Armillotta}, L., {Ostriker}, E.~C., \& {Linzer}, N.~B. 2025, \bibinfo{title}{{Energy-Dependent Transport of Cosmic Rays in the Multiphase, Dynamic Interstellar Medium},} arXiv e-prints, arXiv:2507.00120.
\newblock \doarXiv{2507.00120}

\bibitem[{C. {Bacchini} {et~al.}(2020){Bacchini}, {Fraternali}, {Pezzulli}, \& {Marasco}}]{Bacchini2020}
{Bacchini}, C., {Fraternali}, F., {Pezzulli}, G., \& {Marasco}, A. 2020, \bibinfo{title}{{The volumetric star formation law for nearby galaxies. Extension to dwarf galaxies and low-density regions},} \aap, 644, A125, \dodoi{10.1051/0004-6361/202038962}

\bibitem[{X.-N. {Bai}(2022){Bai}}]{Bai2022}
{Bai}, X.-N. 2022, \bibinfo{title}{{Toward First-principles Characterization of Cosmic-Ray Transport Coefficients from Multiscale Kinetic Simulations},} \apj, 928, 112, \dodoi{10.3847/1538-4357/ac56e1}

\bibitem[{X.-N. {Bai} {et~al.}(2019){Bai}, {Ostriker}, {Plotnikov}, \& {Stone}}]{Bai2019}
{Bai}, X.-N., {Ostriker}, E.~C., {Plotnikov}, I., \& {Stone}, J.~M. 2019, \bibinfo{title}{{Magnetohydrodynamic Particle-in-cell Simulations of the Cosmic-Ray Streaming Instability: Linear Growth and Quasi-linear Evolution},} \apj, 876, 60, \dodoi{10.3847/1538-4357/ab1648}

\bibitem[{C.~J. {Bambic} {et~al.}(2021){Bambic}, {Bai}, \& {Ostriker}}]{Bambic2021}
{Bambic}, C.~J., {Bai}, X.-N., \& {Ostriker}, E.~C. 2021, \bibinfo{title}{{MHD-PIC Simulations of Cosmic-Ray Scattering and Transport in Inhomogeneously Ionized Plasma},} \apj, 920, 141, \dodoi{10.3847/1538-4357/ac0ce7}

\bibitem[{R. {Beck}(2001{\natexlab{a}}){Beck}}]{Beck2001}
{Beck}, R. 2001{\natexlab{a}}, \bibinfo{title}{{Galactic and Extragalactic Magnetic Fields},} \ssr, 99, 243, \dodoi{10.1023/A:1013805401252}

\bibitem[{R. {Beck}(2001{\natexlab{b}}){Beck}}]{Beck01}
{Beck}, R. 2001{\natexlab{b}}, \bibinfo{title}{{Galactic and Extragalactic Magnetic Fields},} \ssr, 99, 243, \dodoi{10.1023/A:1013805401252}

\bibitem[{R. {Beck} \& R. {Wielebinski}(2013){Beck} \& {Wielebinski}}]{Beck2013}
{Beck}, R., \& {Wielebinski}, R. 2013, \bibinfo{title}{{Magnetic Fields in Galaxies},} in Planets, Stars and Stellar Systems. Volume 5: Galactic Structure and Stellar Populations, ed. T.~D. {Oswalt} \& G.~{Gilmore}, Vol.~5, 641, \dodoi{10.1007/978-94-007-5612-0_13}

\bibitem[{A.~R. {Bell}(2004){Bell}}]{Bell2004}
{Bell}, A.~R. 2004, \bibinfo{title}{{Turbulent amplification of magnetic field and diffusive shock acceleration of cosmic rays},} \mnras, 353, 550, \dodoi{10.1111/j.1365-2966.2004.08097.x}

\bibitem[{R.~D. {Blandford} \& J.~P. {Ostriker}(1978){Blandford} \& {Ostriker}}]{Blanford1978}
{Blandford}, R.~D., \& {Ostriker}, J.~P. 1978, \bibinfo{title}{{Particle acceleration by astrophysical shocks.},} \apjl, 221, L29, \dodoi{10.1086/182658}

\bibitem[{A.~S. {Borlaff} {et~al.}(2023){Borlaff}, {Lopez-Rodriguez}, {Beck}, {Clark}, {Ntormousi}, {Tassis}, {Martin-Alvarez}, {Tahani}, {Dale}, {del Moral-Castro}, {Roman-Duval}, {Marcum}, {Beckman}, {Subramanian}, {Eftekharzadeh}, \& {Proudfit}}]{Borlaff2023}
{Borlaff}, A.~S., {Lopez-Rodriguez}, E., {Beck}, R., {et~al.} 2023, \bibinfo{title}{{Extragalactic Magnetism with SOFIA (SALSA Legacy Program). V. First Results on the Magnetic Field Orientation of Galaxies},} \apj, 952, 4, \dodoi{10.3847/1538-4357/acd934}

\bibitem[{A. {Boulares} \& D.~P. {Cox}(1990{\natexlab{a}}){Boulares} \& {Cox}}]{Boulares1990}
{Boulares}, A., \& {Cox}, D.~P. 1990{\natexlab{a}}, \bibinfo{title}{{Galactic Hydrostatic Equilibrium with Magnetic Tension and Cosmic-Ray Diffusion},} \apj, 365, 544, \dodoi{10.1086/169509}

\bibitem[{A. {Boulares} \& D.~P. {Cox}(1990{\natexlab{b}}){Boulares} \& {Cox}}]{Boulares&Cox90}
{Boulares}, A., \& {Cox}, D.~P. 1990{\natexlab{b}}, \bibinfo{title}{{Galactic Hydrostatic Equilibrium with Magnetic Tension and Cosmic-Ray Diffusion},} \apj, 365, 544, \dodoi{10.1086/169509}

\bibitem[{T.~K. {Chan} {et~al.}(2019){Chan}, {Kere{\v{s}}}, {Hopkins}, {Quataert}, {Su}, {Hayward}, \& {Faucher-Gigu{\`e}re}}]{Chan+19}
{Chan}, T.~K., {Kere{\v{s}}}, D., {Hopkins}, P.~F., {et~al.} 2019, \bibinfo{title}{{Cosmic ray feedback in the FIRE simulations: constraining cosmic ray propagation with GeV {\ensuremath{\gamma}}-ray emission},} \mnras, 488, 3716, \dodoi{10.1093/mnras/stz1895}

\bibitem[{A.~C. {Cummings} {et~al.}(2016{\natexlab{a}}){Cummings}, {Stone}, {Heikkila}, {Lal}, {Webber}, {J{\'o}hannesson}, {Moskalenko}, {Orlando}, \& {Porter}}]{Cummings2016}
{Cummings}, A.~C., {Stone}, E.~C., {Heikkila}, B.~C., {et~al.} 2016{\natexlab{a}}, \bibinfo{title}{{Galactic Cosmic Rays in the Local Interstellar Medium: Voyager 1 Observations and Model Results},} \apj, 831, 18, \dodoi{10.3847/0004-637X/831/1/18}

\bibitem[{A.~C. {Cummings} {et~al.}(2016{\natexlab{b}}){Cummings}, {Stone}, {Heikkila}, {Lal}, {Webber}, {J{\'o}hannesson}, {Moskalenko}, {Orlando}, \& {Porter}}]{Cummings+16}
{Cummings}, A.~C., {Stone}, E.~C., {Heikkila}, B.~C., {et~al.} 2016{\natexlab{b}}, \bibinfo{title}{{Galactic Cosmic Rays in the Local Interstellar Medium: Voyager 1 Observations and Model Results},} \apj, 831, 18, \dodoi{10.3847/0004-637X/831/1/18}

\bibitem[{G. {Dashyan} \& Y. {Dubois}(2020){Dashyan} \& {Dubois}}]{Dashyan+20}
{Dashyan}, G., \& {Dubois}, Y. 2020, \bibinfo{title}{{Cosmic ray feedback from supernovae in dwarf galaxies},} \aap, 638, A123, \dodoi{10.1051/0004-6361/201936339}

\bibitem[{B.~T. {Draine}(2011){Draine}}]{Draine2011}
{Draine}, B.~T. 2011, {Physics of the Interstellar and Intergalactic Medium}

\bibitem[{C. {Evoli} {et~al.}(2018){Evoli}, {Blasi}, {Morlino}, \& {Aloisio}}]{Evoli+18}
{Evoli}, C., {Blasi}, P., {Morlino}, G., \& {Aloisio}, R. 2018, \bibinfo{title}{{Origin of the Cosmic Ray Galactic Halo Driven by Advected Turbulence and Self-Generated Waves},} \prl, 121, 021102, \dodoi{10.1103/PhysRevLett.121.021102}

\bibitem[{I.~A. {Grenier} {et~al.}(2015){Grenier}, {Black}, \& {Strong}}]{Grenier2015}
{Grenier}, I.~A., {Black}, J.~H., \& {Strong}, A.~W. 2015, \bibinfo{title}{{The Nine Lives of Cosmic Rays in Galaxies},} \araa, 53, 199, \dodoi{10.1146/annurev-astro-082214-122457}

\bibitem[{M. {Hanasz} {et~al.}(2021){Hanasz}, {Strong}, \& {Girichidis}}]{Hanasz2021}
{Hanasz}, M., {Strong}, A.~W., \& {Girichidis}, P. 2021, \bibinfo{title}{{Simulations of cosmic ray propagation},} Living Reviews in Computational Astrophysics, 7, 2, \dodoi{10.1007/s41115-021-00011-1}

\bibitem[{C.~R. Harris {et~al.}(2020)Harris, Millman, van~der Walt, Gommers, Virtanen, Cournapeau, Wieser, Taylor, Berg, Smith, Kern, Picus, Hoyer, van Kerkwijk, Brett, Haldane, del R{\'{i}}o, Wiebe, Peterson, G{\'{e}}rard-Marchant, Sheppard, Reddy, Weckesser, Abbasi, Gohlke, \& Oliphant}]{Harris2020}
Harris, C.~R., Millman, K.~J., van~der Walt, S.~J., {et~al.} 2020, \bibinfo{title}{Array programming with {NumPy},} Nature, 585, 357, \dodoi{10.1038/s41586-020-2649-2}

\bibitem[{P.~F. {Hopkins} {et~al.}(2022){Hopkins}, {Squire}, {Butsky}, \& {Ji}}]{Hopkins+22}
{Hopkins}, P.~F., {Squire}, J., {Butsky}, I.~S., \& {Ji}, S. 2022, \bibinfo{title}{{Standard self-confinement and extrinsic turbulence models for cosmic ray transport are fundamentally incompatible with observations},} \mnras, 517, 5413, \dodoi{10.1093/mnras/stac2909}

\bibitem[{P.~F. {Hopkins} {et~al.}(2021){Hopkins}, {Squire}, {Chan}, {Quataert}, {Ji}, {Kere{\v{s}}}, \& {Faucher-Gigu{\`e}re}}]{Hopkins+21}
{Hopkins}, P.~F., {Squire}, J., {Chan}, T.~K., {et~al.} 2021, \bibinfo{title}{{Testing physical models for cosmic ray transport coefficients on galactic scales: self-confinement and extrinsic turbulence at {\ensuremath{\sim}}GeV energies},} \mnras, 501, 4184, \dodoi{10.1093/mnras/staa3691}

\bibitem[{J.~D. Hunter(2007)Hunter}]{Hunter2007}
Hunter, J.~D. 2007, \bibinfo{title}{Matplotlib: A 2D graphics environment,} Computing in Science \& Engineering, 9, 90, \dodoi{10.1109/MCSE.2007.55}

\bibitem[{S. {Ji} {et~al.}(2020){Ji}, {Chan}, {Hummels}, {Hopkins}, {Stern}, {Kere{\v{s}}}, {Quataert}, {Faucher-Gigu{\`e}re}, \& {Murray}}]{Ji2020}
{Ji}, S., {Chan}, T.~K., {Hummels}, C.~B., {et~al.} 2020, \bibinfo{title}{{Properties of the circumgalactic medium in cosmic ray-dominated galaxy haloes},} \mnras, 496, 4221, \dodoi{10.1093/mnras/staa1849}

\bibitem[{Y.-F. {Jiang} \& S.~P. {Oh}(2018){Jiang} \& {Oh}}]{Jiang2018}
{Jiang}, Y.-F., \& {Oh}, S.~P. 2018, \bibinfo{title}{{A New Numerical Scheme for Cosmic-Ray Transport},} \apj, 854, 5, \dodoi{10.3847/1538-4357/aaa6ce}

\bibitem[{P. {Kempski} \& E. {Quataert}(2020){Kempski} \& {Quataert}}]{Kempski2020}
{Kempski}, P., \& {Quataert}, E. 2020, \bibinfo{title}{{Thermal instability of halo gas heated by streaming cosmic rays},} \mnras, 493, 1801, \dodoi{10.1093/mnras/staa385}

\bibitem[{C.-G. {Kim} \& E.~C. {Ostriker}(2017){Kim} \& {Ostriker}}]{Kim2017}
{Kim}, C.-G., \& {Ostriker}, E.~C. 2017, \bibinfo{title}{{Three-phase Interstellar Medium in Galaxies Resolving Evolution with Star Formation and Supernova Feedback (TIGRESS): Algorithms, Fiducial Model, and Convergence},} \apj, 846, 133, \dodoi{10.3847/1538-4357/aa8599}

\bibitem[{C.-G. {Kim} {et~al.}(2020){Kim}, {Ostriker}, {Somerville}, {Bryan}, {Fielding}, {Forbes}, {Hayward}, {Hernquist}, \& {Pandya}}]{Kim2020a}
{Kim}, C.-G., {Ostriker}, E.~C., {Somerville}, R.~S., {et~al.} 2020, \bibinfo{title}{{First Results from SMAUG: Characterization of Multiphase Galactic Outflows from a Suite of Local Star-forming Galactic Disk Simulations},} \apj, 900, 61, \dodoi{10.3847/1538-4357/aba962}

\bibitem[{C.-G. {Kim} {et~al.}(2024){Kim}, {Ostriker}, {Kim}, {Gong}, {Bryan}, {Fielding}, {Hassan}, {Ho}, {Jeffreson}, {Somerville}, \& {Steinwandel}}]{KimCG2024}
{Kim}, C.-G., {Ostriker}, E.~C., {Kim}, J.-G., {et~al.} 2024, \bibinfo{title}{{Metallicity Dependence of Pressure-regulated Feedback-modulated Star Formation in the TIGRESS-NCR Simulation Suite},} \apj, 972, 67, \dodoi{10.3847/1538-4357/ad59ab}

\bibitem[{J.-A. {Kim} {et~al.}(2023){Kim}, {Jones}, \& {Dowell}}]{SOFIA2023}
{Kim}, J.-A., {Jones}, T.~J., \& {Dowell}, C.~D. 2023, \bibinfo{title}{{Exploring the Magnetic Field Geometry in NGC 891 with SOFIA/HAWC+},} \aj, 165, 223, \dodoi{10.3847/1538-3881/acc9c7}

\bibitem[{J.-G. {Kim} {et~al.}(2018){Kim}, {Kim}, \& {Ostriker}}]{Kim2018}
{Kim}, J.-G., {Kim}, W.-T., \& {Ostriker}, E.~C. 2018, \bibinfo{title}{{Modeling UV Radiation Feedback from Massive Stars. II. Dispersal of Star-forming Giant Molecular Clouds by Photoionization and Radiation Pressure},} \apj, 859, 68, \dodoi{10.3847/1538-4357/aabe27}

\bibitem[{W.-T. {Kim} {et~al.}(2020){Kim}, {Kim}, \& {Ostriker}}]{Kim2020b}
{Kim}, W.-T., {Kim}, C.-G., \& {Ostriker}, E.~C. 2020, \bibinfo{title}{{Local Simulations of Spiral Galaxies with the TIGRESS Framework. I. Star Formation and Arm Spurs/Feathers},} \apj, 898, 35, \dodoi{10.3847/1538-4357/ab9b87}

\bibitem[{W.-T. {Kim} {et~al.}(2014){Kim}, {Kim}, \& {Kim}}]{Kimwt2014}
{Kim}, W.-T., {Kim}, Y., \& {Kim}, J.-G. 2014, \bibinfo{title}{{Nature of the Wiggle Instability of Galactic Spiral Shocks},} \apj, 789, 68, \dodoi{10.1088/0004-637X/789/1/68}

\bibitem[{W.-T. {Kim} \& E.~C. {Ostriker}(2002){Kim} \& {Ostriker}}]{Kimwt2002}
{Kim}, W.-T., \& {Ostriker}, E.~C. 2002, \bibinfo{title}{{Formation and Fragmentation of Gaseous Spurs in Spiral Galaxies},} \apj, 570, 132, \dodoi{10.1086/339352}

\bibitem[{W.-T. {Kim} \& E.~C. {Ostriker}(2006){Kim} \& {Ostriker}}]{Kimwt2006}
{Kim}, W.-T., \& {Ostriker}, E.~C. 2006, \bibinfo{title}{{Formation of Spiral-Arm Spurs and Bound Clouds in Vertically Stratified Galactic Gas Disks},} \apj, 646, 213, \dodoi{10.1086/504677}

\bibitem[{Y. {Kim} {et~al.}(2015){Kim}, {Kim}, \& {Elmegreen}}]{Kimy2015}
{Kim}, Y., {Kim}, W.-T., \& {Elmegreen}, B.~G. 2015, \bibinfo{title}{{Wiggle Instability of Galactic Spiral Shocks: Effects of Magnetic Fields},} \apj, 809, 33, \dodoi{10.1088/0004-637X/809/1/33}

\bibitem[{H. {Koyama} \& S.-i. {Inutsuka}(2002){Koyama} \& {Inutsuka}}]{Koyama2002}
{Koyama}, H., \& {Inutsuka}, S.-i. 2002, \bibinfo{title}{{An Origin of Supersonic Motions in Interstellar Clouds},} \apjl, 564, L97, \dodoi{10.1086/338978}

\bibitem[{P. {Kroupa}(2001){Kroupa}}]{Kroupa2001}
{Kroupa}, P. 2001, \bibinfo{title}{{On the variation of the initial mass function},} \mnras, 322, 231, \dodoi{10.1046/j.1365-8711.2001.04022.x}

\bibitem[{R. {Kulsrud} \& W.~P. {Pearce}(1969){Kulsrud} \& {Pearce}}]{Kulsrud1969}
{Kulsrud}, R., \& {Pearce}, W.~P. 1969, \bibinfo{title}{{The Effect of Wave-Particle Interactions on the Propagation of Cosmic Rays},} \apj, 156, 445, \dodoi{10.1086/149981}

\bibitem[{R.~M. {Kulsrud}(2005){Kulsrud}}]{Kulsrud2005}
{Kulsrud}, R.~M. 2005, {Plasma Physics for Astrophysics}

\bibitem[{R.~M. {Kulsrud} \& C.~J. {Cesarsky}(1971){Kulsrud} \& {Cesarsky}}]{Kulsrud1971}
{Kulsrud}, R.~M., \& {Cesarsky}, C.~J. 1971, \bibinfo{title}{{The Effectiveness of Instabilities for the Confinement of High Energy Cosmic Rays in the Galactic Disk},} \aplett, 8, 189

\bibitem[{C. {Leitherer} {et~al.}(1999){Leitherer}, {Schaerer}, {Goldader}, {Delgado}, {Robert}, {Kune}, {de Mello}, {Devost}, \& {Heckman}}]{Leitherer1999}
{Leitherer}, C., {Schaerer}, D., {Goldader}, J.~D., {et~al.} 1999, \bibinfo{title}{{Starburst99: Synthesis Models for Galaxies with Active Star Formation},} \apjs, 123, 3, \dodoi{10.1086/313233}

\bibitem[{N.~B. {Linzer} {et~al.}(2025){Linzer}, {Armillotta}, {Ostriker}, \& {Quataert}}]{Nora2025}
{Linzer}, N.~B., {Armillotta}, L., {Ostriker}, E.~C., \& {Quataert}, E. 2025, \bibinfo{title}{{Modeling Cosmic Ray Electron Spectra and Synchrotron Emission in the Multiphase ISM},} arXiv e-prints, arXiv:2507.00142.
\newblock \doarXiv{2507.00142}

\bibitem[{N.~B. {Linzer} {et~al.}(2024){Linzer}, {Kim}, {Kim}, \& {Ostriker}}]{Linzer2024}
{Linzer}, N.~B., {Kim}, J.-G., {Kim}, C.-G., \& {Ostriker}, E.~C. 2024, \bibinfo{title}{{Ultraviolet Radiation Fields in Star-forming Disk Galaxies: Numerical Simulations with TIGRESS-NCR},} \apj, 975, 173, \dodoi{10.3847/1538-4357/ad7733}

\bibitem[{Y.~S. {Lu} {et~al.}(2025){Lu}, {Kere{\v{s}}}, {Hopkins}, {Ponnada}, {Faucher-Gigu{\'e}re}, \& {Hummels}}]{Lu2025}
{Lu}, Y.~S., {Kere{\v{s}}}, D., {Hopkins}, P.~F., {et~al.} 2025, \bibinfo{title}{{Constraining cosmic ray transport models using circumgalactic medium properties and observables},} arXiv e-prints, arXiv:2505.13597, \dodoi{10.48550/arXiv.2505.13597}

\bibitem[{G. {Morlino} \& D. {Caprioli}(2012){Morlino} \& {Caprioli}}]{Morlino2012}
{Morlino}, G., \& {Caprioli}, D. 2012, \bibinfo{title}{{Strong evidence for hadron acceleration in Tycho's supernova remnant},} \aap, 538, A81, \dodoi{10.1051/0004-6361/201117855}

\bibitem[{E.~C. {Ostriker} \& C.-G. {Kim}(2022){Ostriker} \& {Kim}}]{Ostriker2022}
{Ostriker}, E.~C., \& {Kim}, C.-G. 2022, \bibinfo{title}{{Pressure-regulated, Feedback-modulated Star Formation in Disk Galaxies},} \apj, 936, 137, \dodoi{10.3847/1538-4357/ac7de2}

\bibitem[{E.~C. {Ostriker} {et~al.}(2010){Ostriker}, {McKee}, \& {Leroy}}]{Ostriker2010}
{Ostriker}, E.~C., {McKee}, C.~F., \& {Leroy}, A.~K. 2010, \bibinfo{title}{{Regulation of Star Formation Rates in Multiphase Galactic Disks: A Thermal/Dynamical Equilibrium Model},} \apj, 721, 975, \dodoi{10.1088/0004-637X/721/2/975}

\bibitem[{M. {Padovani} {et~al.}(2018){Padovani}, {Ivlev}, {Galli}, \& {Caselli}}]{Padovani+18}
{Padovani}, M., {Ivlev}, A.~V., {Galli}, D., \& {Caselli}, P. 2018, \bibinfo{title}{{Cosmic-ray ionisation in circumstellar discs},} \aap, 614, A111, \dodoi{10.1051/0004-6361/201732202}

\bibitem[{M. {Padovani} {et~al.}(2020){Padovani}, {Ivlev}, {Galli}, {Offner}, {Indriolo}, {Rodgers-Lee}, {Marcowith}, {Girichidis}, {Bykov}, \& {Kruijssen}}]{Padovani2020}
{Padovani}, M., {Ivlev}, A.~V., {Galli}, D., {et~al.} 2020, \bibinfo{title}{{Impact of Low-Energy Cosmic Rays on Star Formation},} \ssr, 216, 29, \dodoi{10.1007/s11214-020-00654-1}

\bibitem[{ {Planck Collaboration} {et~al.}(2016){Planck Collaboration}, {Adam}, {Ade}, {Aghanim}, {Akrami}, {Alves}, {Arg{\"u}eso}, {Arnaud}, {Arroja}, {Ashdown}, {Aumont}, {Baccigalupi}, {Ballardini}, {Banday}, {Barreiro}, {Bartlett}, {Bartolo}, {Basak}, {Battaglia}, {Battaner}, {Battye}, {Benabed}, {Beno{\^\i}t}, {Benoit-L{\'e}vy}, {Bernard}, {Bersanelli}, {Bertincourt}, {Bielewicz}, {Bikmaev}, {Bock}, {B{\"o}hringer}, {Bonaldi}, {Bonavera}, {Bond}, {Borrill}, {Bouchet}, {Boulanger}, {Bucher}, {Burenin}, {Burigana}, {Butler}, {Calabrese}, {Cardoso}, {Carvalho}, {Casaponsa}, {Castex}, {Catalano}, {Challinor}, {Chamballu}, {Chary}, {Chiang}, {Chluba}, {Chon}, {Christensen}, {Church}, {Clemens}, {Clements}, {Colombi}, {Colombo}, {Combet}, {Comis}, {Contreras}, {Couchot}, {Coulais}, {Crill}, {Cruz}, {Curto}, {Cuttaia}, {Danese}, {Davies}, {Davis}, {de Bernardis}, {de Rosa}, {de Zotti}, {Delabrouille}, {Delouis}, {D{\'e}sert}, {Di Valentino}, {Dickinson}, {Diego}, {Dolag}, {Dole}, {Donzelli}, {Dor{\'e}},
  {Douspis}, {Ducout}, {Dunkley}, {Dupac}, {Efstathiou}, {Eisenhardt}, {Elsner}, {En{\ss}lin}, {Eriksen}, {Falgarone}, {Fantaye}, {Farhang}, {Feeney}, {Fergusson}, {Fernandez-Cobos}, {Feroz}, {Finelli}, {Florido}, {Forni}, {Frailis}, {Fraisse}, {Franceschet}, {Franceschi}, {Frejsel}, {Frolov}, {Galeotta}, {Galli}, {Ganga}, {Gauthier}, {G{\'e}nova-Santos}, {Gerbino}, {Ghosh}, {Giard}, {Giraud-H{\'e}raud}, {Giusarma}, {Gjerl{\o}w}, {Gonz{\'a}lez-Nuevo}, {G{\'o}rski}, {Grainge}, {Gratton}, {Gregorio}, {Gruppuso}, {Gudmundsson}, {Hamann}, {Handley}, {Hansen}, {Hanson}, {Harrison}, {Heavens}, {Helou}, {Henrot-Versill{\'e}}, {Hern{\'a}ndez-Monteagudo}, {Herranz}, {Hildebrandt}, {Hivon}, {Hobson}, {Holmes}, {Hornstrup}, {Hovest}, {Huang}, {Huffenberger}, {Hurier}, {Ili{\'c}}, {Jaffe}, {Jaffe}, {Jin}, {Jones}, {Juvela}, {Karakci}, {Keih{\"a}nen}, {Keskitalo}, {Khamitov}, {Kiiveri}, {Kim}, {Kisner}, {Kneissl}, {Knoche}, {Knox}, {Krachmalnicoff}, {Kunz}, {Kurki-Suonio}, {Lacasa}, {Lagache}, {L{\"a}hteenm{\"a}ki},
  {Lamarre}, {Langer}, {Lasenby}, {Lattanzi}, {Lawrence}, {Le Jeune}, {Leahy}, {Lellouch}, {Leonardi}, {Le{\'o}n-Tavares}, {Lesgourgues}, {Levrier}, {Lewis}, {Liguori}, {Lilje}, {Lilley}, {Linden-V{\o}rnle}, {Lindholm}, {Liu}, {L{\'o}pez-Caniego}, {Lubin}, {Ma}, {Mac{\'\i}as-P{\'e}rez}, {Maggio}, {Maino}, {Mak}, {Mandolesi}, {Mangilli}, {Marchini}, {Marcos-Caballero}, {Marinucci}, {Maris}, {Marshall}, {Martin}, {Martinelli}, {Mart{\'\i}nez-Gonz{\'a}lez}, {Masi}, {Matarrese}, {Mazzotta}, {McEwen}, {McGehee}, {Mei}, {Meinhold}, {Melchiorri}, {Melin}, {Mendes}, {Mennella}, {Migliaccio}, {Mikkelsen}, {Millea}, {Mitra}, {Miville-Desch{\^e}nes}, {Molinari}, {Moneti}, {Montier}, {Moreno}, {Morgante}, {Mortlock}, {Moss}, {Mottet}, {M{\"u}nchmeyer}, {Munshi}, {Murphy}, {Narimani}, {Naselsky}, {Nastasi}, {Nati}, {Natoli}, {Negrello}, {Netterfield}, {N{\o}rgaard-Nielsen}, {Noviello}, {Novikov}, {Novikov}, {Olamaie}, {Oppermann}, {Orlando}, {Oxborrow}, {Paci}, {Pagano}, {Pajot}, {Paladini}, {Pandolfi}, {Paoletti},
  {Partridge}, {Pasian}, {Patanchon}, {Pearson}, {Peel}, {Peiris}, {Pelkonen}, {Perdereau}, {Perotto}, {Perrott}, {Perrotta}, {Pettorino}, {Piacentini}, {Piat}, {Pierpaoli}, {Pietrobon}, {Plaszczynski}, {Pogosyan}, {Pointecouteau}, {Polenta}, {Popa}, {Pratt}, {Pr{\'e}zeau}, {Prunet}, {Puget}, {Rachen}, {Racine}, {Reach}, {Rebolo}, {Reinecke}, {Remazeilles}, {Renault}, {Renzi}, {Ristorcelli}, {Rocha}, {Roman}, {Romelli}, {Rosset}, {Rossetti}, {Rotti}, {Roudier}, {Rouill{\'e} d'Orfeuil}, {Rowan-Robinson}, {Rubi{\~n}o-Mart{\'\i}n}, {Ruiz-Granados}, {Rumsey}, {Rusholme}, {Said}, {Salvatelli}, {Salvati}, {Sandri}, {Sanghera}, {Santos}, {Saunders}, {Sauv{\'e}}, {Savelainen}, {Savini}, {Schaefer}, {Schammel}, {Scott}, {Seiffert}, {Serra}, {Shellard}, {Shimwell}, {Shiraishi}, {Smith}, {Souradeep}, {Spencer}, {Spinelli}, {Stanford}, {Stern}, {Stolyarov}, {Stompor}, {Strong}, {Sudiwala}, {Sunyaev}, {Sutter}, {Sutton}, {Suur-Uski}, {Sygnet}, {Tauber}, {Tavagnacco}, {Terenzi}, {Texier}, {Toffolatti}, {Tomasi},
  {Tornikoski}, {Tramonte}, {Tristram}, {Troja}, {Trombetti}, {Tucci}, {Tuovinen}, {T{\"u}rler}, {Umana}, {Valenziano}, {Valiviita}, {Van Tent}, {Vassallo}, {Vibert}, {Vidal}, {Viel}, {Vielva}, {Villa}, {Wade}, {Walter}, {Wandelt}, {Watson}, {Wehus}, {Welikala}, {Weller}, {White}, {White}, {Wilkinson}, {Yvon}, {Zacchei}, {Zibin}, \& {Zonca}}]{Planck2016}
{Planck Collaboration}, {Adam}, R., {Ade}, P.~A.~R., {et~al.} 2016, \bibinfo{title}{{Planck 2015 results. I. Overview of products and scientific results},} \aap, 594, A1, \dodoi{10.1051/0004-6361/201527101}

\bibitem[{I. {Plotnikov} {et~al.}(2021){Plotnikov}, {Ostriker}, \& {Bai}}]{Plotnikov2021}
{Plotnikov}, I., {Ostriker}, E.~C., \& {Bai}, X.-N. 2021, \bibinfo{title}{{Influence of Ion-Neutral Damping on the Cosmic-Ray Streaming Instability: Magnetohydrodynamic Particle-in-cell Simulations},} \apj, 914, 3, \dodoi{10.3847/1538-4357/abf7b3}

\bibitem[{E. {Quataert} \& P.~F. {Hopkins}(2025){Quataert} \& {Hopkins}}]{Quataert2025}
{Quataert}, E., \& {Hopkins}, P.~F. 2025, \bibinfo{title}{{Cosmic Ray Feedback in Massive Halos: Implications for the Distribution of Baryons},} The Open Journal of Astrophysics, 8, 66, \dodoi{10.33232/001c.138772}

\bibitem[{T.-E. {Rathjen} {et~al.}(2023){Rathjen}, {Naab}, {Walch}, {Seifried}, {Girichidis}, \& {W{\"u}nsch}}]{Rathjen+23}
{Rathjen}, T.-E., {Naab}, T., {Walch}, S., {et~al.} 2023, \bibinfo{title}{{SILCC - VII. Gas kinematics and multiphase outflows of the simulated ISM at high gas surface densities},} \mnras, 522, 1843, \dodoi{10.1093/mnras/stad1104}

\bibitem[{S. Recchia(2020)Recchia}]{Recchia2020}
Recchia, S. 2020, \bibinfo{title}{Cosmic ray driven galactic winds,} International Journal of Modern Physics D, 29, 2030006, \dodoi{10.1142/S0218271820300062}

\bibitem[{P. {Reichherzer} {et~al.}(2022){Reichherzer}, {Merten}, {D{\"o}rner}, {Becker Tjus}, {Pueschel}, \& {Zweibel}}]{Reichenherzer2022}
{Reichherzer}, P., {Merten}, L., {D{\"o}rner}, J., {et~al.} 2022, \bibinfo{title}{{Regimes of cosmic-ray diffusion in Galactic turbulence},} SN Applied Sciences, 4, 15, \dodoi{10.1007/s42452-021-04891-z}

\bibitem[{W.~W. {Roberts}(1969){Roberts}}]{Roberts1969}
{Roberts}, W.~W. 1969, \bibinfo{title}{{Large-Scale Shock Formation in Spiral Galaxies and its Implications on Star Formation},} \apj, 158, 123, \dodoi{10.1086/150177}

\bibitem[{M. {Ruszkowski} \& C. {Pfrommer}(2023){Ruszkowski} \& {Pfrommer}}]{Ruszkowski+23}
{Ruszkowski}, M., \& {Pfrommer}, C. 2023, \bibinfo{title}{{Cosmic ray feedback in galaxies and galaxy clusters},} \aapr, 31, 4, \dodoi{10.1007/s00159-023-00149-2}

\bibitem[{R. {Shetty} {et~al.}(2007){Shetty}, {Vogel}, {Ostriker}, \& {Teuben}}]{Shetty2007}
{Shetty}, R., {Vogel}, S.~N., {Ostriker}, E.~C., \& {Teuben}, P.~J. 2007, \bibinfo{title}{{Kinematics of Spiral-Arm Streaming in M51},} \apj, 665, 1138, \dodoi{10.1086/520037}

\bibitem[{B. {Sike} {et~al.}(2024){Sike}, {Thomas}, {Ruszkowski}, {Pfrommer}, \& {Weber}}]{Sike+24}
{Sike}, B., {Thomas}, T., {Ruszkowski}, M., {Pfrommer}, C., \& {Weber}, M. 2024, \bibinfo{title}{{Cosmic Ray-Driven Galactic Winds with Resolved ISM and Ion-Neutral Damping},} arXiv e-prints, arXiv:2410.06988, \dodoi{10.48550/arXiv.2410.06988}

\bibitem[{M.~A. {Skinner} \& E.~C. {Ostriker}(2013){Skinner} \& {Ostriker}}]{Skinner2013}
{Skinner}, M.~A., \& {Ostriker}, E.~C. 2013, \bibinfo{title}{{A Two-moment Radiation Hydrodynamics Module in Athena Using a Time-explicit Godunov Method},} \apjs, 206, 21, \dodoi{10.1088/0067-0049/206/2/21}

\bibitem[{J.~M. {Stone} {et~al.}(2008){Stone}, {Gardiner}, {Teuben}, {Hawley}, \& {Simon}}]{Stone2008}
{Stone}, J.~M., {Gardiner}, T.~A., {Teuben}, P., {Hawley}, J.~F., \& {Simon}, J.~B. 2008, \bibinfo{title}{{Athena: A New Code for Astrophysical MHD},} \apjs, 178, 137, \dodoi{10.1086/588755}

\bibitem[{J.~M. {Stone} {et~al.}(2020){Stone}, {Tomida}, {White}, \& {Felker}}]{Stone2020}
{Stone}, J.~M., {Tomida}, K., {White}, C.~J., \& {Felker}, K.~G. 2020, \bibinfo{title}{{The Athena++ Adaptive Mesh Refinement Framework: Design and Magnetohydrodynamic Solvers},} \apjs, 249, 4, \dodoi{10.3847/1538-4365/ab929b}

\bibitem[{R.~S. {Sutherland} \& M.~A. {Dopita}(1993){Sutherland} \& {Dopita}}]{Sutherland1993}
{Sutherland}, R.~S., \& {Dopita}, M.~A. 1993, \bibinfo{title}{{Cooling Functions for Low-Density Astrophysical Plasmas},} \apjs, 88, 253, \dodoi{10.1086/191823}

\bibitem[{T. {Thomas} {et~al.}(2023){Thomas}, {Pfrommer}, \& {Pakmor}}]{Thomas+23}
{Thomas}, T., {Pfrommer}, C., \& {Pakmor}, R. 2023, \bibinfo{title}{{Cosmic-ray-driven galactic winds: transport modes of cosmic rays and Alfv{\'e}n-wave dark regions},} \mnras, 521, 3023, \dodoi{10.1093/mnras/stad472}

\bibitem[{T. {Thomas} {et~al.}(2024){Thomas}, {Pfrommer}, \& {Pakmor}}]{Thomas+24}
{Thomas}, T., {Pfrommer}, C., \& {Pakmor}, R. 2024, \bibinfo{title}{{Why are thermally- and cosmic ray-driven galactic winds fundamentally different?},} arXiv e-prints, arXiv:2405.13121, \dodoi{10.48550/arXiv.2405.13121}

\bibitem[{M. Turk {et~al.}(2025)Turk, Goldbaum, ZuHone, Hummels, Ji, Lang, Munk, Smith, Kowalik, de~Val-Borro, {et~al.}}]{Turk2025}
Turk, M., Goldbaum, N.~J., ZuHone, J.~A., {et~al.} 2025, Introducing yt 4.0: Analysis and Visualization of Volumetric Data, Tech. rep., Manubot

\bibitem[{A. {Vijayan} {et~al.}(2020){Vijayan}, {Kim}, {Armillotta}, {Ostriker}, \& {Li}}]{Vijayan+20}
{Vijayan}, A., {Kim}, C.-G., {Armillotta}, L., {Ostriker}, E.~C., \& {Li}, M. 2020, \bibinfo{title}{{Kinematics and Dynamics of Multiphase Outflows in Simulations of the Star-forming Galactic Interstellar Medium},} \apj, 894, 12, \dodoi{10.3847/1538-4357/ab8474}

\bibitem[{K. {Wada} \& J. {Koda}(2004){Wada} \& {Koda}}]{Wada2004}
{Wada}, K., \& {Koda}, J. 2004, \bibinfo{title}{{Instabilities of spiral shocks - I. Onset of wiggle instability and its mechanism},} \mnras, 349, 270, \dodoi{10.1111/j.1365-2966.2004.07484.x}

\bibitem[{D.~G. {Wentzel}(1974){Wentzel}}]{Wentzel1974}
{Wentzel}, D.~G. 1974, \bibinfo{title}{{Cosmic-ray propagation in the Galaxy: collective effects.},} \araa, 12, 71, \dodoi{10.1146/annurev.aa.12.090174.000443}

\bibitem[{J. {Wiener} {et~al.}(2019){Wiener}, {Zweibel}, \& {Ruszkowski}}]{Wiener2019}
{Wiener}, J., {Zweibel}, E.~G., \& {Ruszkowski}, M. 2019, \bibinfo{title}{{Cosmic ray acceleration of cool clouds in the circumgalactic medium},} \mnras, 489, 205, \dodoi{10.1093/mnras/stz2007}

\bibitem[{H. {Yan} \& A. {Lazarian}(2002){Yan} \& {Lazarian}}]{Yan2002}
{Yan}, H., \& {Lazarian}, A. 2002, \bibinfo{title}{{Scattering of Cosmic Rays by Magnetohydrodynamic Interstellar Turbulence},} \prl, 89, 281102, \dodoi{10.1103/PhysRevLett.89.281102}

\bibitem[{E.~G. {Zweibel}(2017){Zweibel}}]{Zweibel17}
{Zweibel}, E.~G. 2017, \bibinfo{title}{{The basis for cosmic ray feedback: Written on the wind},} Physics of Plasmas, 24, 055402, \dodoi{10.1063/1.4984017}

\end{thebibliography}

\end{document}